\documentclass[12pt, notitlepage]{article}
\pdfoutput=1

\usepackage{float}
\usepackage{subfig}
\usepackage{graphicx}
\usepackage[T1]{fontenc}
\usepackage[utf8]{inputenc}
\usepackage{authblk}
\usepackage{amsmath,bm}
\usepackage[left=2cm,right=2cm,top=2cm,bottom=2cm]{geometry}
\renewcommand\footnotemark{}
\usepackage[dvipsnames]{xcolor}

\newcommand{\rf}[1]{(\ref{#1})}
\newcommand{\beq}{\begin{equation}}
\newcommand{\beql}[1]{\beq\label{#1}}
\newcommand{\eeq}{\end{equation}}
\newcommand{\bea}{\begin{eqnarray}}
\newcommand{\eea}{\end{eqnarray}}

\begin{document}

\title{Critical Phenomena in Causal Dynamical Triangulations}

\author[a,b]{J.~Ambj\o rn}
\author[a]{D.~Coumbe}
\author[c]{J.~Gizbert-Studnicki}
\author[c]{A.~G\"orlich}
\author[c]{J.~Jurkiewicz}
\affil[a]{\small{The Niels Bohr Institute, Copenhagen University, \authorcr Blegdamsvej 17, DK-2100 Copenhagen Ø, Denmark. \authorcr E-mail: ambjorn@nbi.dk, daniel.coumbe@nbi.ku.dk.\vspace{+2ex}}} 

\affil[b]{\small{IMAPP, Radboud University, \authorcr Nijmegen, PO Box 9010, The Netherlands.\vspace{+2ex}}}

\affil[c]{\small{The M. Smoluchowski Institute of Physics, Jagiellonian University, \authorcr \L ojasiewicza 11, Krak\'ow, PL 30-348, Poland. \authorcr Email: jakub.gizbert-studnicki@uj.edu.pl, andrzej.goerlich@uj.edu.pl, jerzy.jurkiewicz@uj.edu.pl.}}

\date{\small({Dated: \today})}          
\maketitle


\begin{abstract}

Four-dimensional CDT (causal dynamical triangulations) is  a 
lattice theory of geometries which one might use in an attempt to define quantum gravity non-perturbatively,
following the standard procedures of lattice field theory. Being a theory of geometries, the phase
transitions which in usual lattice field theories are used to  define the continuum limit of the lattice 
theory will in the CDT case be transitions between different types of geometries. This picture is interwoven 
with the topology of space which is kept fixed in the lattice theory, the reason being that ``classical''
geometries around which the quantum fluctuations take place depend crucially on the imposed topology.
Thus it is possible that the topology of space can influence the  phase transitions and the corresponding 
critical phenomena used to extract continuum physics. In this article we perform a systematic comparison
between a CDT phase transition where space has spherical topology and the ``same'' transition
where space has toroidal topology.  The ``classical'' geometries around which the systems fluctuate are 
very different it the two cases, but we find that the order of phase transition is not affected by this.

\vspace{1cm}

\end{abstract}

\begin{section}{Introduction}\label{intro}

Quantum gravity is the attempt to consistently describe gravity as a quantum field theory. Such a description is expected for a number of reasons, including the fact that the other fundamental interactions have all been successfully formulated as quantum field theories. However, treating general relativity, our best description of gravity, as a perturbative quantum field theory spawns a proliferation of divergences at high energies that cannot be removed using standard renormalization techniques~\cite{thooft,two-loop}. Thus, at least the simplest attempt at quantum gravity is known to fail.

Since the usual perturbative procedure does not work for gravity, a generalised nonperturbative approach has gained traction in recent years, in which couplings are not required to be small or tend to zero in the high-energy limit, but are only required to approach a finite constant. The scale dependence of couplings, as dictated by the renormalization group, must then approach a fixed point at high energies that is precisely scale-invariant, which by construction guarantees a constant limit. This nonperturbative approach to quantum gravity is known as the asymptotic safety scenario~\cite{weinberg}.
In particular, the so-called exact renormalization group approach has been to calculate an effective action of 
quantum gravity and search for non-perturbative fixed points \cite{renorm-group}.
However, the actual calculations using the exact renormalisation group always includes a truncation of the 
effective action and it makes reliably establishing the existence of a high-energy fixed point difficult. This motivates a complementary lattice approach, where the fixed point would appear as a continuum limit that can be studied with controlled systematic errors. Studying quantum gravity on the lattice thus provides a powerful complementary tool for testing the asymptotic safety scenario.

To date the most successful lattice formulation of quantum gravity is that of causal dynamical triangulations (CDT),
\footnote{For the original articles see \cite{origcdt}, for a review \cite{physrep}.} at least in the sense that it has a rich phase structure where some of the transitions might be second order 
transitions which potentially can be used to define continuum limits. Of course, even the existence of a continuum limit
does not ensure that the continuum theory is (a quantum version of) general relativity. It has to be proven, but a discussion of whether or not that is the case will not be a topic of this article\footnote{Since  CDT has a built in 
time foliation, it has been argued \cite{jordan} that a natural continuum limit could also be a version of 
Horava-Lifshitz gravity \cite{horava}.  }.

 In CDT continuous spacetime is approximated by a network of locally flat $d$-dimensional triangles subject to a causality condition, in which the lattice is foliated into space-like hypersurfaces each with a fixed topology. Using Regge calculus \cite{regge} one can write a discretised Einstein-Hilbert action in terms of bulk variables associated with the lattice regularisation for a given CDT triangulation as 

\begin{equation} \label{eq:GeneralEinstein-ReggeAction}
S_{EH}=-\left(\kappa_{0}+6\Delta\right)N_{0}+\kappa_{4}\left(N_{4,1}+N_{3,2}\right)+\Delta\left(2N_{4,1}+N_{3,2}\right),
\end{equation}

\noindent where $N_{i,j}$ denotes the number of $4$-dimensional simplicial building blocks with $i$ vertices on hypersurface $t$ and $j$ vertices on hypersurface $t+1$, and $N_0$ is the number of vertices in the triangulation. $\kappa_{0}$, $\Delta$ and $\kappa_{4}$ are bare coupling constants. $\kappa_{0}$ and $\kappa_{4}$ are related to Newton's constant and the cosmological constant, respectively, and $\Delta$ defines the ratio of the length of space-like and time-like links on the lattice. The parameter $\kappa_{4}$, which is proportional to the cosmological constant, is tuned such that one can take an infinite-volume limit.

There are two varieties of elementary building blocks in CDT, the $(4,1)$ and $(3,2)$ simplices~\cite{physrep}. The local Monte Carlo moves used in CDT simulations do not always preserve the numbers $N_{4,1}$ and $N_{3,2}$ of these building blocks, or their sum. However, the total system volume can be controlled by a volume-fixing term

\begin{equation}
\delta V = \epsilon (N_{4,1} - N_{4}^{\rm{target}})^2,
\end{equation}

\noindent such that it is sharply peaked about a chosen value, with a well-defined range of fluctuations. $N_{4}^{\rm{target}}$ is the chosen target volume about which it fluctuates, and $\epsilon$ is a numerical constant that controls the magnitude of volume fluctuations. Prior to starting numerical simulations one has the freedom to fix either the total number of $4$-dimensional simplicial building blocks $N_{4}=N_{4,1}+N_{3,2}$ or just the number of $N_{4,1}$ simplices, where the latter volume fixing is often chosen merely for technical convenience.  

In CDT simulations, the fixed topology of spatial slices and that of the proper time axis must be chosen from the outset. The vast majority of CDT Monte Carlo simulations have been performed with the spatial topology of a three-sphere $S^{3}$, and the temporal topology of $S^{1}$, yielding a global spacetime topology $\mathcal{T}=S^{3} \times S^{1}$. However, one is not only restricted to spherical topology. In fact, one is free to choose any suitable topology (see~\cite{kevin} for a recent study of CDT with toroidal topology). 

Since $\kappa_{4}$ is tuned to its critical value in the simulations, the only parameters that need to be varied in order to explore the CDT parameter space are $\kappa_{0}$ and $\Delta$. The CDT parameter space has now been studied fairly extensively, with the central features depicted in Fig.~\ref{pdnew}. Phases $A$ and $B$ have geometrical properties that make them unlikely to model our macroscopic universe, and are typically thought to be lattice artifacts. Phase $C_{b}$ is the recently discovered bifurcation phase, with a number of peculiar geometric features (see
\cite{Ambjorn:2015qja,Coumbe:2015oaa,Ambjorn:2015fza,Ambjorn:2017tnl,Ambjorn:2016mnn} for more details). Phase $C$ is the physically interesting phase of CDT, since it exhibits a semiclassical geometry that closely resembles Euclidean de Sitter space in $4$-dimensions at large distances~\cite{physrep,Coumbe:2014noa}. 

\begin{figure}[H]
\centering
\includegraphics[width=0.6\linewidth]{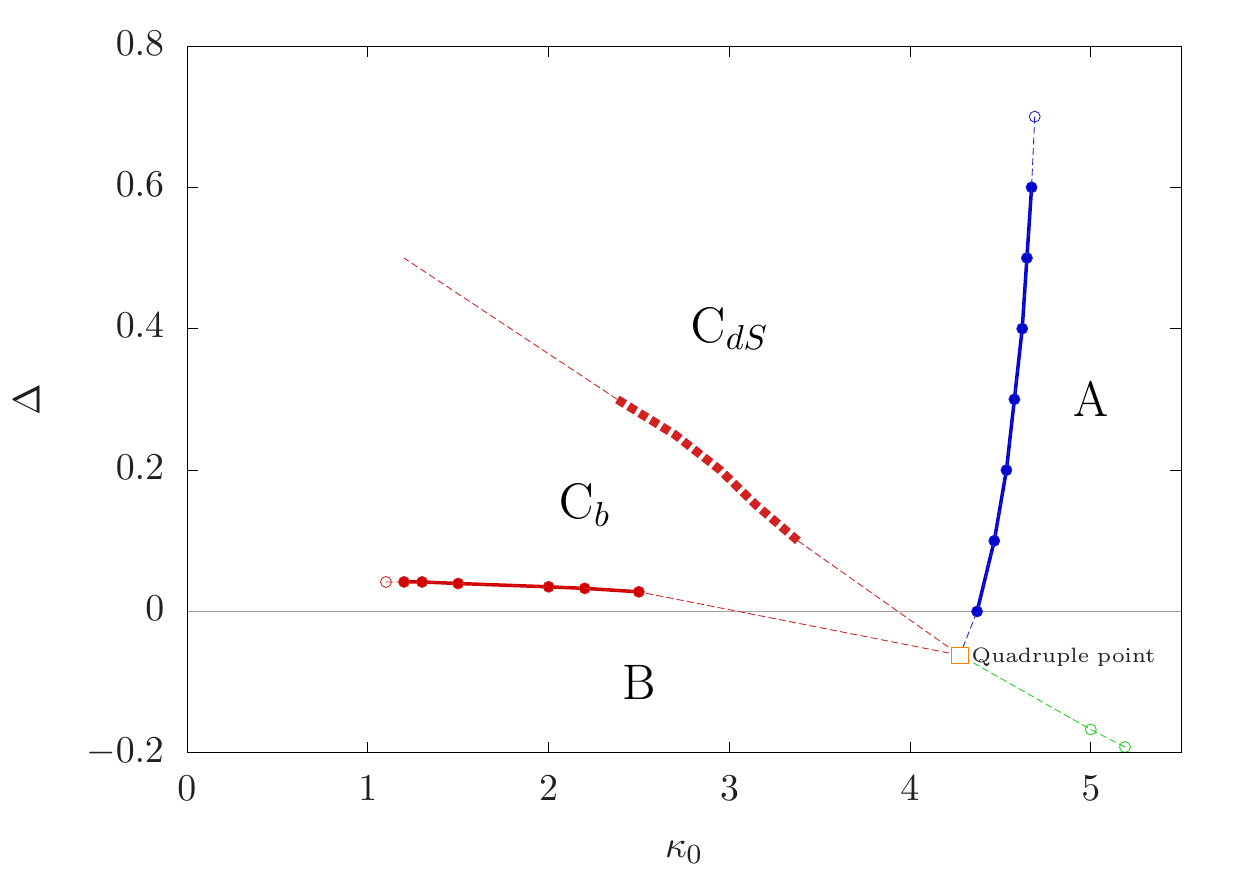}
\caption{\small The phase structure of $4$-dimensional CDT.}
\label{pdnew}
\end{figure}

There exist a number of previous studies on the phase transition lines bordering the physically interesting phase C. The transition between phases $A$ and $C$ was found to be almost certainly first order, while the $B-C_{b}$ transition likely second order~\cite{Ambjorn:2011cg,Ambjorn:2012ij}. A recent study also found the $C-C_{b}$ transition to be a likely second order transition~\cite{Ambjorn:2017tnl}. Definitively establishing the order of all transitions in the CDT parameter space is important since proving the existence of a second order transition may establish a continuum limit and facilitate contact with the asymptotic safety scenario~\cite{Ambjorn:2016cpa}. 

However, given the freedom in how the numerical CDT simulations can be setup, such as the topology of each spatial slice, the number of time slices and the particular volume-fixing method, we want to investigate the possible impact that these variables may have on CDT phase transitions. In this work, we therefore explore the potential impact topology, time slicing and volume-fixing may have on critical phenomena in CDT. Here, we restrict our attention to the $A-C$ transition only, since numerical Monte Carlo simulations run relatively fast in this part of the parameter space which makes the large data collections needed for such a study more feasible. The lessons learnt from this study are expected to aid investigations of other CDT transitions in future works. 

In order to facilitate Monte Carlo simulations near the phase transition, we tested a new numerical code aimed at decreasing autocorrelation time of the measured data for some systems. This code is based on the so-called parallel tempering/replica exchange method. The technical details of this new code will be published elsewhere.


\end{section}

\begin{section}{Order parameters and observables}\label{OrderOfTrans}

Phase transitions are usually triggered by breaking of some symmetry. The symmetry difference between the disordered and the ordered phase can  be captured by an order parameter $OP$, which is usually zero (or constant) in the disordered phase and changes (usually rises) in the ordered phase.
There is  some freedom in defining an order parameter and thus order parameters should be tailored to the nature of a given phase transition. The choice of a  "good" order parameter is especially important in numerical studies, such as Monte Carlo simulations  used to investigate  Causal Dynamical Triangulations. 

In the numerical study of four-dimensional CDT, one defines the pseudo-critical points by looking for peaks in the susceptibility 
\beql{susc}
\chi_{OP} \equiv \langle OP^2 \rangle - \langle OP \rangle^2,
\eeq
within the $(\kappa_0, \Delta)$ parameter space.  As can be seen in the phase diagram of Fig. \ref{pdnew} the $A$-$C$  transition line is almost parallel with the $\Delta$-axis, thus one usually fixes $\Delta$ and measures $\chi_{OP}$ as a function of $\kappa_0$, looking for a maximum in $\chi_{OP}$ at $\kappa_0^{crit}$ (see e.g. Fig. \ref{Fig1}).
Typically, the position of the pseudo-critical points  moves in the parameter space when the system size is increased. Therefore, in order to quantify finite size effects, one  measures systems with higher and higher lattice volume $N_{4,1}$ (or alternatively  $N_{4}$, depending on the volume fixing method). One then performs a finite size scaling analysis by fitting the power-law behaviour
\begin{equation}\label{powerlaw}
\kappa_{0}^{crit}(N_{4,1})=\kappa_{0}^{crit}(\infty) - CN_{4,1}^{-\frac{1}{\gamma}},
\end{equation}
 where $\kappa_{0}^{crit}(N_{4,1})$ is the pseudo-critical value for a finite system size $N_{4,1}$, $\kappa_{0}^{crit}(\infty)$ is the true critical value in the infinite volume limit and $C$ is a constant of proportionality. A critical exponent of $\gamma=1$ indicates a first order transition, while $\gamma > 1$ suggests a higher order transition.
 
 One can also analyse the behaviour of the order parameter $OP$ measured at the pseudo-critical point $\kappa_{0}^{crit}$. First order transitions are usually characterised by two distinct metastable states of $OP$ and thus the order parameter  jumps (in the Monte Carlo simulation time) between the two states, which can be observed as two separate peaks in the measured $OP$ histograms (see e.g. Fig. \ref{Fig2}).  For a first order transition, the probability of tunnelling between the two states should decrease when the lattice volume $N_{4,1}$ increases,  and thus the two peaks should become easier to distinguish. At some lattice size, the separation of the peaks is so large (relative to the amplitude of fluctuations within a single peak) that the jumps of the $OP$ between the metastates are suppressed such that  the system gets  frozen in one of the metastates and one observes hysteresis around the critical point. If instead, the transition is second (or higher) order then only one state is present, and thus only one  peak in the $OP$ histogram is observed at the critical point.\footnote{In some  cases, e.g. for the $B$-$C_b$ phase transition \cite{Ambjorn:2012ij, Ambjorn:2011cg}, one could observe the appearance of two metastates also for a higher order transition measured for finite lattice sizes, but the separate peaks would merge into a single peak in the infinite volume limit. The reason for this atypical behaviour was analysed in detail in \cite{ Ambjorn:2016mnn}.} The critical point is then characterised by the maximal amplitude of $OP$ fluctuations, which is captured by the maximum of the susceptibility $\chi_{OP}(\kappa_0)$ observed at $\kappa_{0}^{crit}$.

Another quantity of interest is the Binder cumulant\footnote{Note that here we use a definition of the Binder cumulant which is shifted (by a  $-2/3$ constant) versus the original Binder's formulation  \cite{BinderPhysRevLett}: $B_x=1- \frac{1}{3} \frac{\langle x^{4} \rangle}{\langle x^2\rangle^{2}}$. The definition \rf{binder} was also used in previous CDT phase transition studies \cite{Ambjorn:2012ij} and thus we keep it in order to ease comparison with these results. The virtue of using our definition is that, as explained in the text, the  deviation of (critical) $B_{OP}$ from zero with rising lattice volume may signal a first order transition, while the convergence to zero is  characteristic of a higher order transition. One could as well use the original Binder's definition and look at the deviation from $2/3$. }
\begin{equation}\label{binder}
B_{OP}\equiv\frac{1}{3}\left(1-\frac{\langle OP^{4} \rangle}{\langle OP^2\rangle^{2}}\right),
\end{equation}
which (similar to $\chi_{OP}$) can be  used to find   the position of the $A$-$C$ phase transition line. In this case,  pseudo-critical points are signalled by local minima of $B_{OP}$  in the $(\kappa_0, \Delta)$ parameter space. Again, one  defines   the $A$-$C$ pseudo-critical points  by fixing $\Delta$ and analyses $B_{OP}$ as a function of $\kappa_0$ looking for 
 minima appearing at  $\kappa_0^{crit}$. The result will again depend on lattice volume and one can use equation \rf{powerlaw} to estimate the $\kappa_0^{crit}(\infty)$  and the critical exponent $\gamma$, and it is believed that finite size effects are smaller compared to the susceptibility method discussed above. One can also look at scaling of the (minimum) level of $B_{OP}$ measured at the phase transition point: 
\beql{Bmin}
B_{OP}^{min}\equiv B_{OP}(\kappa_0^{crit}) .
\eeq
For a first order transition, where the $OP$ histograms measured at the critical point $\kappa_0^{crit}$ have two shifted peaks, the $B_{OP}^{min}(N_{4,1}\to\infty)$ will move away from zero as the two peaks become more and more apparent when the histograms  approach two shifted Dirac-delta functions placed at  positions of the peaks' centres. If instead, the transition is second (or higher) order, then the histograms have only one peak and $B_{OP}^{min}(N_{4,1}\to\infty)$ should tend to zero.

The above observables and their characteristic behaviour for  the first and the higher order transitions are summarised in Table \ref{Table1}.\footnote{One could also analyse scaling of $\chi_{OP}^{max} \equiv \chi_{OP}(\kappa_0^{crit})$, but  $\chi_{OP}$ and  $B_{OP}$ are strongly related (see e.g. Eqs. \rf{relations} and \rf{relations2}) and thus one can focus just on the latter. Another possibility is to check  the connected Binder cumulant $B^c_{OP}\equiv \frac{1}{3}\left(1-\frac{\langle (OP-\langle OP \rangle)^{4} \rangle}{\langle (OP-\langle OP \rangle)^2\rangle^{2}}\right)$.  $B_{OP}^c(N_{4,1}\to\infty)$ measured at the transition point should tend to zero for a first order transition and, as a consequence of the positive kurtosis in the OP probability distribution at the higher order transition, it should be deflected from zero for the higher order transition. Unfortunately, this quantity does not give any statistically significant signals for the data discussed.}

\begin{table}
\begin{center}
\begin{tabular} {|c|c|c|c|}
\hline
{OBSERVABLE}	& {1st order transition}	& {Higher order transition} \\ \hline
\hline
$OP$ histograms measured at	&  double peaks	& single peak or  \\ 
 pseudo-critical points $\kappa_o^{crit}$	&  peak separation $\uparrow$ with $N_{4,1}\to\infty$	&  peaks merging with $N_{4,1}\to\infty$  \\ \hline
Pseudo-critical point 	&  $\gamma$	&  $\gamma$ \\
scaling $\kappa_o^{crit}(N_{4,1})$, eq. \rf{powerlaw}	&  $=1$	& $>1$  \\ \hline
Binder cumulant \rf{binder} scaling	&   $B_{OP}^{min}(N_{4,1}\to\infty)$	&   $B_{OP}^{min}(N_{4,1}\to\infty)$ \\
at pseudo-critical points	$\kappa_o^{crit}$ &  	$ < 0 $ &  $ =0$ \\ \hline
\end{tabular}
\end{center}
\caption{\small 
Characteristics of the first and the higher order phase transitions.}
\label{Table1}
\end{table}

In a previous study of the $A$-$C$  transition in CDT with spherical spatial topology \cite{Ambjorn:2012ij} and in a recent paper about the phase structure of CDT with toroidal spatial topology \cite{Ambjorn:2018qbf}, a number of order parameters were defined and used to study the phase transition in question. They were mainly related to very global properties of  triangulations, such as the (extensive) total number of vertices  $N_0$  or the (intensive) ratios $OP_1\equiv \frac{N_0}{N_4}$ and $OP_2\equiv \frac{N_{3,2}}{N_{4,1}}$. In fact, when one couldn't observe the phase transition signal using $OP$, one could try to use  some monotonic functions of the above order parameters, e.g. $f(OP) = \sqrt{OP}$ or $f(OP) = \ln {OP}$. Such a choice was useful when  an order parameter was changing by a few orders of magnitude at the phase transition, as used in Ref. \cite{Ambjorn:2018qbf}. 

Now, we want to choose "the best" order parameters (or their monotonic functions) giving the strongest signal-to-noise ratios irrespective of the simulation details, i.e. the ones being "critical" no matter how we choose the topology, time slicing or the volume-fixing method. Specifically, we will analyse the susceptibility \rf{susc} or the Binder cumulant  \rf{binder} in search of extrema ($\chi_{OP}(\kappa_0)$  maxima  or  $B_{OP}(\kappa_0)$ minima) in the transition region. We therefore define "criticality" via the existence of such extrema. For each $OP$ considered, we first check the "criticality" of $OP$ and, if it is not observed, we also check the "criticality" of $\sqrt{OP}$ and $ \ln {(OP)}$ both for the susceptibility and the Binder cumulant. 
In fact, one may show that these quantities are strongly dependent if  the order parameter's fluctuations are relatively small compared to its mean value, which seems to be the case in all data analysed herein. In this case, the susceptibility of a function of an order parameter $f(OP)$ is very well approximated by
\beql{suscf}
\chi_{f(OP)} \approx  \big(f'(\langle OP \rangle)\big)^2 \chi_{OP} \ ,
\eeq
where $f'$ denotes the derivative of the function $f$.  For the square root and the logarithm one gets (up to a trivial rescaling\footnote{In fact $\chi_{\sqrt{OP}} \approx \frac{1}{4}\frac{\chi_{OP}} {  \langle OP \rangle} $.})
\beql{suscsqrt}
\chi_{\sqrt{OP}} \approx  \frac{\chi_{OP}} {  \langle OP \rangle} ,
\eeq
\beql{susclog}
\chi_{\ln OP} \approx \frac{\chi_{OP}} {  \langle OP \rangle^2} ,
\eeq
respectively. Therefore, from \rf{binder} and \rf{susc}, one immediately obtains
$$
B_{\sqrt{OP}}=\frac{1}{3}\left(1-\frac{\langle \sqrt{OP}^{4} \rangle}{\langle \sqrt{OP}^2\rangle^{2}}\right)=- \frac{1}{3}\left(\frac{\langle {OP}^{2} \rangle -\langle {OP}\rangle^{2} }{\langle {OP}\rangle^{2}}\right ) = - \frac{1}{3}\frac{\chi_{OP}}{\langle {OP}\rangle^{2}}\approx - \frac{1}{3}\chi_{\ln{OP}} \ ,
$$ 
and so (at least up to a rescaling)
\beql{relations} 
B_{\sqrt{OP}}  \approx - \chi_{\ln{OP}}.
\eeq
One can also empirically check that in all cases analysed the maxima of $\chi_{\sqrt{OP}}$ coincide with the minima of $B_{\ln {OP}}$, i.e. the position of the critical $\kappa_0^{crit}$ based on these extrema is identical, although the extremal values are different (cannot be matched by a trivial rescaling) - which we denote via
\beql{relations2}
\chi_{\sqrt{OP}} \sim -B_{\log{OP}} .
\eeq
In a sense the functions $\sqrt{OP}$ and $\ln{OP}$ used to compute the susceptibility and the Binder cumulant  are "conjugate" to each other, and thus it is enough  to look at one of them, e.g. $\sqrt{OP}$. By assuming small $OP$ fluctuations  (relative to $\langle OP \rangle$) and using equations \rf{relations} and \rf{suscf}  one can  easily show that
$$
B_{OP} \approx - \chi_{\ln {OP}^2} =  - \chi_{2 \ln {OP}} \approx  -4  \chi_{ \ln {OP}} \approx 4 B_{\sqrt{OP}}  \ ,
$$ 
so again (up to a rescaling)
\beql{relations3} 
B_{\sqrt{OP}}  \approx B_{{OP}} .
\eeq
Consequently, the susceptibilities and Binder cumulants of $\sqrt{OP}$ capture the  most important information encoded in our data and thus we will focus on that function if we cannot get the phase transition signal for the $OP$ itself.

The question remains, which of the proposed order parameters $N_0$, $OP_1\equiv \frac{N_0}{N_4}$ or $OP_2\equiv \frac{N_{3,2}}{N_{4,1}}$ is best suited for our purposes, i.e. which one is really "critical" independent of the topology, time slicing or the volume-fixing method. To answer this question, in each case we have analysed the behaviour of $\langle OP \rangle$, $\chi_{OP}$ and $\frac{\chi_{OP}} { \langle OP \rangle}\approx\chi_{\sqrt{OP}}$. The exemplary results for the toroidal topology with  $T=4$ proper-time period and two different volume fixing methods are presented in Fig. \ref{Fig01}, Fig. \ref{Fig02} and  Fig. \ref{Fig03}.

\begin{figure}[H]
  \centering
  \scalebox{.6}{\includegraphics{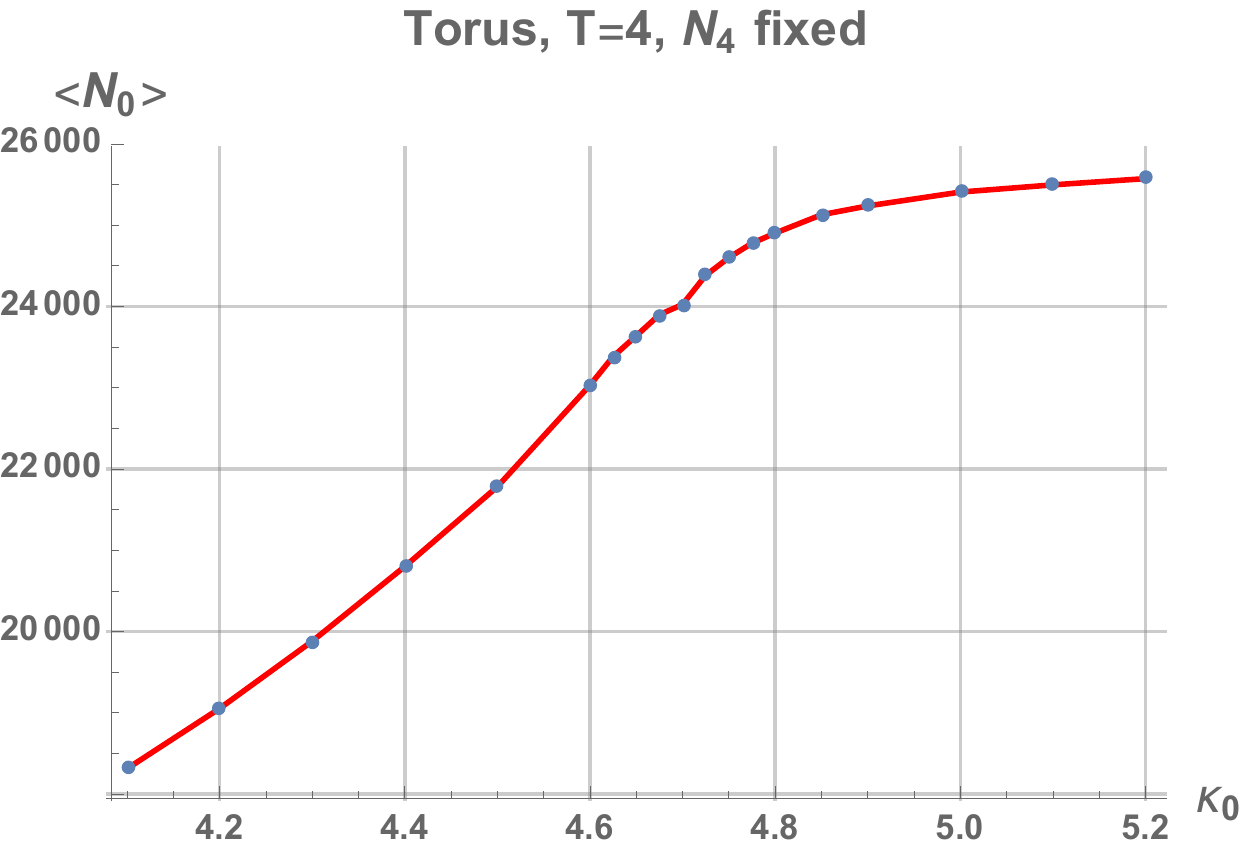}}
  \scalebox{.6}{\includegraphics{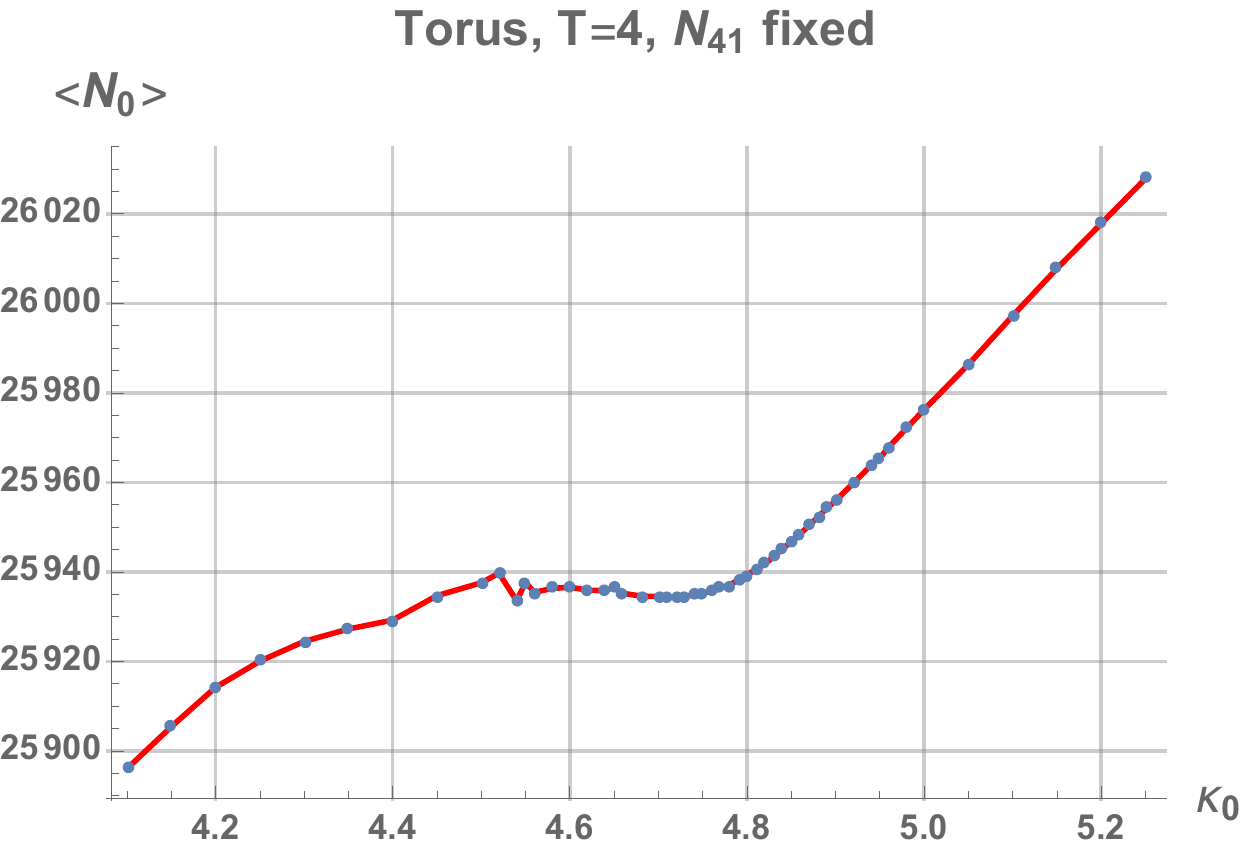}}
  \scalebox{.6}{\includegraphics{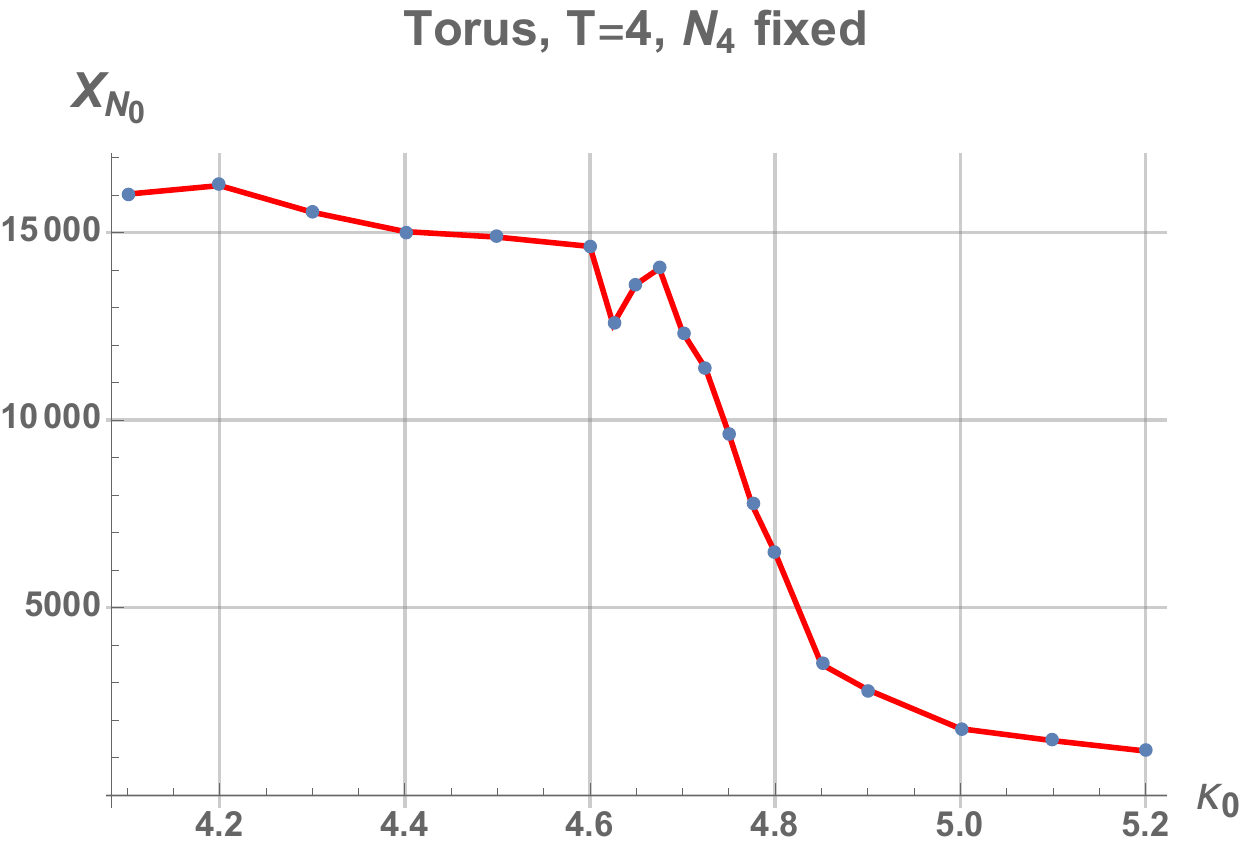}}
  \scalebox{.6}{\includegraphics{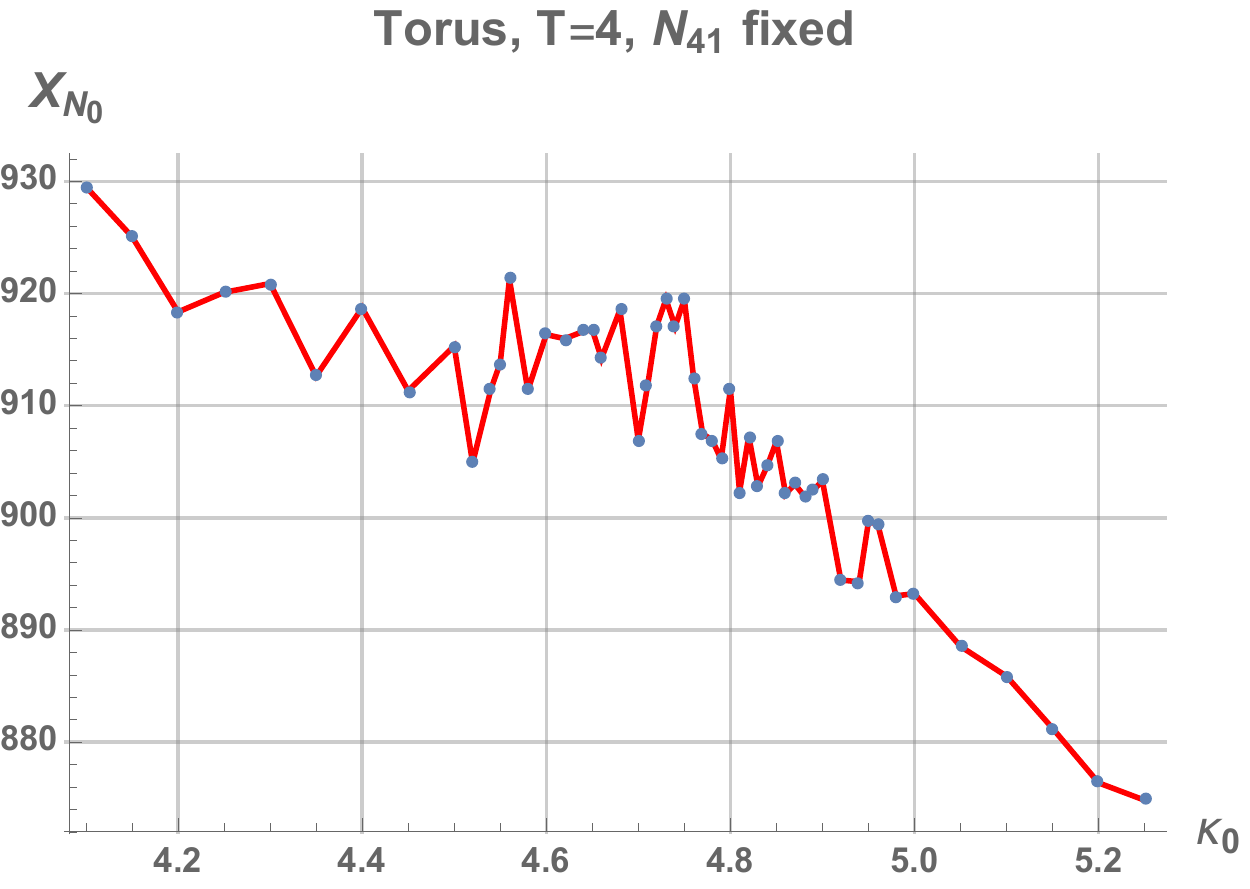}}
  \scalebox{.6}{\includegraphics{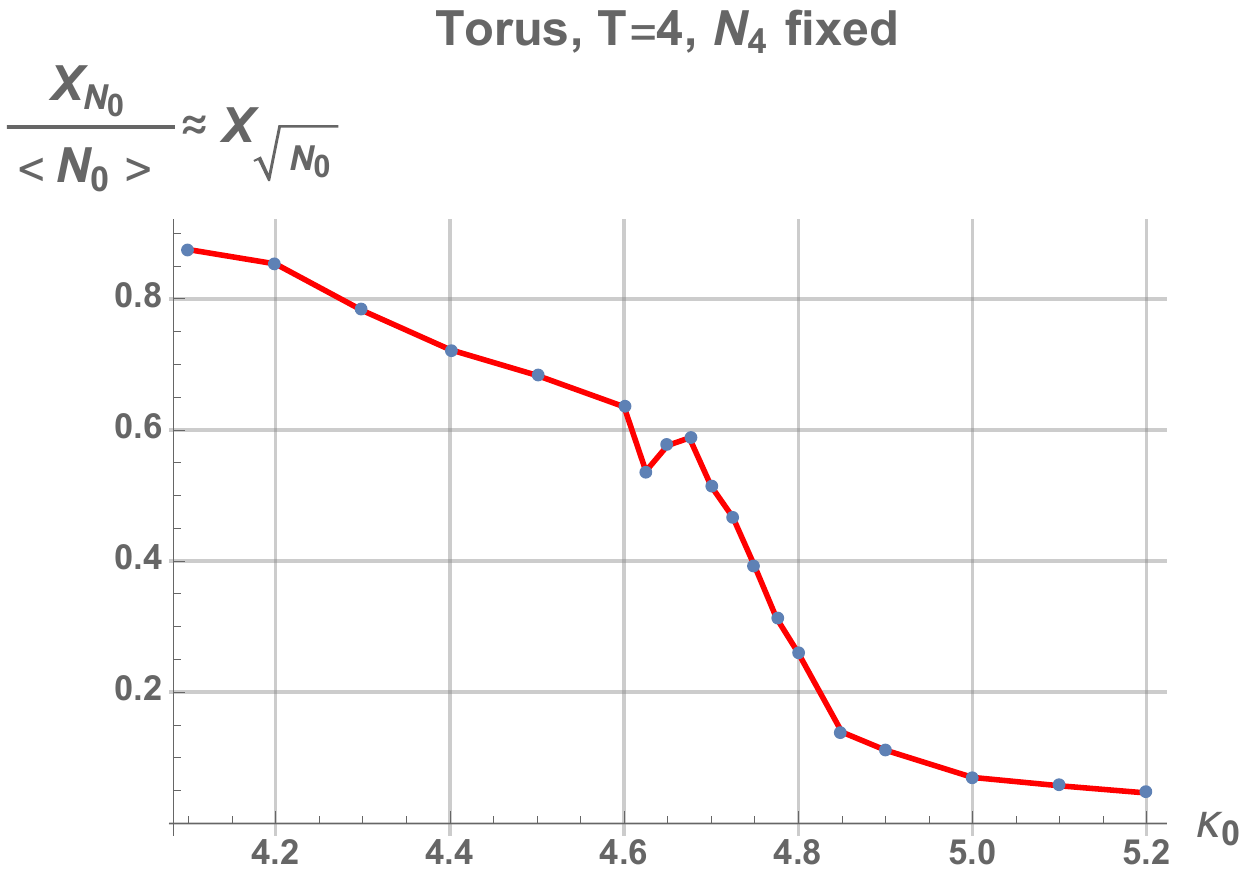}}
  \scalebox{.6}{\includegraphics{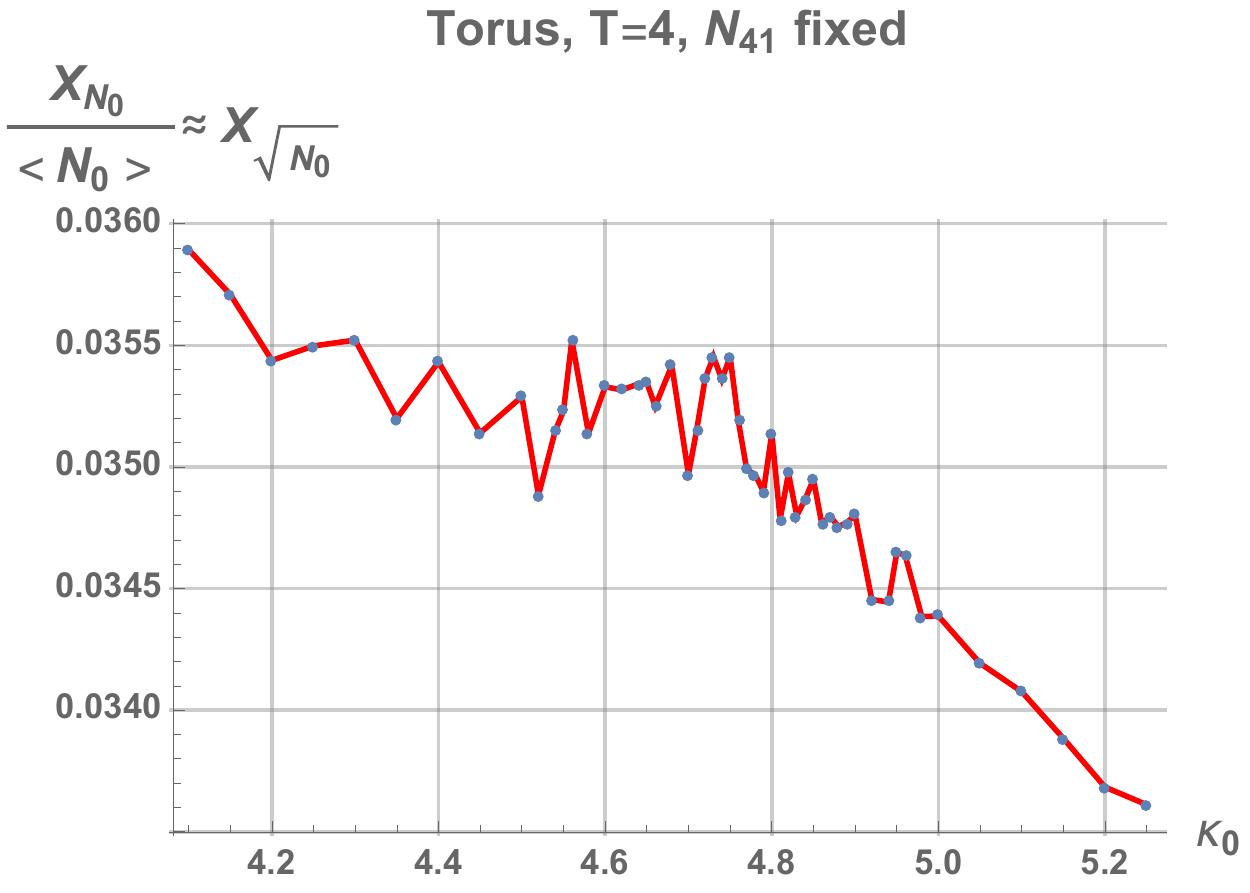}}
\caption{\small{The mean value $\langle OP \rangle$ (top), the susceptibility $\chi_{OP}$ (middle) and the susceptibility $ \chi_{\sqrt{OP}}$ (bottom) of the order parameter $OP=N_0$ measured near the $A$-$C$ transition in CDT with toroidal spatial topology and a proper-time period $T=4$. Left plots are for $N_4$ volume fixing and right plots are for $N_{4,1}$ volume fixing.  None of the susceptibility plots shows "critical" behaviour.}}
\label{Fig01}
\end{figure}

Let us start with the $OP=N_0$, which was used in the original study of the $A$-$C$ phase transition performed for spherical spatial topology \cite{Ambjorn:2012ij} where the $A$-$C$ transition was shown to be first-order. The analysis was done with the total $N_4$ volume  fixed and for a time period of $T=80$.  In this case, the order parameter $\langle N_0 \rangle$ was constant in the "disordered" $A$ phase (appearing at large $\kappa_0$) and decreased in the "ordered" $C$ phase (appearing at smaller $\kappa_0$).\footnote{\label{foot1}For the $A$-$C$ transition, phase $A$ can be treated as the "disordered"  and phase $C$ as the "ordered" one. We use these words in analogy with the Ising 
spin system where the spins fluctuate around a minimum of the Hamiltonian in the ordered phase, while they fluctuate around zero in the disordered phase. In phase $C$ the geometries fluctuate around a non-trivial ``classical'' solution, which in the case where the spatial topology is $S^3$ has the form of $S^4$ with a ``stalk'' of cut off size in the spatial directions connecting the north and south pole of 
$S^4$ in the time direction, and making the topology compatible with $S^1\times S^3$ rather than $S^4$ (see 
\cite{physrep} for details and Fig.\ \ref{FigT300} for a plot of the spatial volume profile, including the stalk). This ``classical'' solution breaks translational symmetry, which is restored when 
the motion of the centre of mass of the $S^4$ part of the geometry in the $S^1$ time direction is included. This is 
in contrast to the situation in phase $A$ where the spatial volumes at different times seems to fluctuate 
around the ``trivial'' mean value $N_{4,1}/T$.}  
 The same type of $\langle N_0 \rangle$ behaviour is  observed in toroidal CDT with $T=4$ time slices when $N_4$ is fixed - see Fig. \ref{Fig01} (left), but this is no longer the case when one chooses to fix $N_{4,1}$ - see Fig. \ref{Fig01} (right). In the latter case $\langle N_0 \rangle$  is also increasing in phase $A$, although there is some visible inflection point occurring at the $A$-$C$ phase transition. The order parameter was definitely "critical" in the spherical topology \cite{Ambjorn:2012ij} but it seems not to be "critical" in the toroidal CDT case, where one cannot observe any statistically  significant peaks in the  susceptibility $\chi_{N_0}$ or in  $\frac{\chi_{N_0}} { \langle N_0 \rangle}\approx\chi_{\sqrt{N_0}}$ for any volume fixing method.

\begin{figure}[H]
  \centering
  \scalebox{.6}{\includegraphics{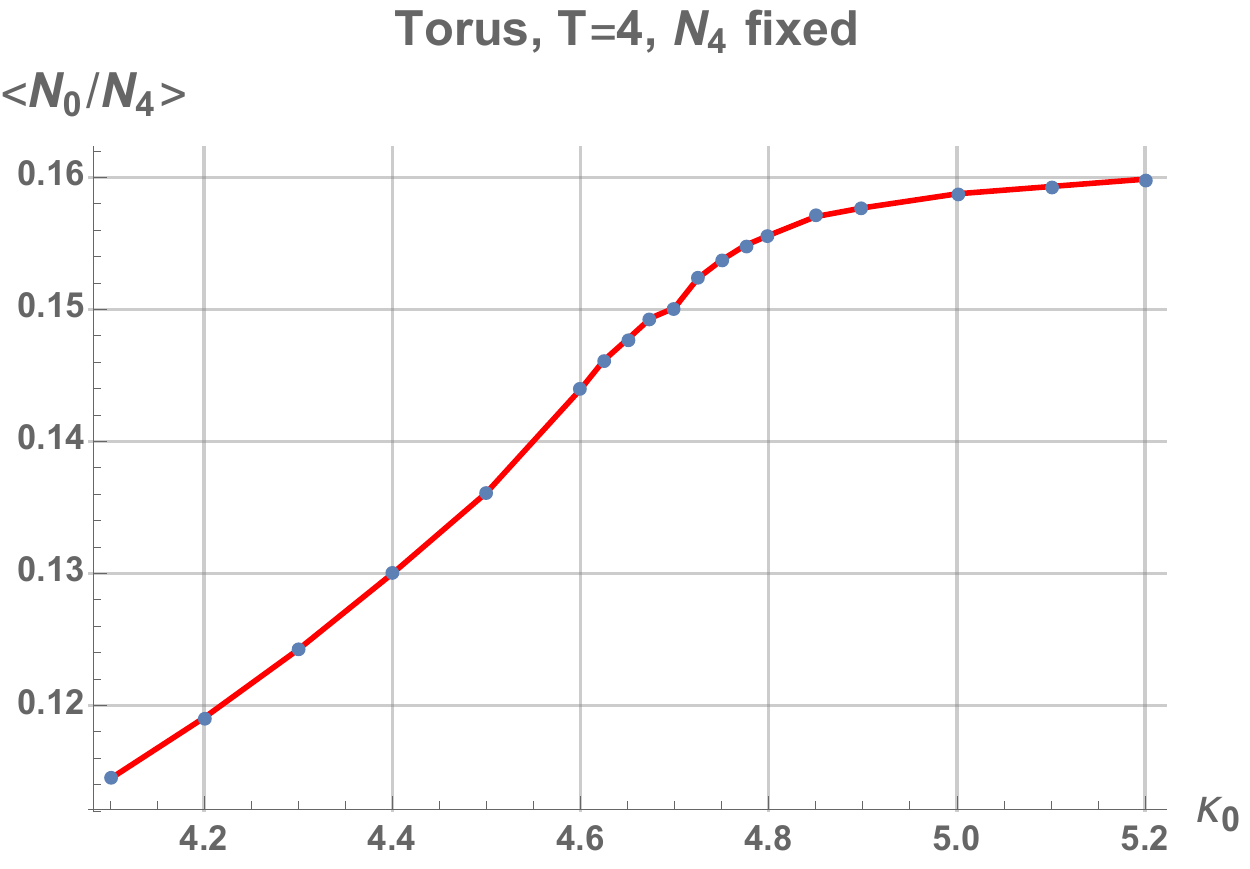}}
  \scalebox{.6}{\includegraphics{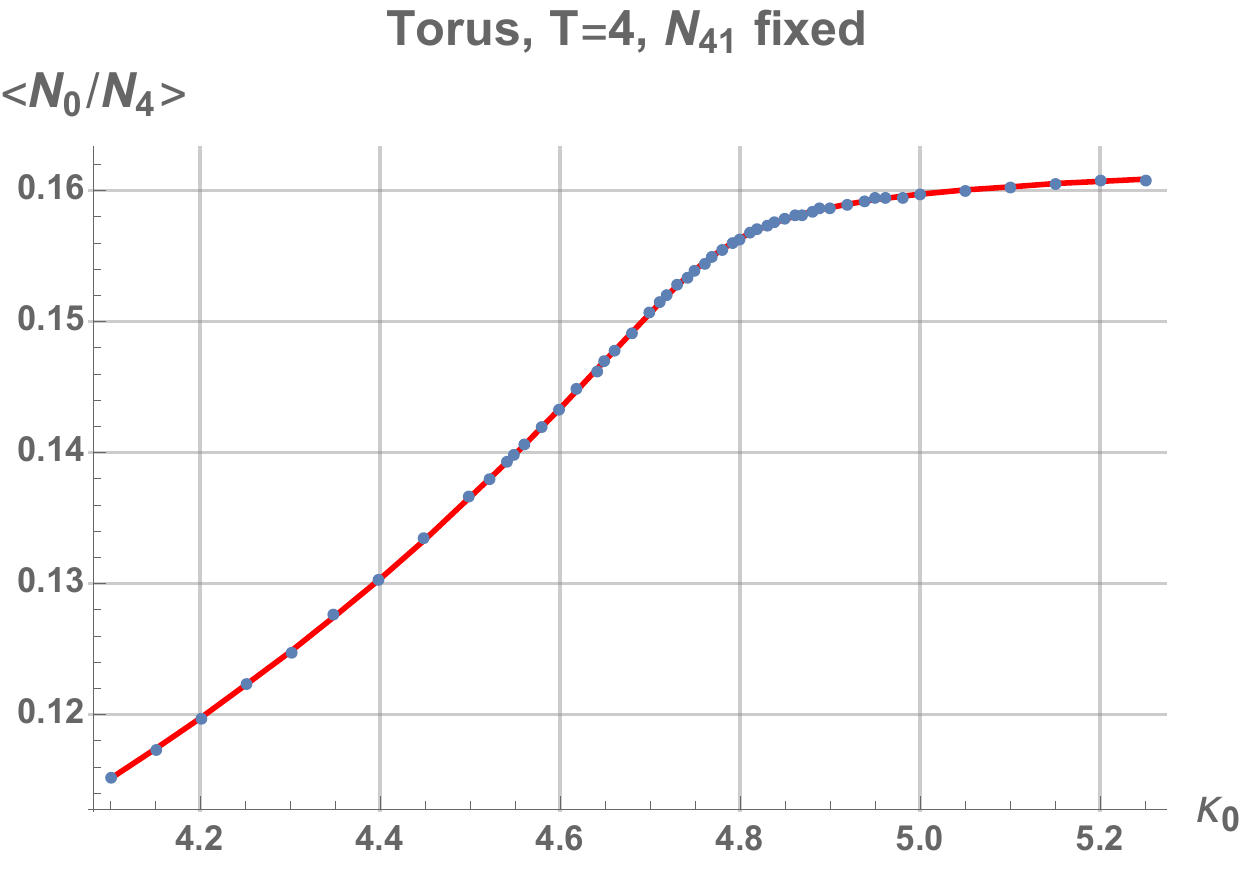}}
  \scalebox{.6}{\includegraphics{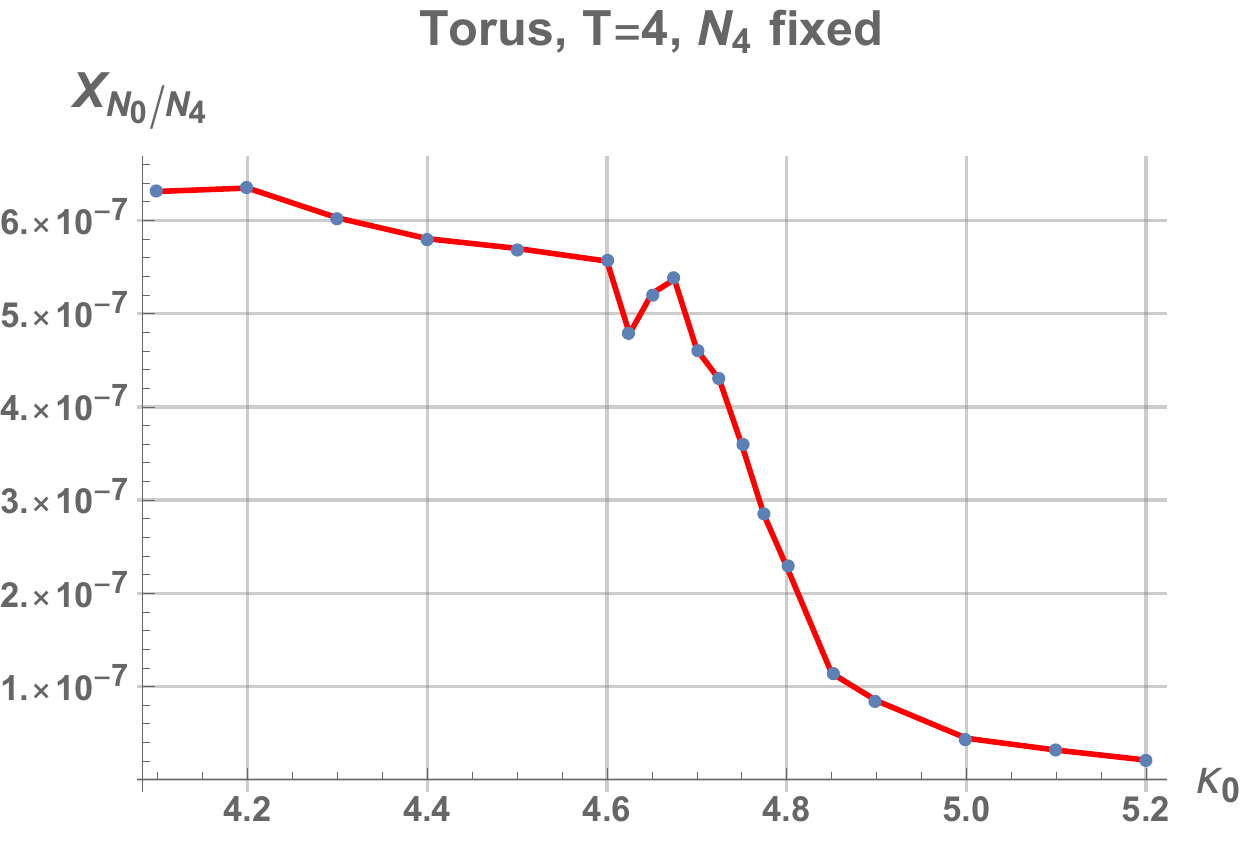}}
  \scalebox{.6}{\includegraphics{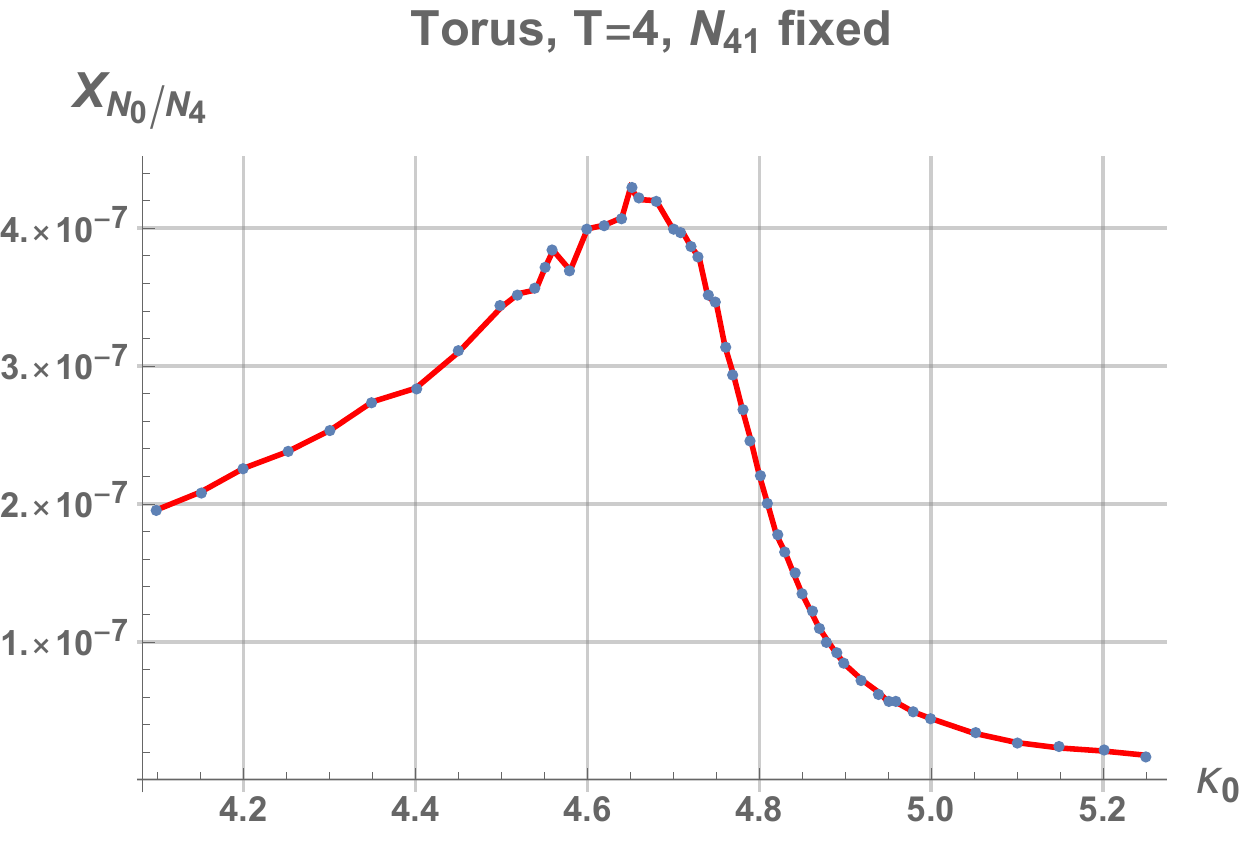}}
  \scalebox{.6}{\includegraphics{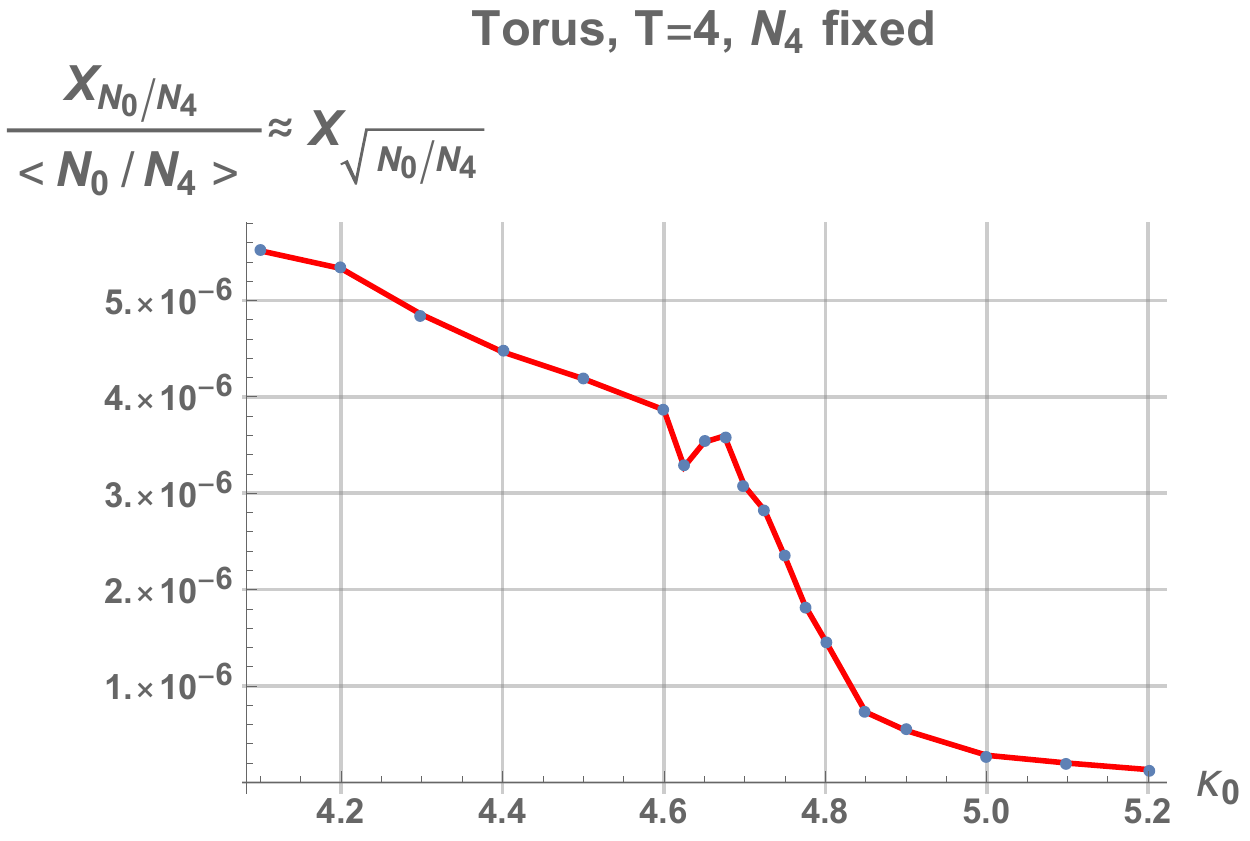}}
  \scalebox{.6}{\includegraphics{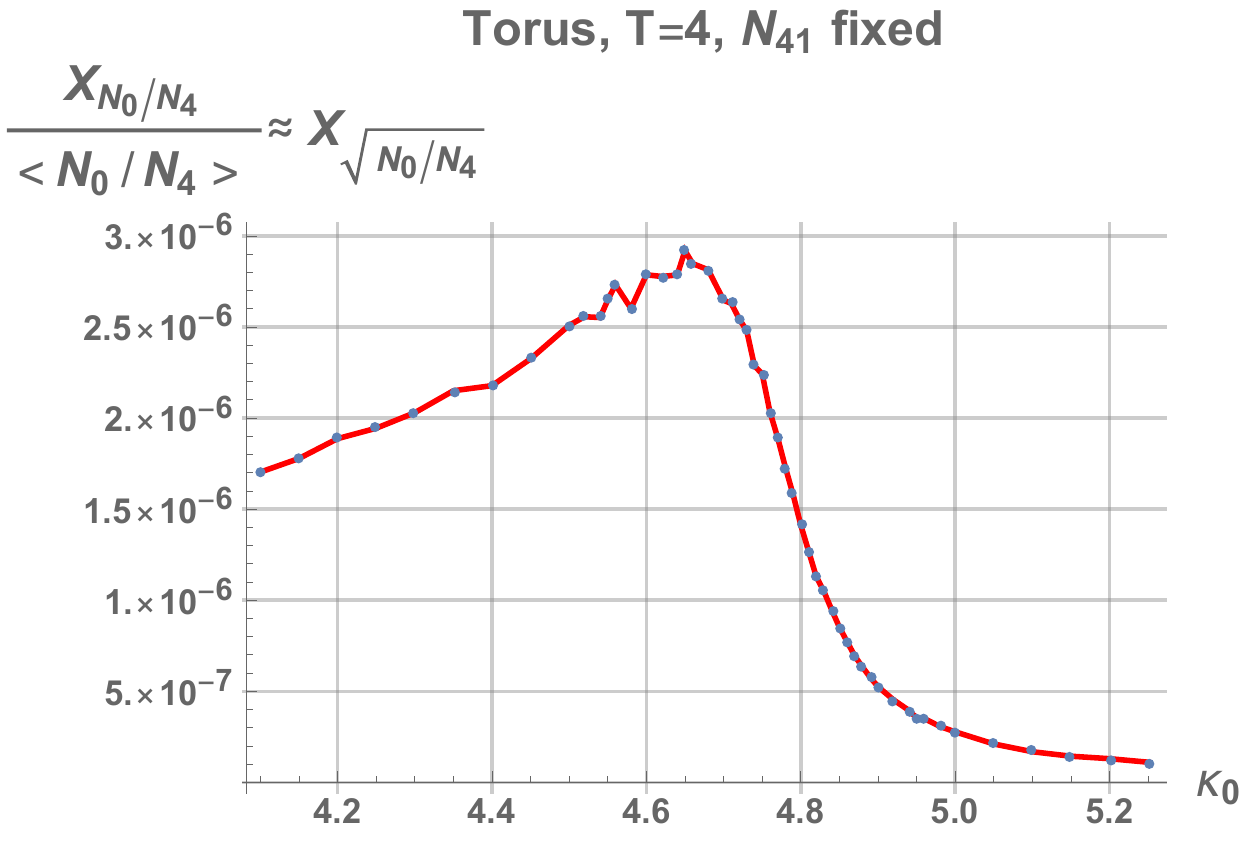}}
\caption{\small{The mean value $\langle OP \rangle$ (top), the susceptibility $\chi_{OP}$ (middle) and the susceptibility $ \chi_{\sqrt{OP}}$ (bottom) of the order parameter $OP_1 \equiv \frac{N_{0}}{N_{4}}$ near the $A$-$C$ transition in CDT with toroidal spatial topology and a $T=4$ proper-time period. Left plots are for $N_4$ volume fixing and the right plots are for $N_{4,1}$ volume fixing.  The susceptibility plots seem to show "critical" behaviour for $N_{4,1}$ volume fixing but the order parameter is not "critical" when $N_{4}$ is fixed.}}
\label{Fig02}
\end{figure}

The situation looks better if one chooses to use the (intensive) $OP=OP_1\equiv \frac{N_0}{N_4}$. This was obviously "critical" in the original study \cite{Ambjorn:2012ij} of spherical CDT, when $N_4$ was kept fixed. $OP_1$ again is {\it not} "critical" in the toroidal case for the same volume fixing method - see Fig. \ref{Fig02} (left) - but it is seems to be "critical" for toroidal CDT with $N_{4,1}$ fixed - see Fig. \ref{Fig02} (right), where one can observe peaks both for $\chi_{OP_1}$ and $\chi_{\sqrt{OP_1}}$.

\begin{figure}[H]
  \centering
 \scalebox{.6}{\includegraphics{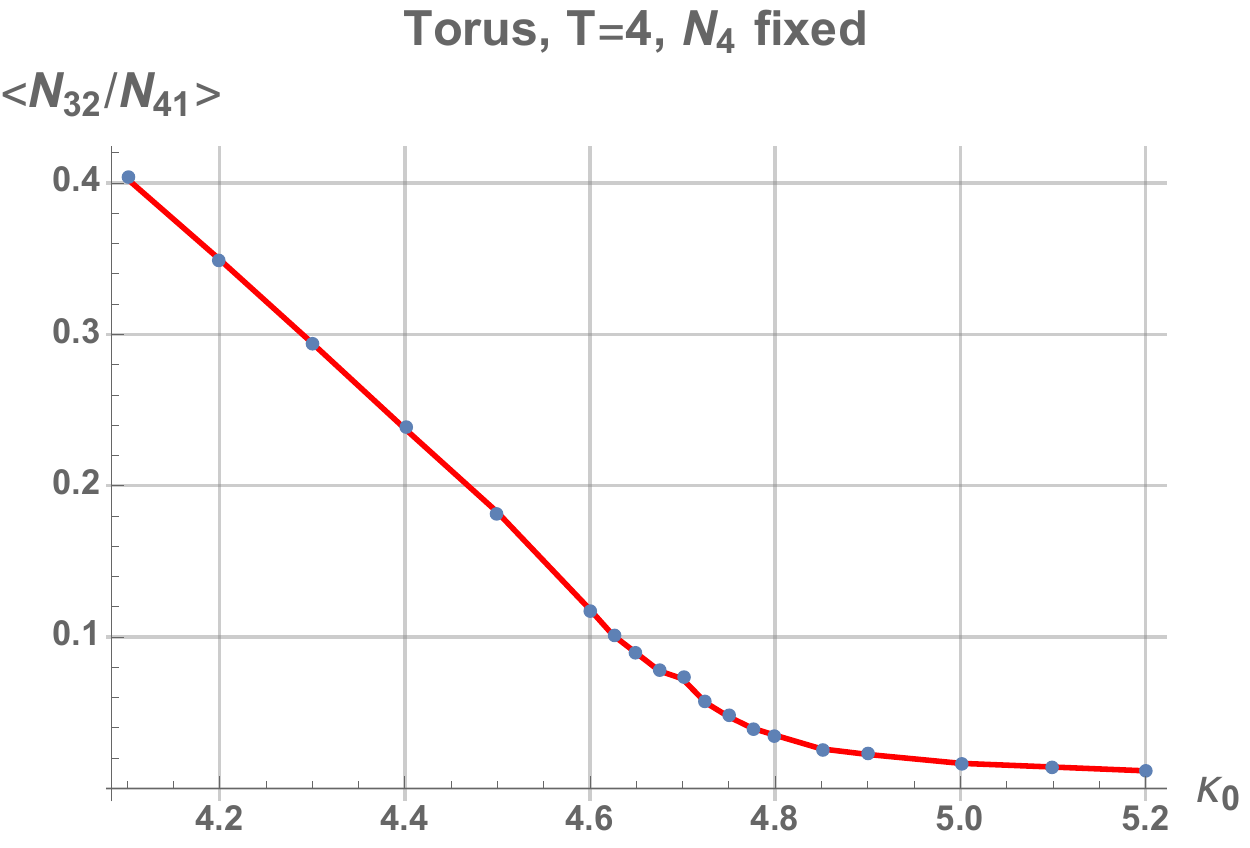}}
  \scalebox{.6}{\includegraphics{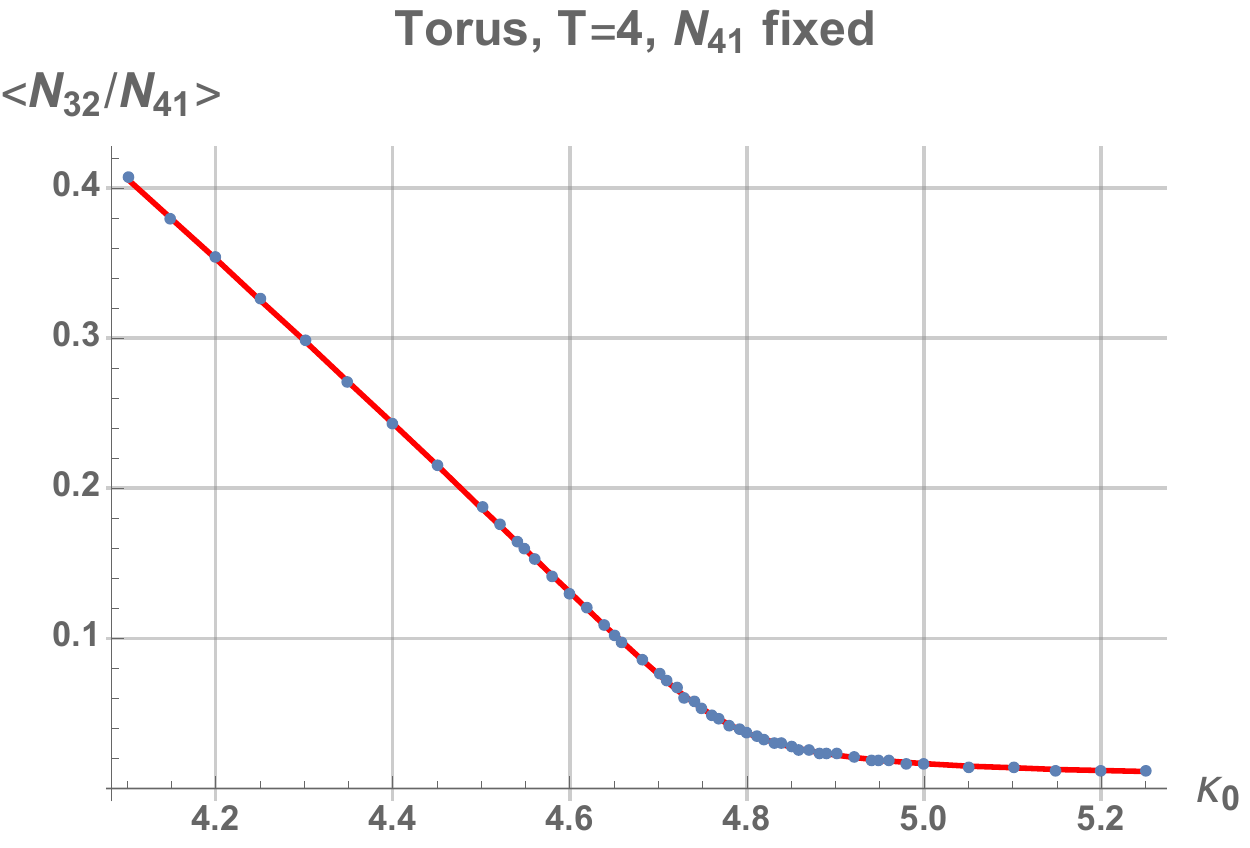}}
  \scalebox{.6}{\includegraphics{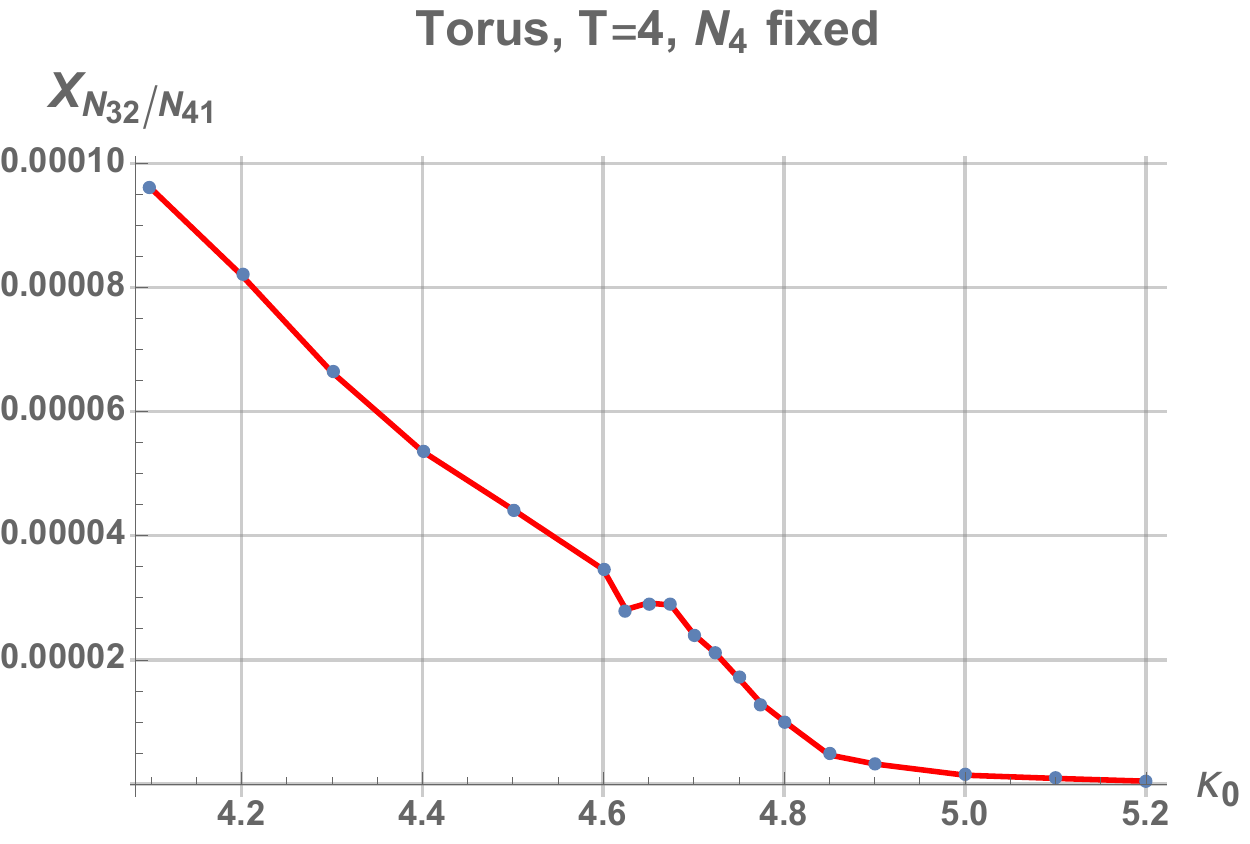}}
  \scalebox{.6}{\includegraphics{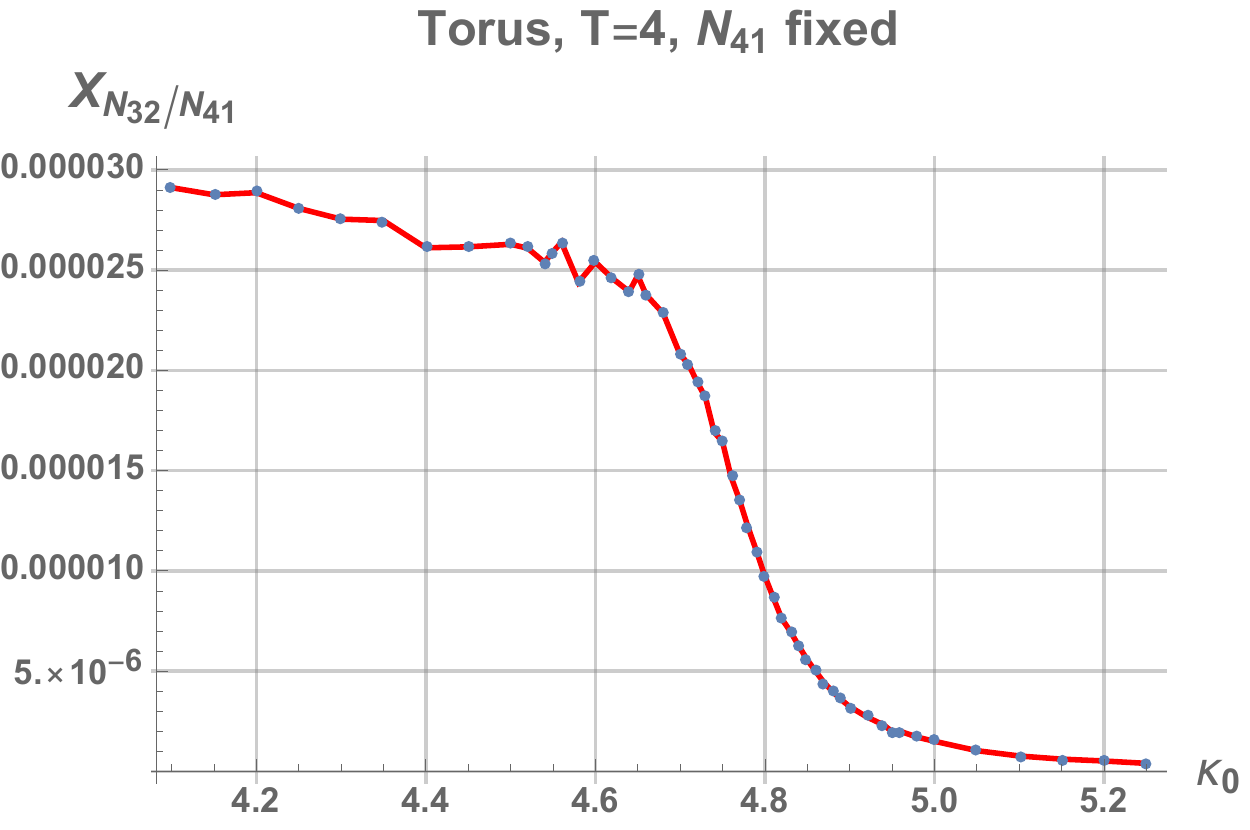}}
  \scalebox{.6}{\includegraphics{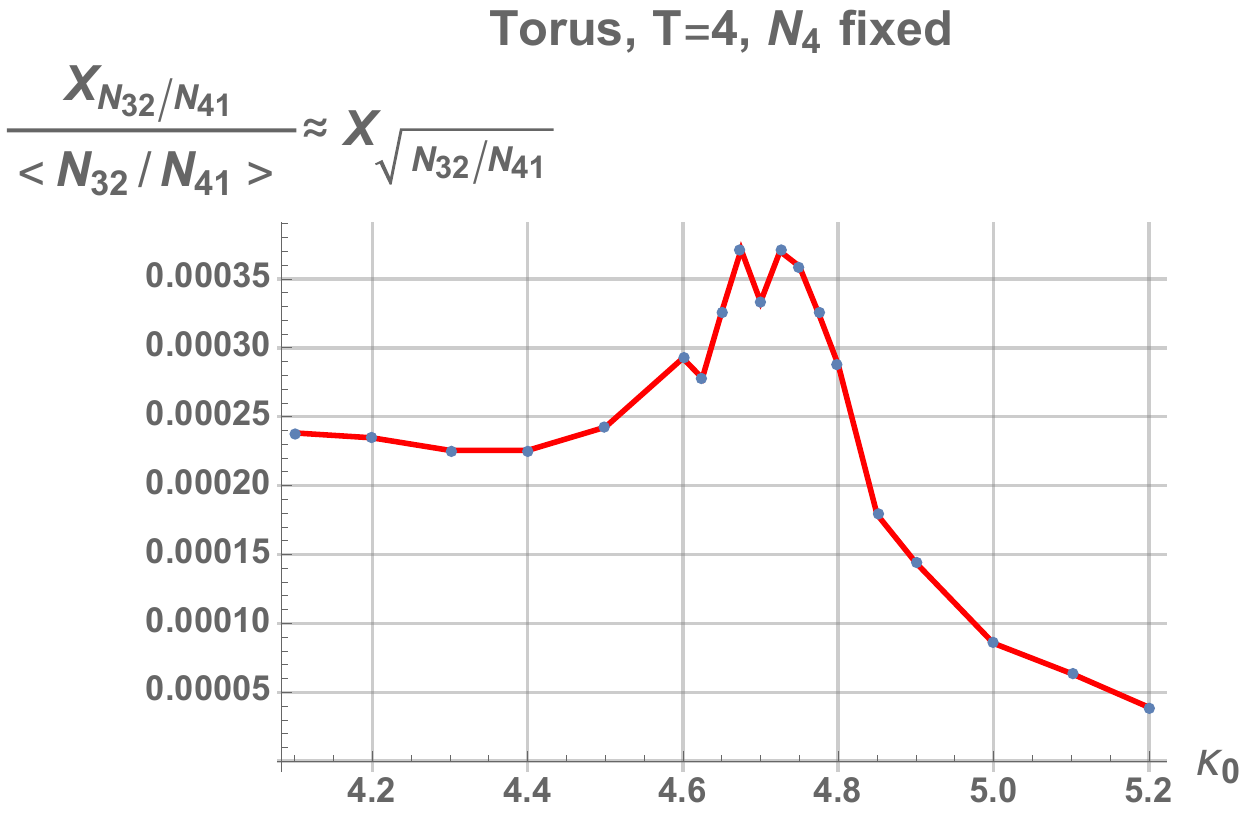}}
  \scalebox{.6}{\includegraphics{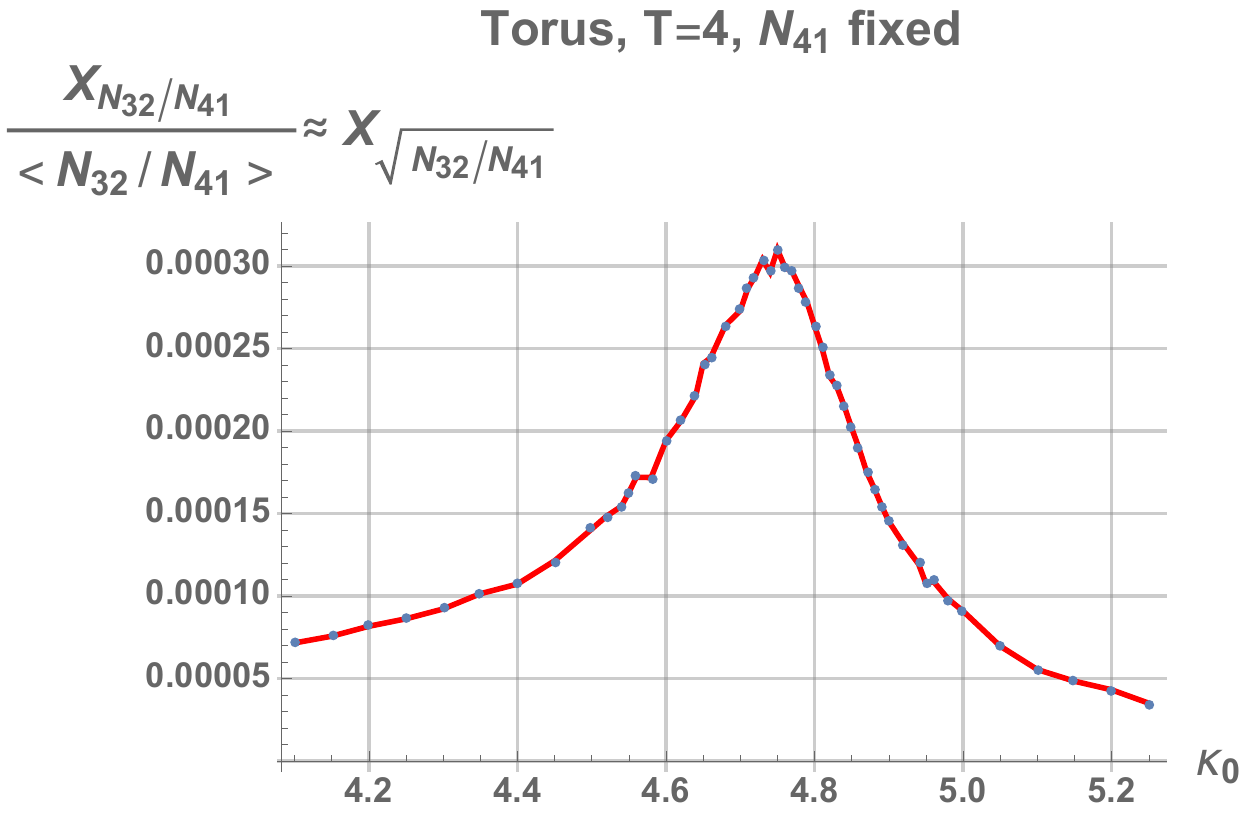}}  
\caption{\small{The mean value $\langle OP \rangle$ (top), the susceptibility $\chi_{OP}$ (middle) and the susceptibility $ \chi_{\sqrt{OP}}$ (bottom) of the order parameter $OP_2 \equiv \frac{N_{3,2}}{N_{4,1}}$ measured near the $A$-$C$ transition in CDT with toroidal spatial topology and a $T=4$ proper-time period. Left plots are for $N_4$ volume fixing and the right plots are for $N_{4,1}$ volume fixing. $ \chi_{\sqrt{OP}}$ shows "critical" behaviour in all cases analysed.}}
\label{Fig03}
\end{figure}

\begin{figure}[H]
  \centering
  \scalebox{.6}{\includegraphics{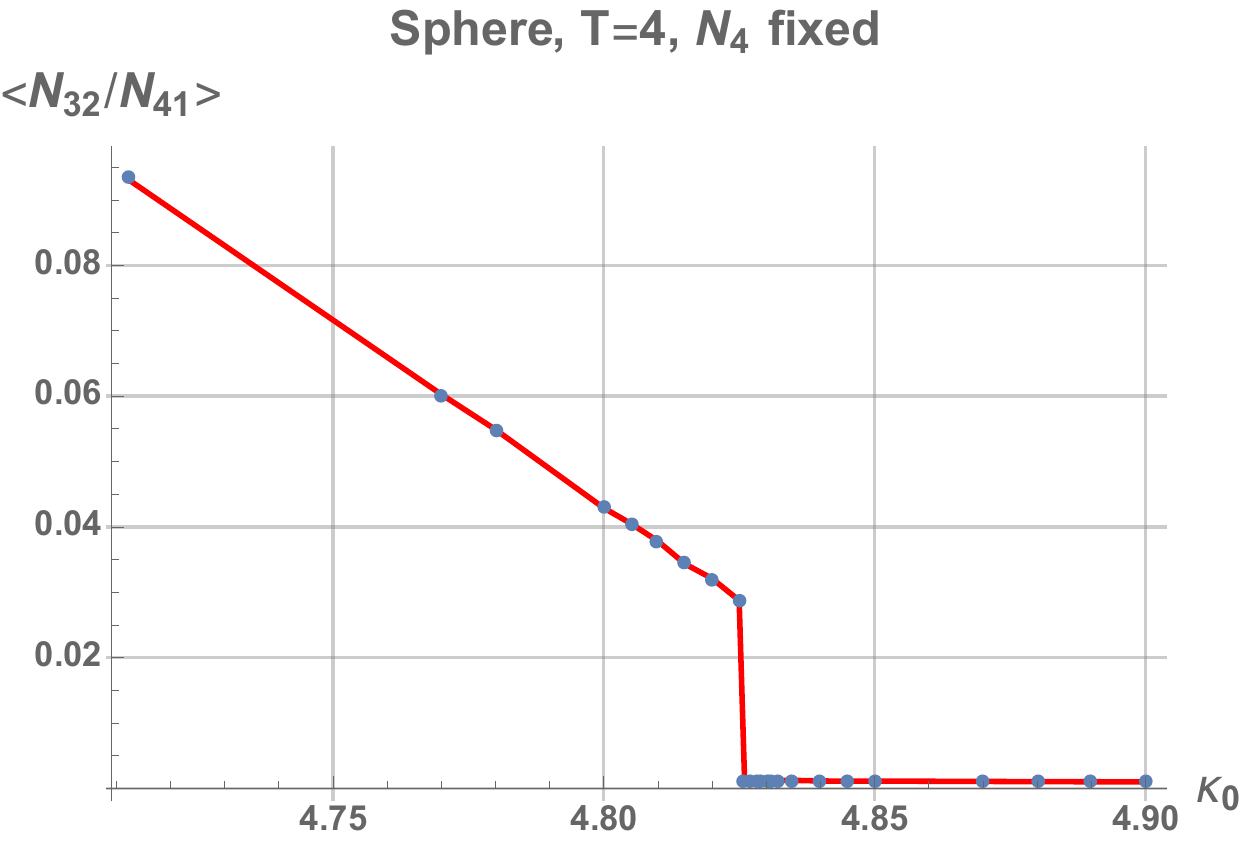}}
  \scalebox{.6}{\includegraphics{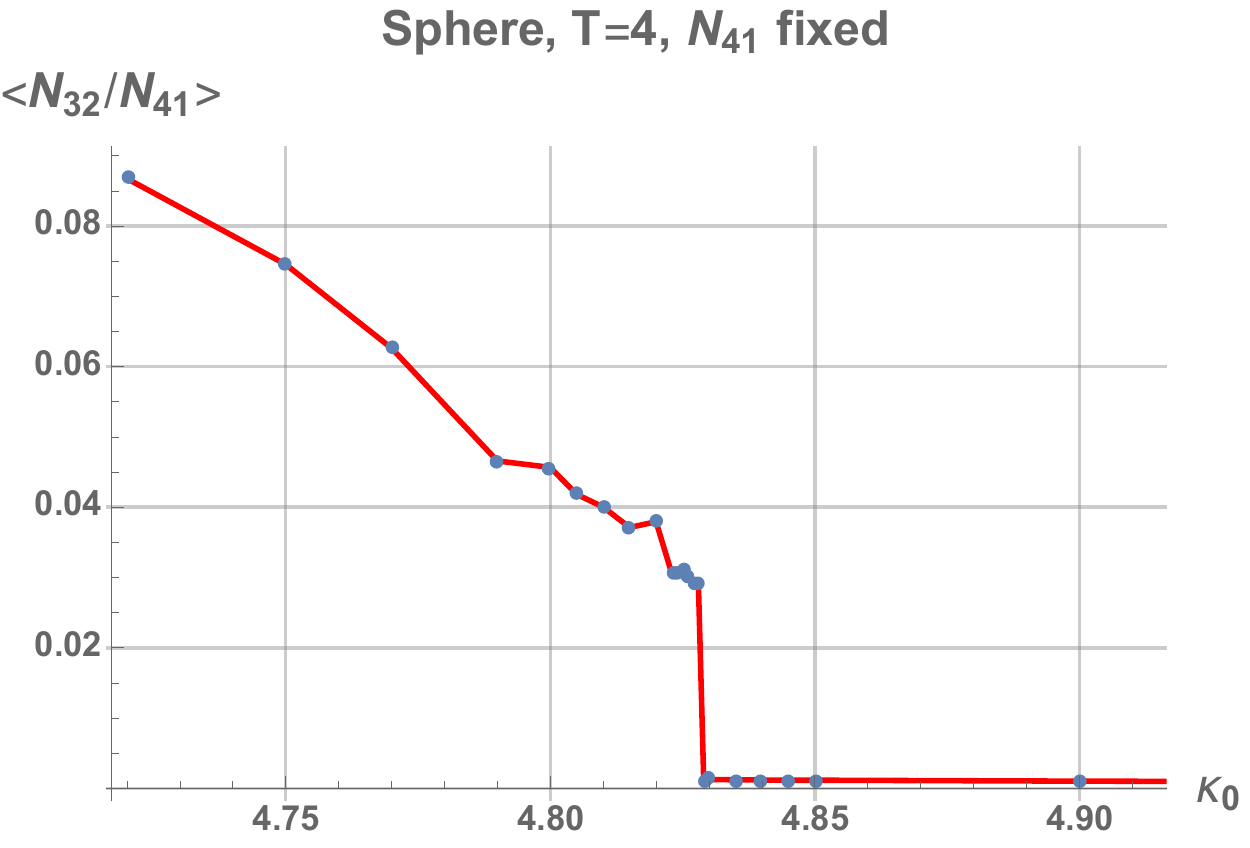}}
  \scalebox{.6}{\includegraphics{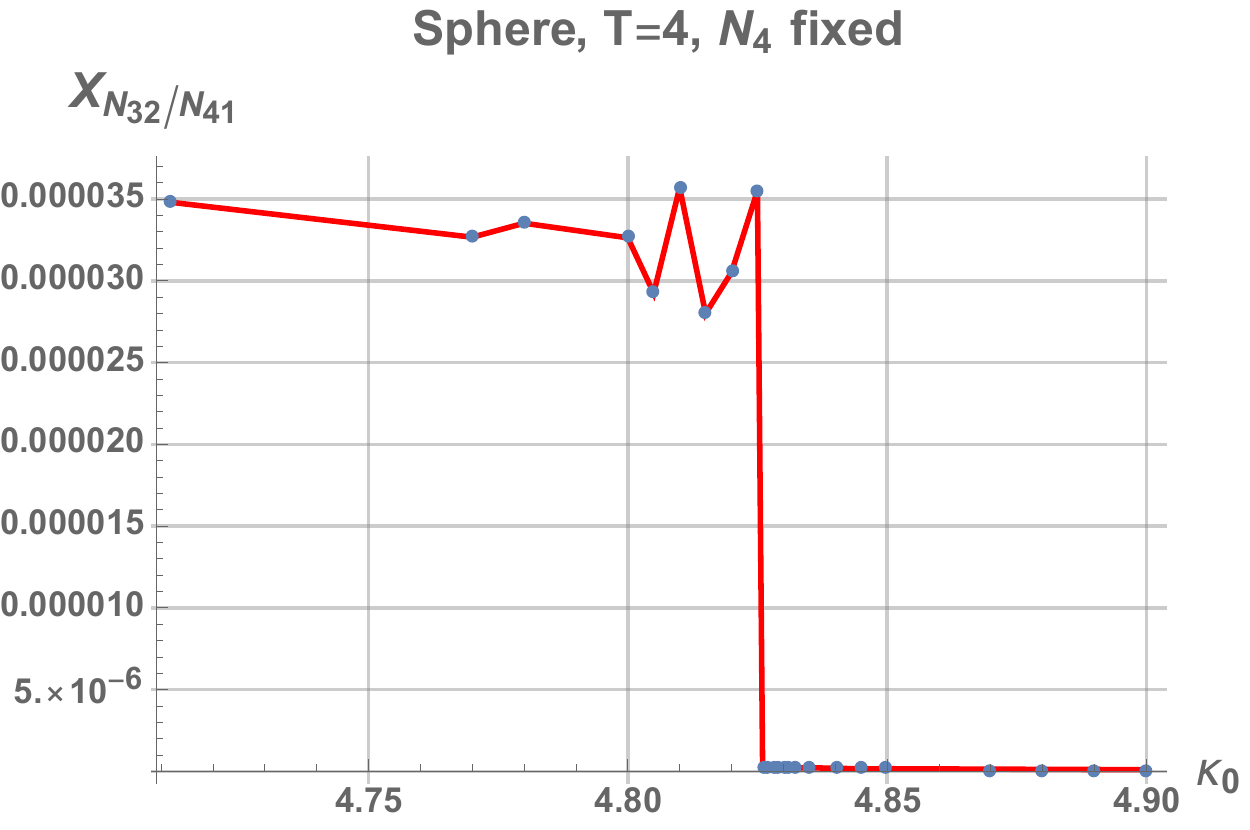}}
  \scalebox{.6}{\includegraphics{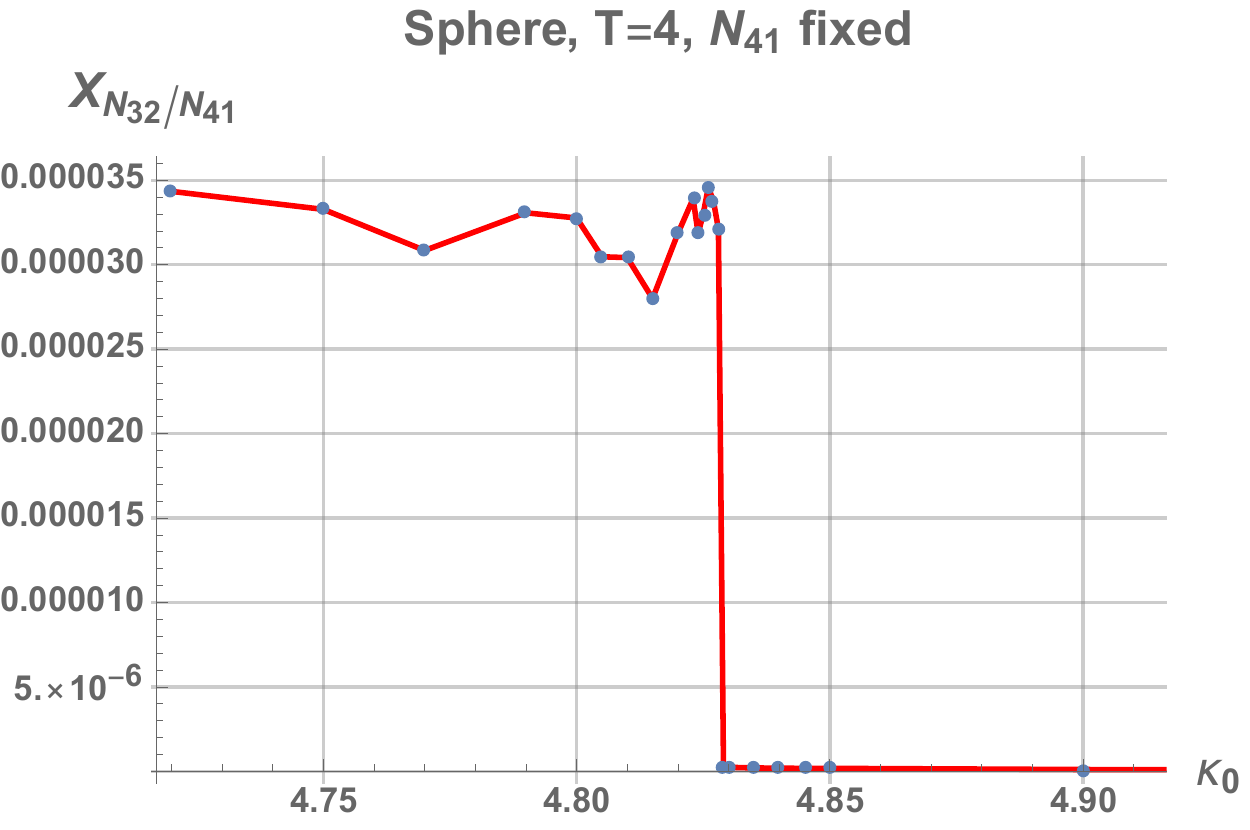}}
  \scalebox{.6}{\includegraphics{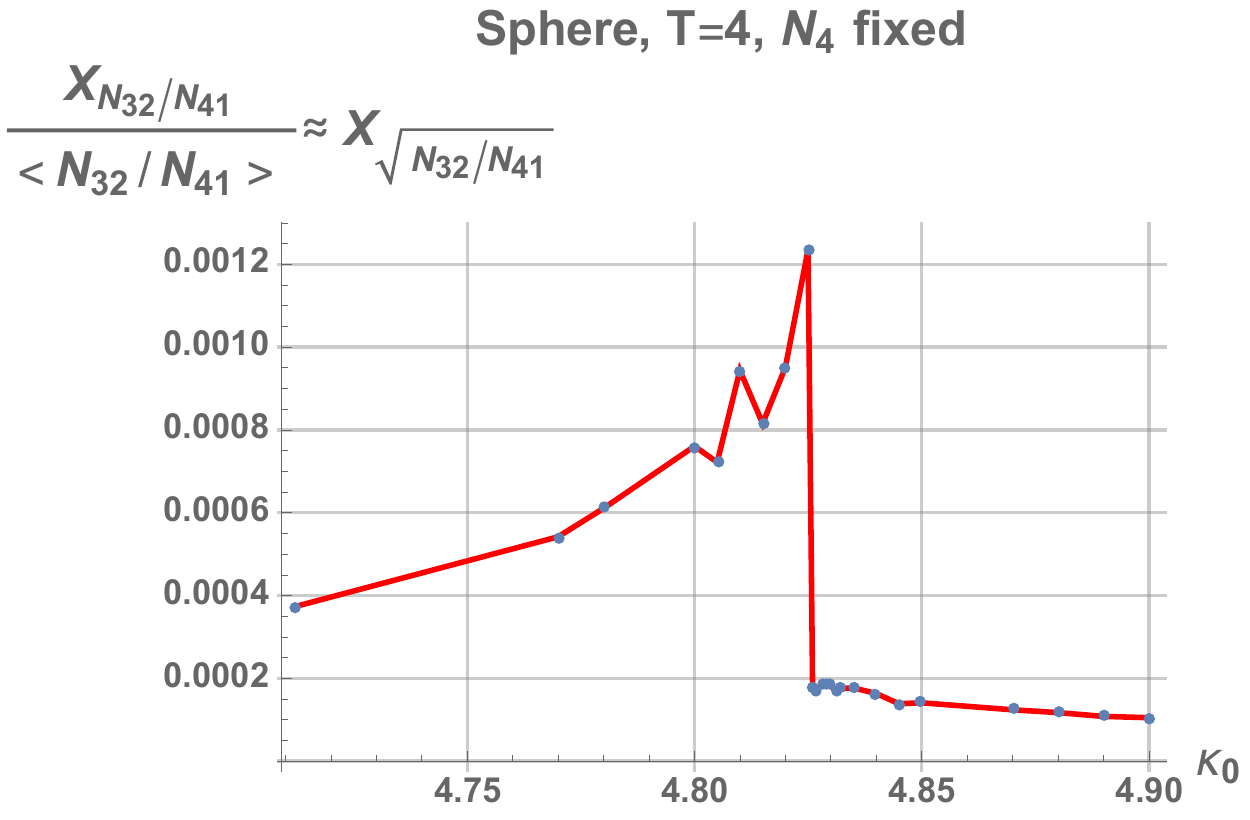}}
  \scalebox{.6}{\includegraphics{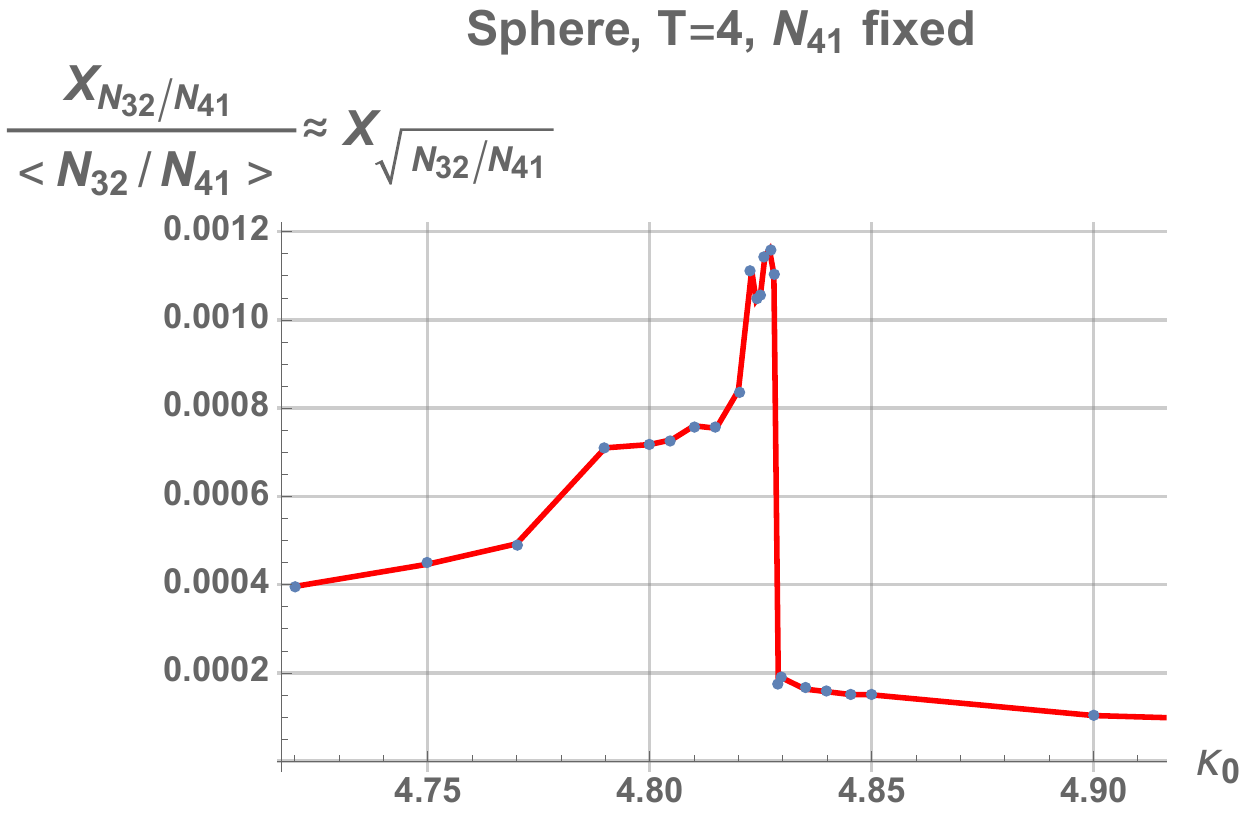}}
\caption{\small{The mean value $\langle OP \rangle$ (top), the susceptibility $\chi_{OP}$ (middle) and the susceptibility $ \chi_{\sqrt{OP}}$ (bottom) of the order parameter $OP_2 \equiv \frac{N_{3,2}}{N_{4,1}}$ measured near the$A$-$C$ transition in CDT with spherical spatial topology and a $T=4$ proper-time period. Left plots are for $N_4$ fixed and the right plots are for $N_{4,1}$ fixed. Both $ \chi_{{OP}}$ and $ \chi_{\sqrt{OP}}$ show "critical" behaviour in all cases analysed.}}
\label{Fig03s}
\end{figure}

As already discussed, the $N_0$ susceptibility $\chi_{N_0}$ does not show any critical behaviour for any volume fixing method for the toroidal CDT case, so the observed peak in susceptibility  $\chi_{\frac{N_0}{N_4}}$ cannot  come from $N_0$ but from $N_4=N_{4,1}+N_{3,2}$ fluctuations. For fixed $N_{4,1}$, the only fluctuations possible are for $N_{3,2}$, which can be quantified by the ratio  $OP_2 \equiv \frac{N_{3,2}}{N_{4,1}}$. This order parameter  (or, more precisely, its function $\sqrt{OP_2} \equiv \sqrt{\frac{N_{3,2}}{N_{4,1}}}$) is {\it truly critical} in all cases analysed here - see Fig. \ref{Fig03} and Fig. \ref{Fig03s}, where we plot the data for $T=4$ time slices for the toroidal and the spherical spatial topology, respectively. Similar results were also obtained  for larger $T$.  In fact one can show that the  spurious "criticality" of $OP_1 \equiv \frac{N_0}{N_4}$, observed for toroidal CDT when $N_{4,1}$ was fixed, is illusory and can be explained by the behaviour of the "truly critical" $\sqrt{OP_2}$. For fixed $N_{4,1}=\overline N_{4,1}$, assuming constant $N_0\approx \langle N_0 \rangle$ and using the
small fluctuations approximation \rf{suscf} one has
\beql{chiApprox}
\chi_{\frac{N_0}{N_4}} \approx \chi_{\frac{\langle N_0 \rangle}{\overline N_{4,1}+N_{3,2}}}=\chi_{\frac{\langle N_0 \rangle}{\overline N_{4,1}}\left(1+\sqrt{OP_2}^2\right)^{-1}}\approx \left( \frac{2 \langle N_0 \rangle\langle \sqrt{OP_2}\rangle }{\overline N_{4,1} \left(1+ \langle \sqrt{OP_2}\rangle^2\right)^2}\right)^2 \chi_{\sqrt{OP_2}} \ .
\eeq
In Figure \ref{Fig04} we again plot the susceptibility $\chi_{\frac{N_0}{N_4}}$ calculated for the toroidal CDT with fixed $N_{4,1}$, the same as in Fig. \ref{Fig02} right, together with its approximation by the susceptibility $\chi_{\sqrt{OP_2}}$ \rf{chiApprox}  (the dashed-black line). The spurious "criticality" peak of $\chi_{\frac{N_0}{N_4}}$ is perfectly explained by the truly "critical" peak of 
$\chi_{\sqrt{OP_2}}$ (the green line) multiplied by the prefactor (the dashed-green line). 

\begin{figure}[H]
  \centering
  \scalebox{.75}{\includegraphics{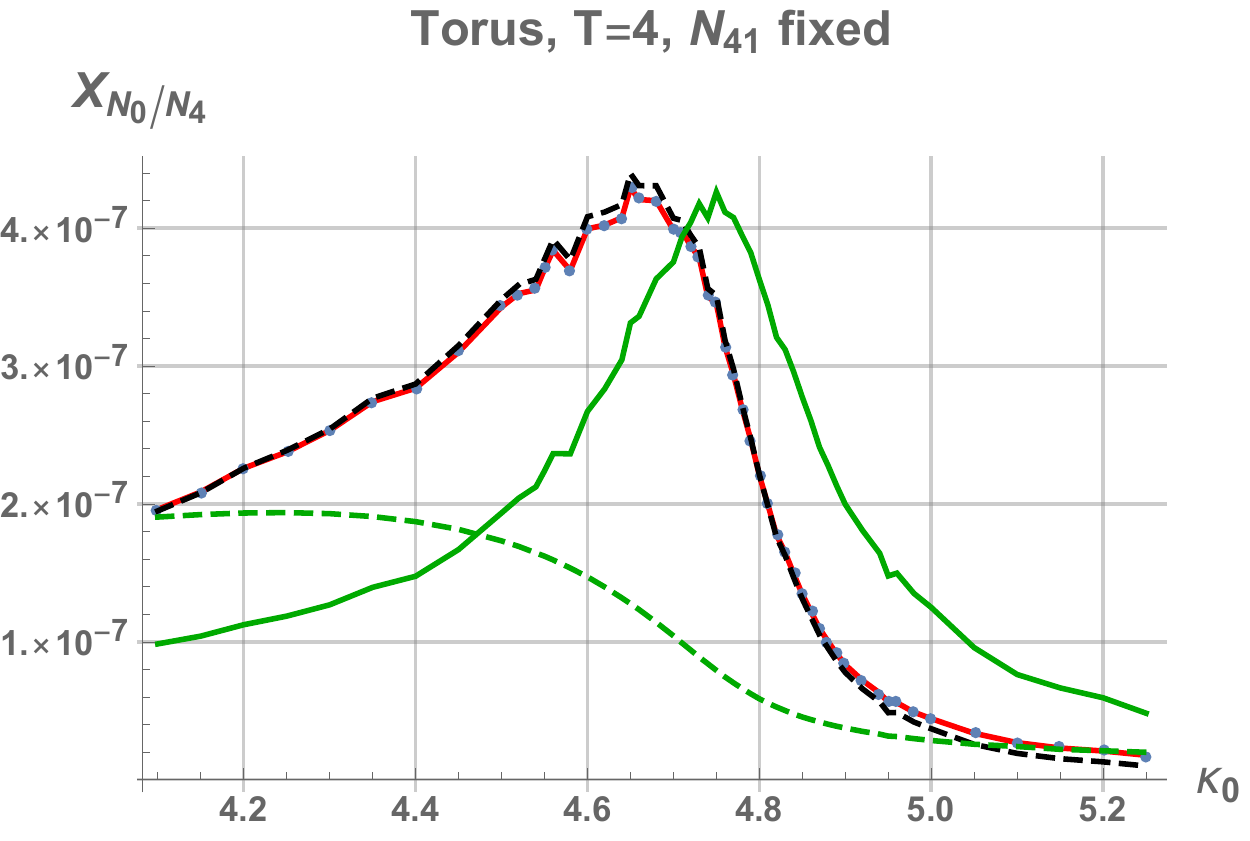}}
\caption{\small{The peak of susceptibility  $\chi_{\frac{N_0}{N_4}}$ is perfectly explained by the peak of the "really critical" $\chi_{\sqrt{OP_2}}$ (green line) multiplied by a prefactor of equation \rf{chiApprox} (dashed-green line). The (small-fluctuations) approximation \rf{chiApprox} is plotted as the dashed-black line and it is very close to the measured $\chi_{\frac{N_0}{N_4}}$ data (red line). }}
\label{Fig04}
\end{figure}

Summing up, among all the order parameters analysed
\beql{op2}
OP_2 \equiv \frac{N_{3,2}}{N_{4,1}},
\eeq
(or more precisely its function $\sqrt{OP_2}$) shows  "truly critical" behaviour irrespective of the topology, time slicing or the volume-fixing method. Therefore, in the following sections we will focus on this order parameter. Specifically, we will try to repeat the results of the previous phase transition study  \cite{Ambjorn:2012ij} for the spherical spatial topology, originally done using $N_0$ or $OP_1 \equiv \frac{N_0}{N_4}$, now using $OP_2$. We will then investigate the impact of the volume-fixing method ($N_{4,1}$ vs $N_4$), the time-slicing (large $T=80$ vs small $T=4$) and spatial topology (the spherical topology vs the toroidal topology) on our results. 

\end{section}

\begin{section}{Spherical topology}\label{fields}

%

\begin{subsection}{The order of the $A$-$C$ transition}\label{sphere1}

The $A$-$C$ transition in four-dimensional CDT with spherical spatial topology was earlier analysed in Ref. \cite{Ambjorn:2012ij} where it was found to likely be a first order transition. The study was based on the  analysis of the $N_0$ order parameter and using a $N_4$ volume fixing method applied to triangulations with $T=80$ time slices. The data showed all characteristics of the first order transition summarised in Table \ref{Table1}, namely one could observe double peaks in the $N_0$ histograms measured at the pseudo-critical points, the critical  exponent of equation \rf{powerlaw} was consistent with $\gamma = 1$ and the (minimum of the) Binder cumulant $B_{N_0}^{min}$ was diverging from zero  for large lattice volumes $N_4\to \infty$ \cite{Ambjorn:2012ij}.

Now we will use the $OP_2$ parameter defined in formula \rf{op2}. Measurements presented in this section were made at $\Delta=0.6$, for a maximum of 8 different lattice volumes ranging from $N_{4,1}=20k$ to $N_{4,1}=160k$ in increments of $20k$ with $T=80$ time slices. For the $N_{4,1}=20k$ and $40k$ ensembles a double peak 
cannot be distinguished from a histogram analysis because the lattice volumes are too small to exhibit clear transitions between metastable phases. However, the pseudo-critical $\kappa_{0}^{crit}$ values can still be estimated by finding the maximum in susceptibility $\chi_{OP_2}$ of the order parameter as a function of $\kappa_{0}$ (see Fig.~\ref{Fig1}), albeit with reduced accuracy.\footnote{Alternatively, one can  look at the maximum of the susceptibility $\chi_{\sqrt{OP_2}}$, which gives similar results to $\chi_{OP_2}$ but with a slightly higher pseudo-critical $\kappa_{0}$ value.}

The same procedure of locating the peak in the susceptibility $\chi_{OP_2}$ is performed for the larger system sizes, where a full histogram analysis is possible for lattice volumes between $N_{4,1}=60k$ and $N_{4,1}=160k$ at the transition points. For system sizes equal to or greater than $N_{4,1}=60k$ a histogram with a double Gaussian structure is clearly observed for each point, as shown in Fig.~\ref{Fig2}. The pseudo-critical $\kappa_{0}^{crit}$ values at which the transition occurs exhibit a clear volume dependency, with the separation of the double Gaussian peaks becoming more pronounced with increasing system size (see Fig.~\ref{Fig2}). 

\begin{figure}[H]
  \centering
  \scalebox{.6}{\includegraphics{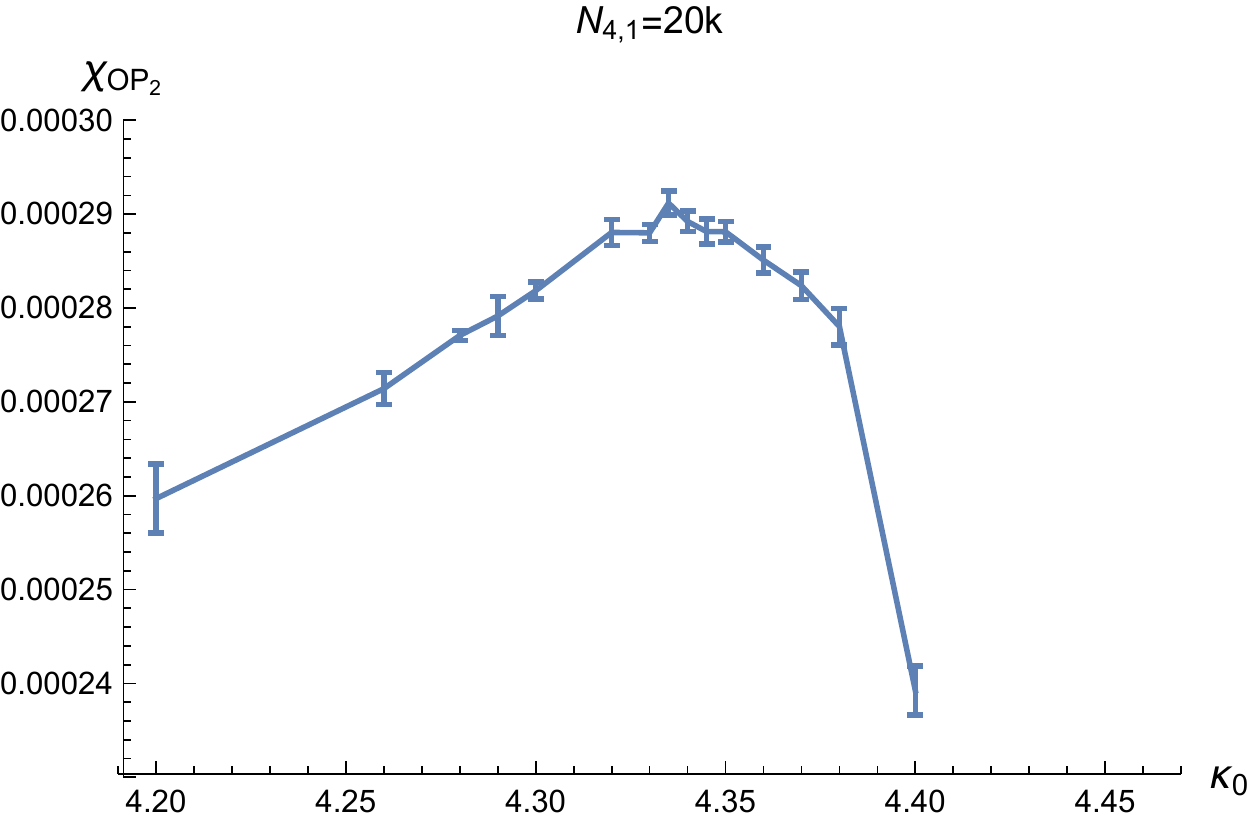}}
  \scalebox{.6}{\includegraphics{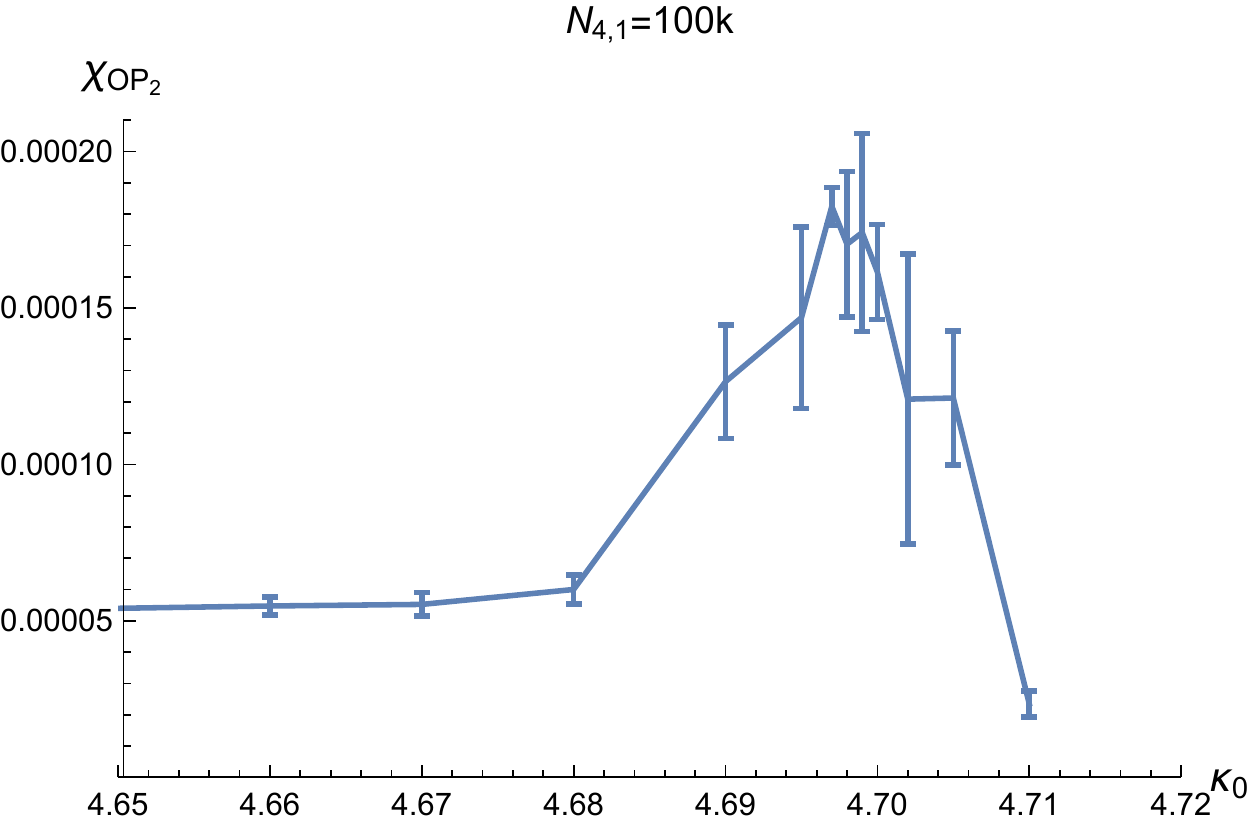}}
  \scalebox{.6}{\includegraphics{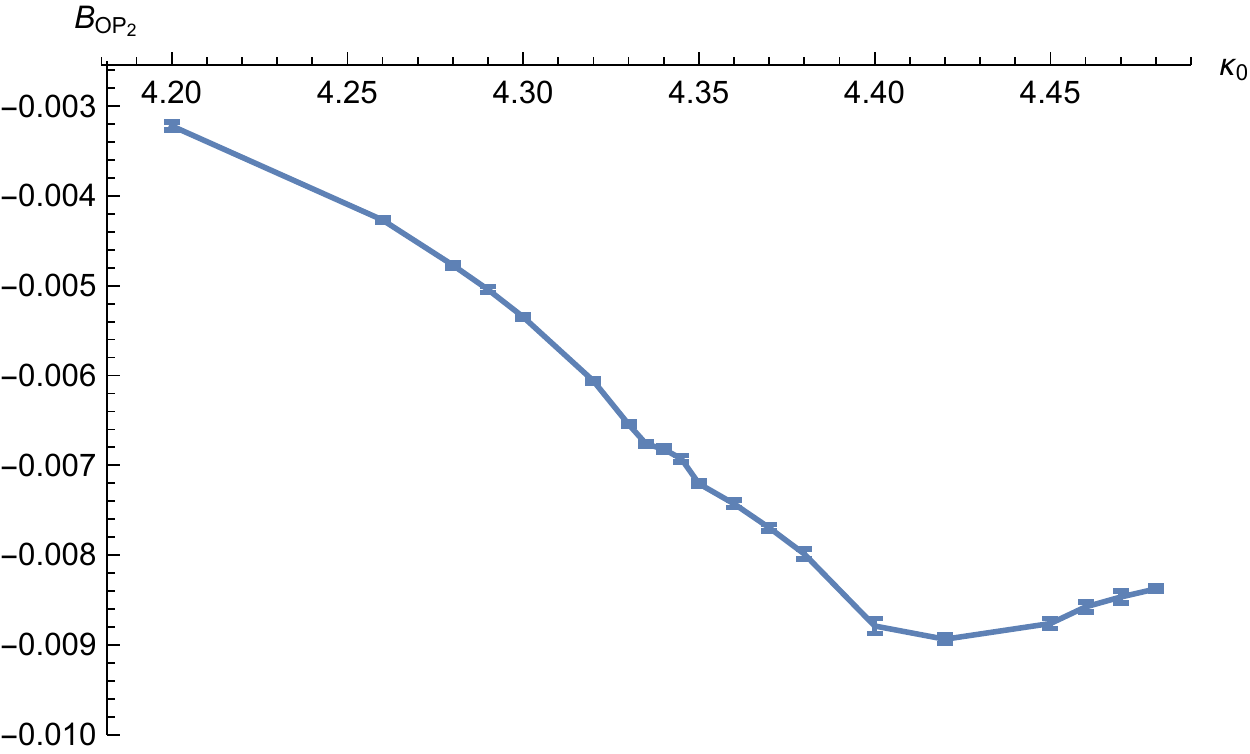}}
  \scalebox{.6}{\includegraphics{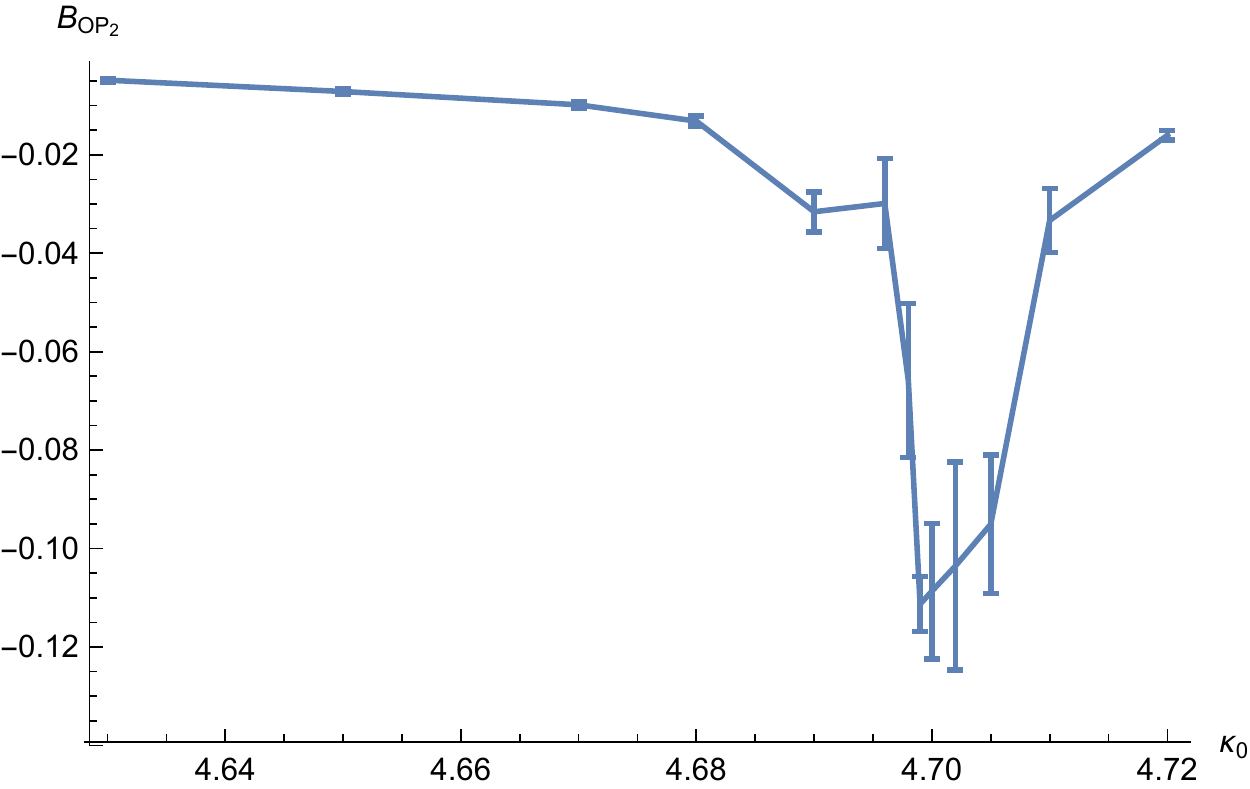}}
\caption{\small  {The susceptibility $\chi_{OP_2}$ as defined by Eq.~(\ref{susc}) (top charts) 
and the  Binder cumulant $B_{OP_2}$ as defined by Eq.~(\ref{binder}) (bottom charts) as a function of $\kappa_{0}$. Left charts are  for $N_{4,1}=20k$  and right charts are for  $N_{4,1}=100k$. } }
\label{Fig1}
\end{figure}

\begin{figure}[H]
  \centering
  \scalebox{.6}{\includegraphics{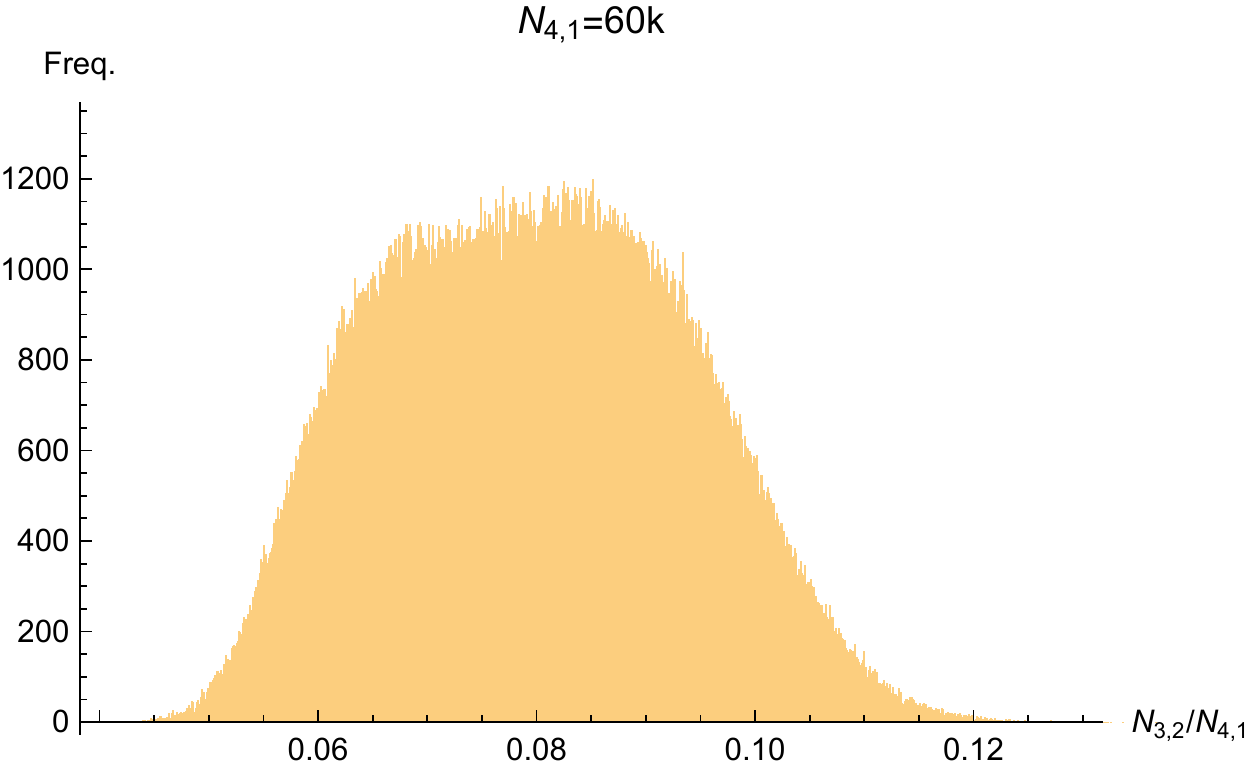}}
  \scalebox{.6}{\includegraphics{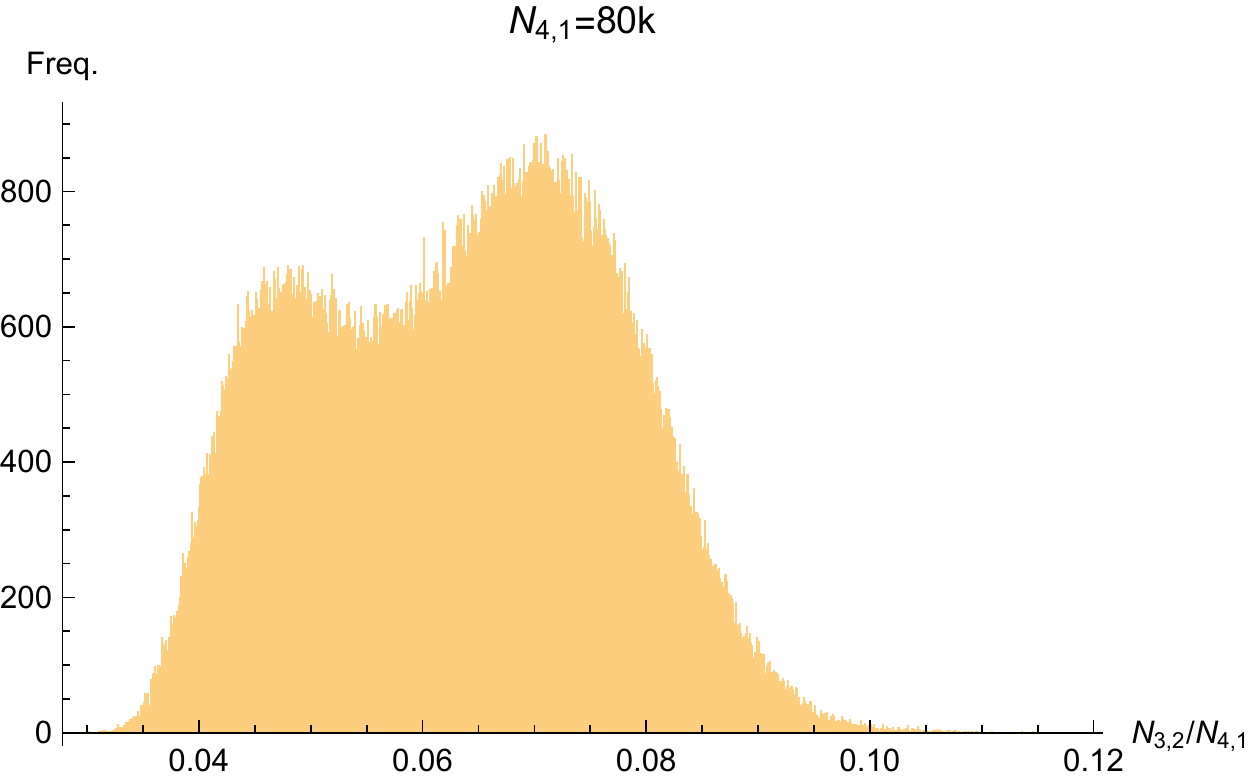}}
  \scalebox{.6}{\includegraphics{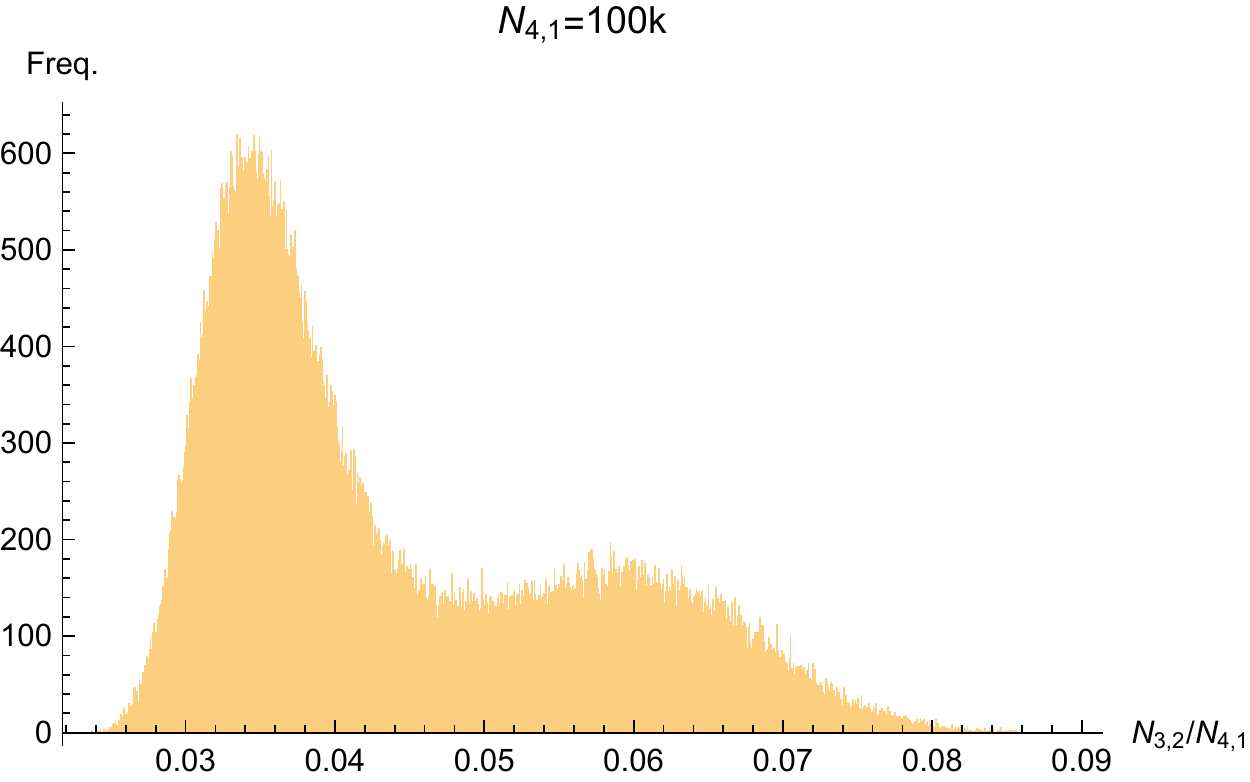}}
  \scalebox{.6}{\includegraphics{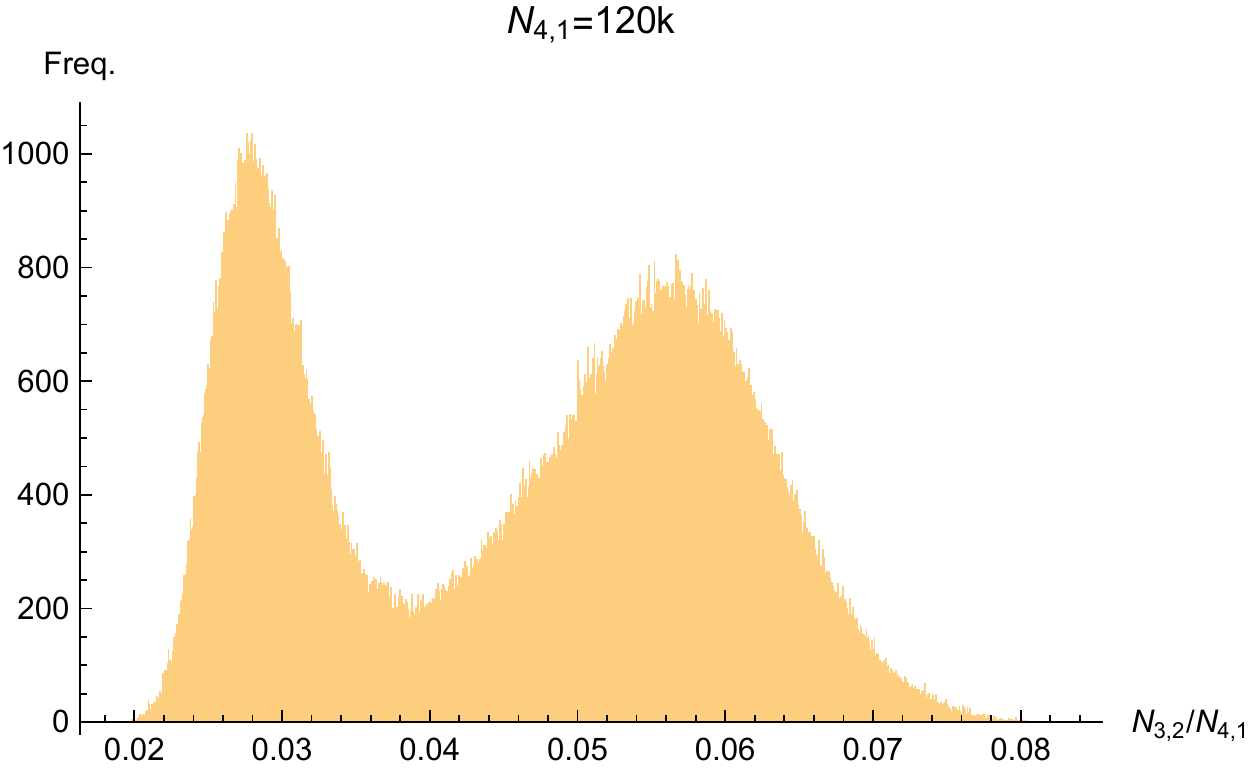}}
  \scalebox{.6}{\includegraphics{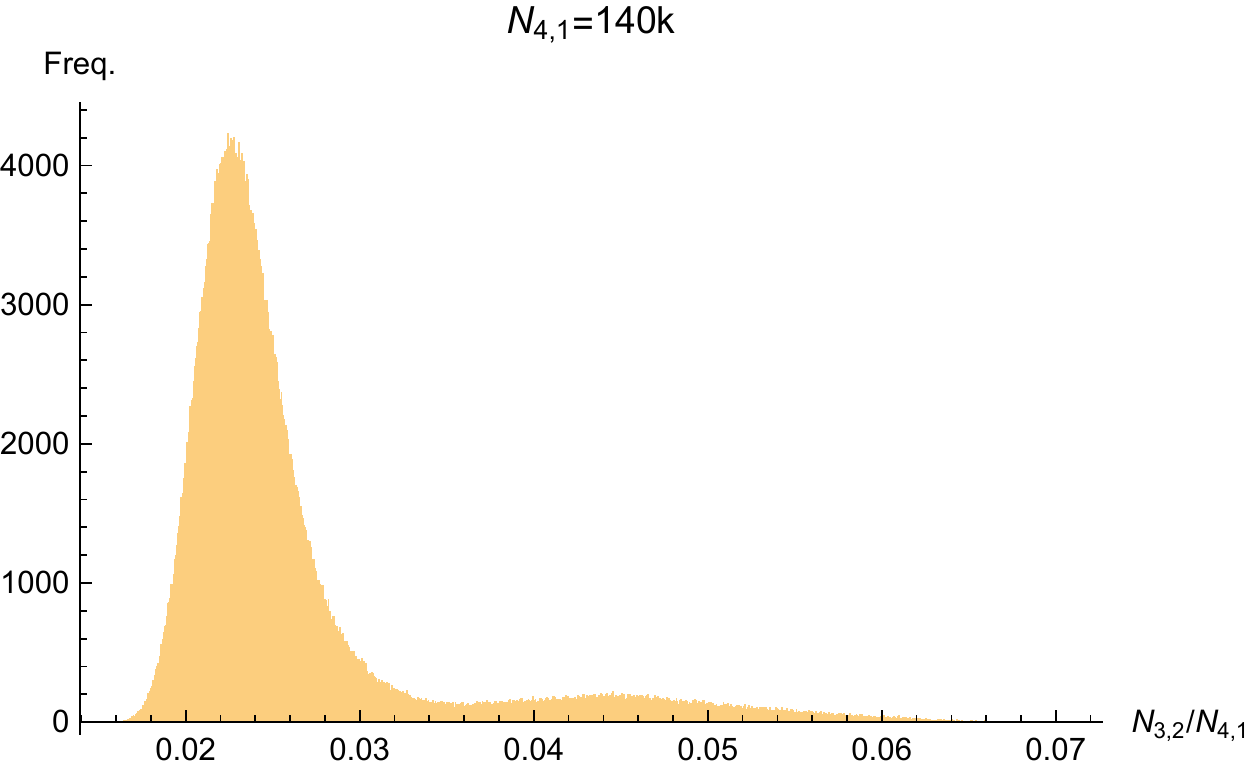}}
  \scalebox{.6}{\includegraphics{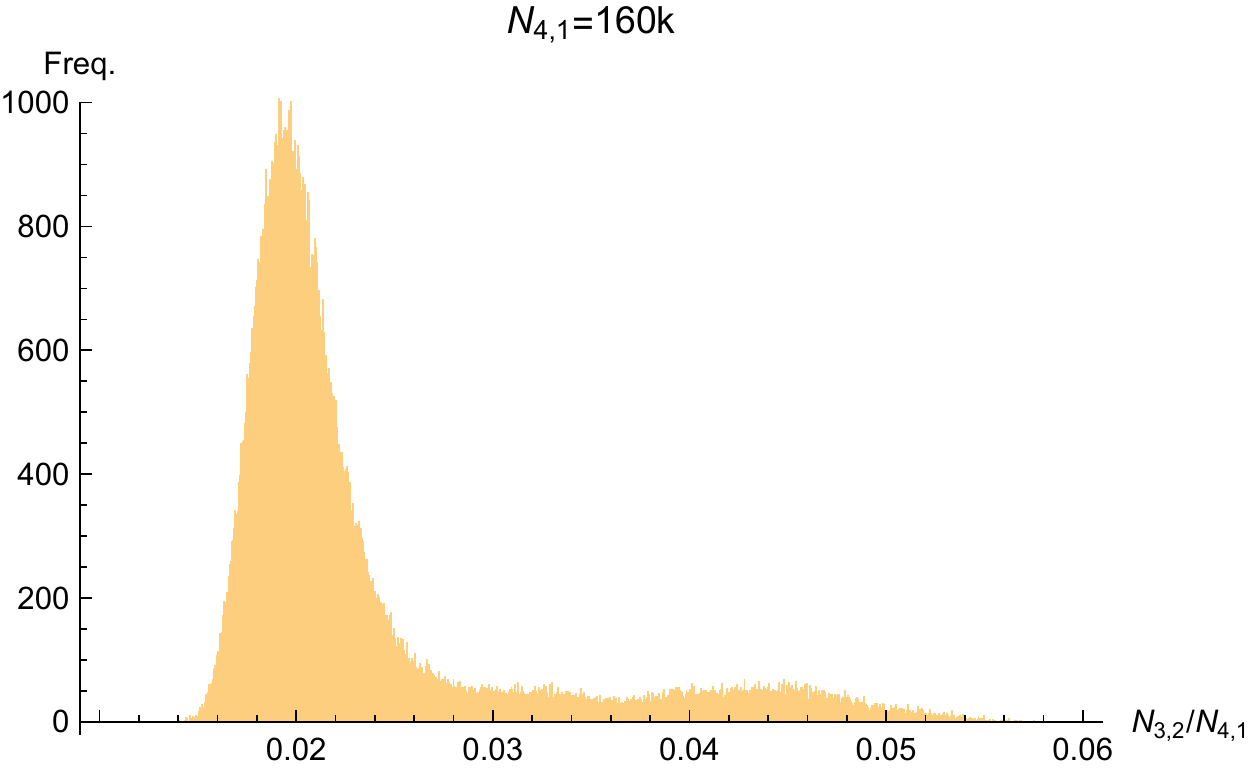}}
  \caption{\small Histogram of $OP_2\equiv N_{3,2}/N_{4,1}$ at pseudo-critical $\kappa_{0}^{crit}$ values for $\Delta=0.6$ with fixed lattice volumes $N_{4,1}=60k, 80k, 100k, 120k, 140k, 160k$. The height of the two peaks is not exactly the same as (due to the $\kappa_0$ resolution applied) some data were measured slightly away from the real pseudo-critical points.}
\label{Fig2}
\end{figure}

The order of a given transition can be quantified by the critical 
exponent $\gamma$ of Eq.\rf{powerlaw} which details how exactly the pseudo-critical values $\kappa_{0}^{crit}$ scale with system size. 
%
%
Applying Eq.~(\ref{powerlaw}) to $\kappa_0^{crit}$ defined by the maxima of susceptibility $\chi_{OP2}$  measured for
all system sizes (see Fig. \ref{Fig3}) yields a shift exponent of $\gamma=1.16 \pm 0.07$, confirming the likely first order nature of the $C$-$A$ transition reported previously in Ref.~\cite{Ambjorn:2012ij}.

 An alternative method for estimating the critical exponent $\gamma$ is to determine the pseudo-critical $\kappa_0^{crit}$ values by locating the minima of $B_{OP_2}$ for different lattice volumes (see Fig. \ref{Fig1}). The $\kappa_0^{crit}$ values determined in this way are shifted to higher values compared with $\kappa_0^{crit}$ based on peaks in $\chi_{OP_2}$, however this discrepancy appears to reduce for larger lattice volumes (see Fig.  \ref{Fig1} and Fig. \ref{Fig3}).\footnote{The minimum of $B_{OP_2}$ could not be determined for $N_{4,1}=60k$ based on the available data due to this shift.} Applying Eq.~(\ref{powerlaw}) to the $\kappa_0^{crit}$ values determined by locating the minima of $B_{OP_2}$ yields $\gamma=1.64 \pm 0.18$, which is statistically inconsistent with $\gamma=1$.   


\begin{figure}[H]
\centering
\scalebox{.6}{\includegraphics{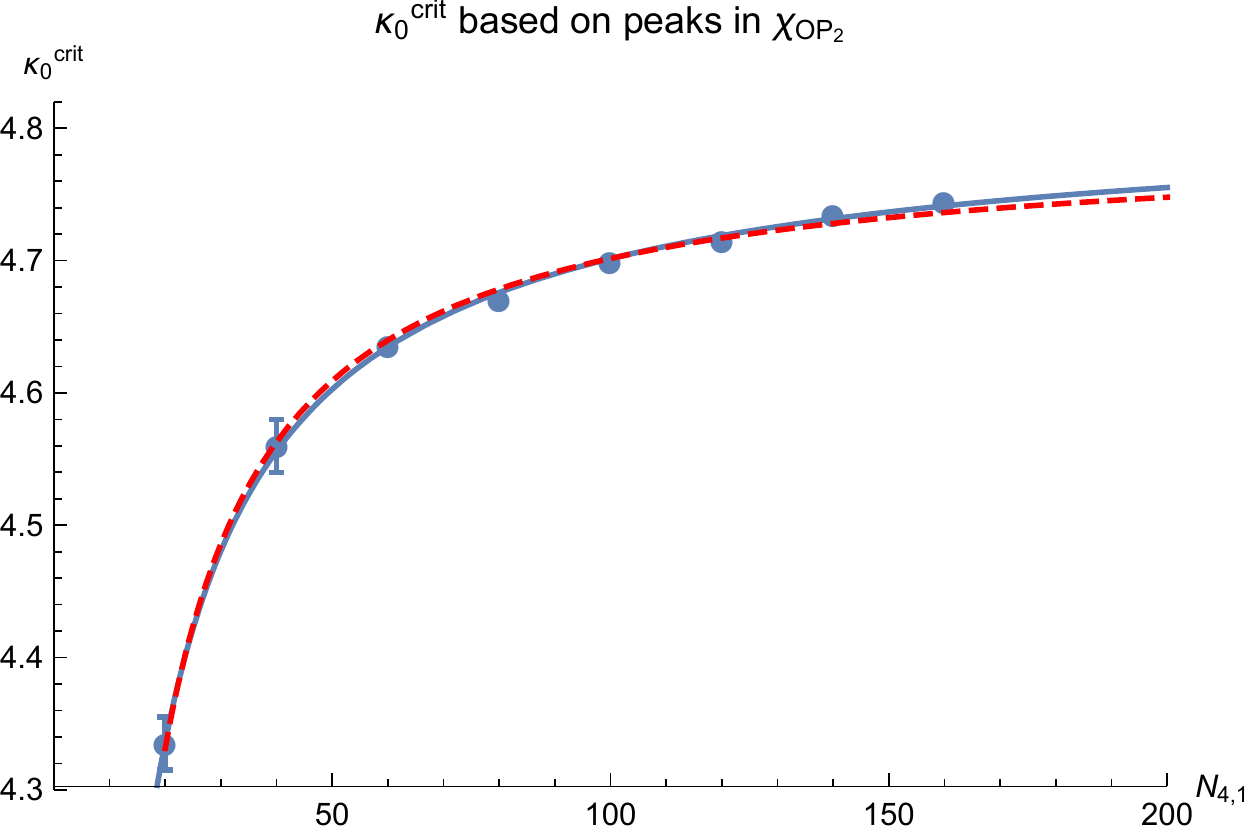}}
\scalebox{.6}{\includegraphics{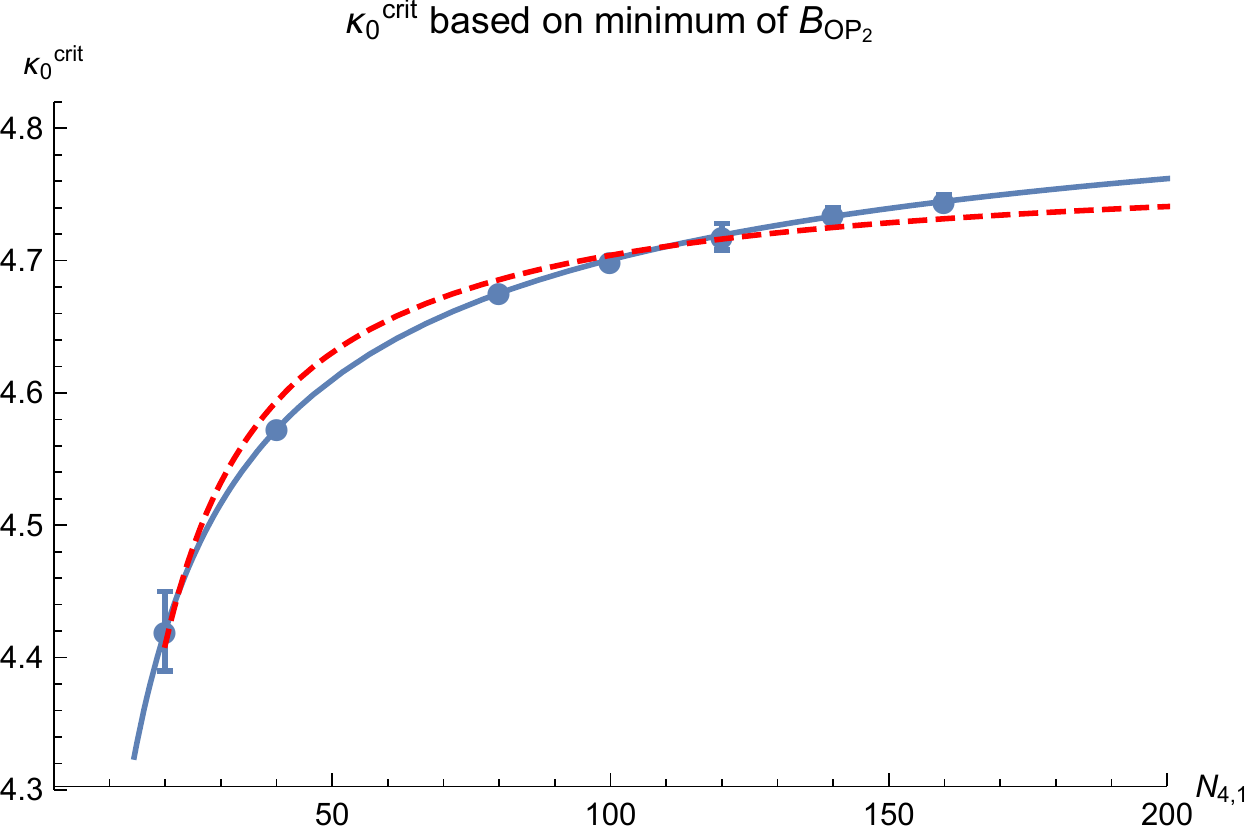}}
\caption{\small The left plot shows the lattice volume dependence of pseudo-critical points $\kappa^{crit}_{0}$ as determined by locating the peak values of $\chi_{OP_{2}}$ for lattice volumes $N_{4,1}=20k, 40k, 60k, 80k, 100k, 120k, 140k, 160k$ together with a fit to Eq.~(\ref{powerlaw}) (solid blue line) for which $\gamma=1.16 \pm 0.07$ and the same fit with a forced value of $\gamma=1$ (dashed red line). The right plot shows the lattice volume dependence of pseudo-critical points $\kappa^{crit}_{0}$ as determined by locating the minimum of $B_{OP_{2}}$ for lattice volumes $N_{4,1}=20k, 40k, 80k, 100k, 120k, 140k, 160k$ together with a fit to Eq.~(\ref{powerlaw}) (solid blue line) for which 
$\gamma=1.64 \pm 0.18$ and the same fit with a forced value of $\gamma=1$ (dashed red line).}
\label{Fig3}
\end{figure}

%
%


One can also measure the minimum (critical) value $B_{OP_2}^{min}$, defined by Eq. \rf{Bmin}, as shown in Fig.~\ref{Fig4}. As the lattice volume increases the value of $B_{OP_2}^{min}$ moves away from zero, which bolsters the conclusion that the $A$-$C$ transition is first order.

\begin{figure}[H]
\centering
\includegraphics[width=0.6\linewidth]{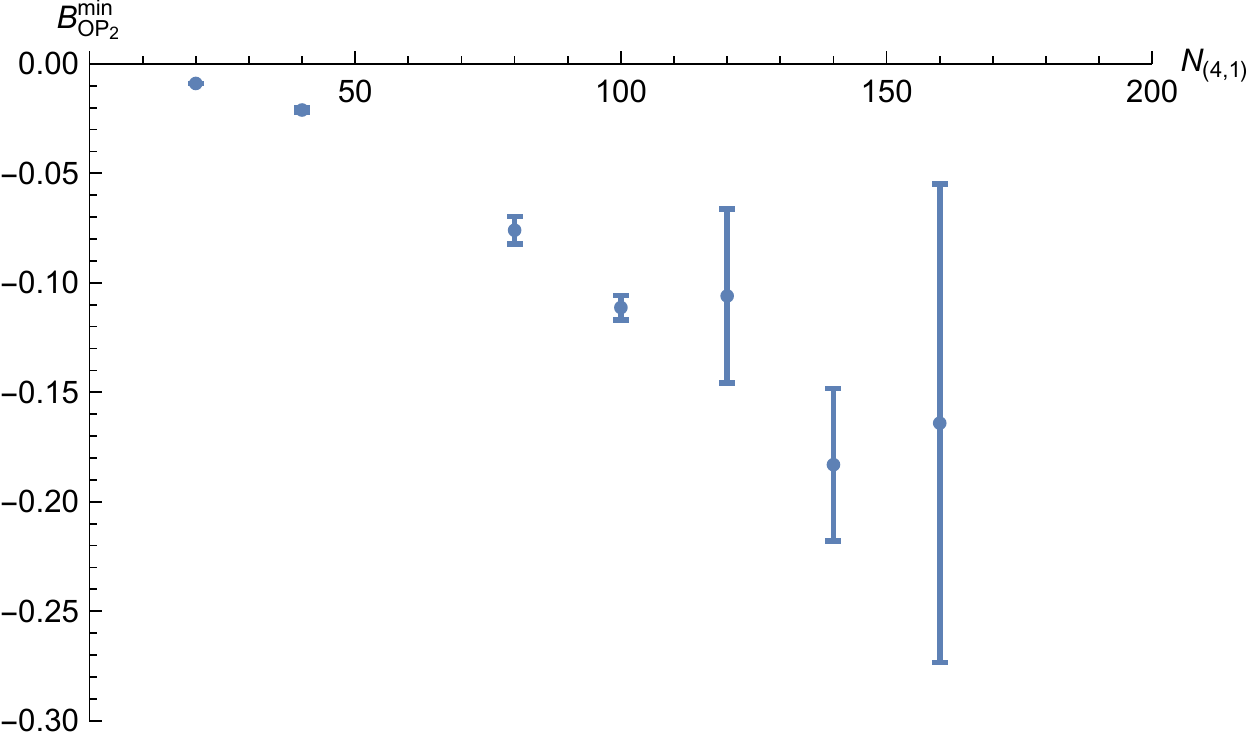}
\caption{\small The dependence of the minimum of the Binder cumulant 
$B_{OP_2}^{min}$ as defined by Eqs.~\rf{binder} and \rf{Bmin} on the system size $N_{4,1}$ for $\Delta=0.6$.}
\label{Fig4}
\end{figure}

\end{subsection}


\begin{subsection}{Impact of the volume fixing method}
\label{t80-1}

In this section we keep the number of time slices fixed at $T=80$ and investigate what impact, if any, the choice of fixing $N_{4}$ or $N_{4,1}$ has on critical phenomena at the $A$-$C$ transition, and in particular whether double peaks 
are present in a histogram analysis using either volume fixing method. 

\begin{subsubsection}{Fixed $N_{4}=120k$ ($T=80$)}\label{SecN4T80}

Here, we successfully reproduce, albeit for a different order parameter, the result published in Ref.~\cite{Ambjorn:2012ij} for the $A$-$C$ transition with $\kappa_{0}=4.710$, $\Delta=0.6$ for fixed $N_{4}=120k$ and using $T=80$ time slices, as shown in Figs.~\ref{N4-1} and~\ref{N4-2}. As can be seen in Fig.~\ref{N4-1} the Monte Carlo time history exhibits near-discontinuous jumps between two metastable 
states characterised by distinct values of $OP_2\equiv N_{3,2}/N_{4,1}$, with the histogram of this data (Fig.~\ref{N4-2}) yielding a clear double 
peak structure. 

\begin{figure}[H]
\centering
\includegraphics[width=0.6\linewidth]{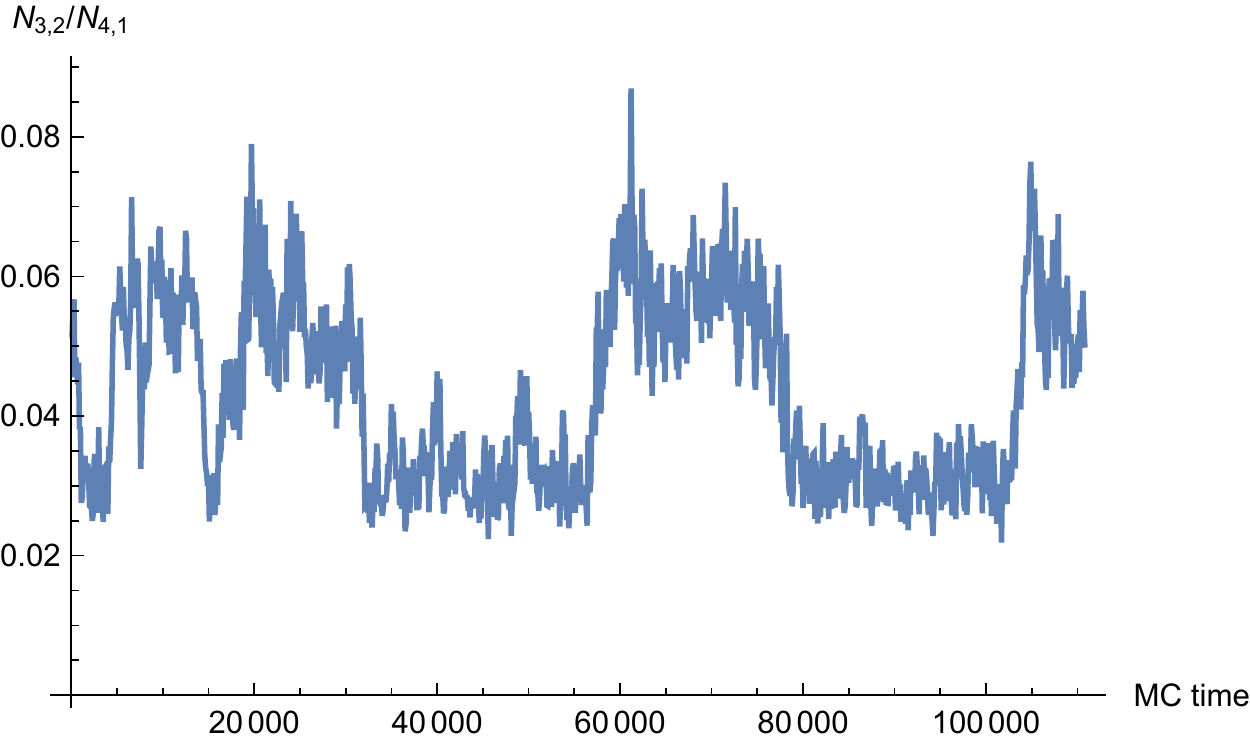}
\caption{\small Monte Carlo time history of $OP_2\equiv N_{3,2}/N_{4,1}$ at $\kappa_{0}=4.710$, $\Delta=0.6$ for fixed $N_{4}=120k$ and with $T=80$ time slices.}
\label{N4-1}
\end{figure}

\begin{figure}[H]
\centering
\includegraphics[width=0.6\linewidth]{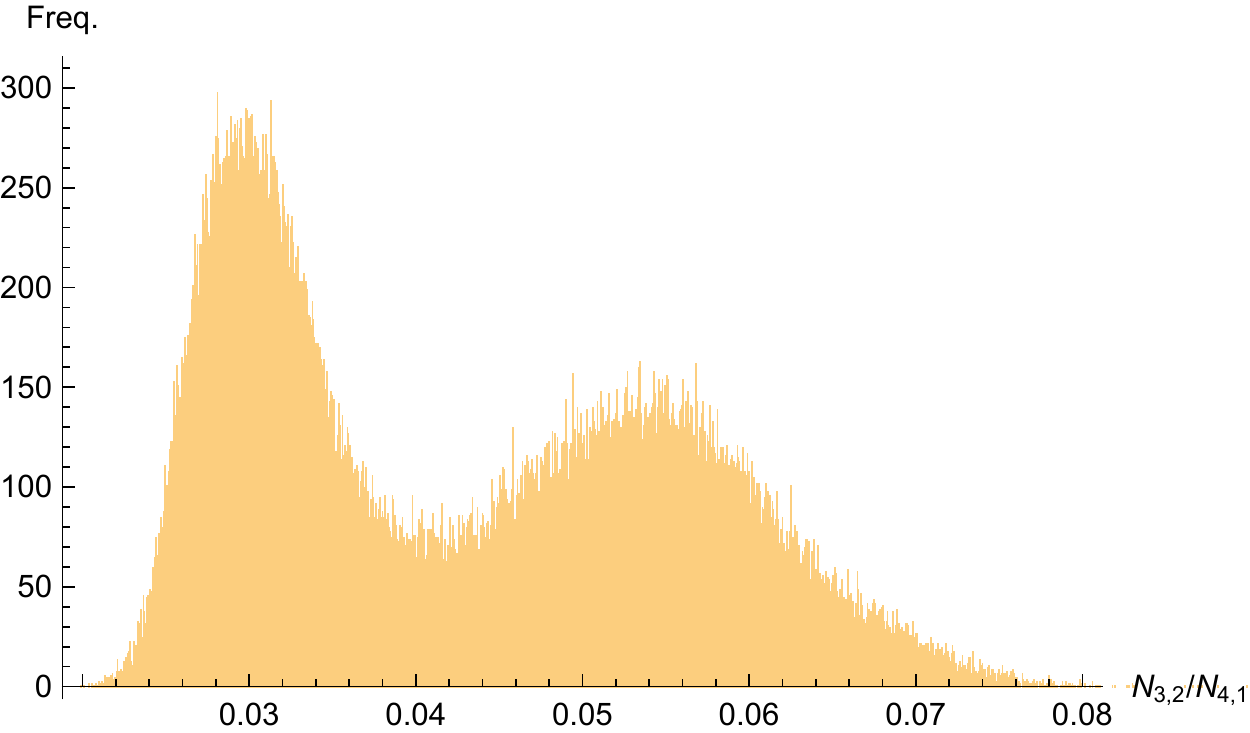}
\caption{\small Histogram of $OP_2\equiv N_{3,2}/N_{4,1}$ values at $\kappa_{0}=4.710$, $\Delta=0.6$ for fixed $N_{4}=120k$ and with $T$=80 time slices.}
\label{N4-2}
\end{figure}

\end{subsubsection}


\begin{subsubsection}{Fixed $ N_{4,1}=115.2k$ ($T=80$)}\label{SecN41T80}

In order to make a more direct comparison between the two volume fixing methods we keep the bare couplings the same as in subsection~\ref{SecN4T80}, namely $\kappa_{0}=4.710$ and $\Delta=0.6$. 
Reading off the value of $\overline {OP_2} \approx 0.04$ that is equidistant between the two Gaussian peaks in the  histogram in Fig. \ref{N4-2} , measured for $\overline{N_4}=120,000$ and using

%
\begin{equation}
\overline{N_{4}}=\overline{N_{4,1}} + \overline{N_{3,2}} =\overline{ N_{4,1} }\left( 1+ \overline{OP_2 }\right),
\end{equation}

\noindent where the over-line denotes an average quantity, 
allows us to determine the average number of $N_{4,1}$ simplices to be $\overline{N_{4,1}}=115,200$ at the transition point. We can therefore keep the same bare couplings as before but fix the number of $N_{4,1}$ simplices at $\overline{N_{4,1}}=115.2k$. The results of this study are shown in Figs.~\ref{N41-1} and~\ref{N41-2}, where it is clear that a double 
peak structure in the histogram is present for the exact same bare couplings when 
$N_{4,1}$ is fixed instead of $N_4$. Thus, a double 
peak in the histogram of $OP_2\equiv N_{3,2}/N_{4,1}$ is present at the transition point regardless of the particular volume fixing method.

\begin{figure}[H]
\centering
\includegraphics[width=0.6\linewidth]{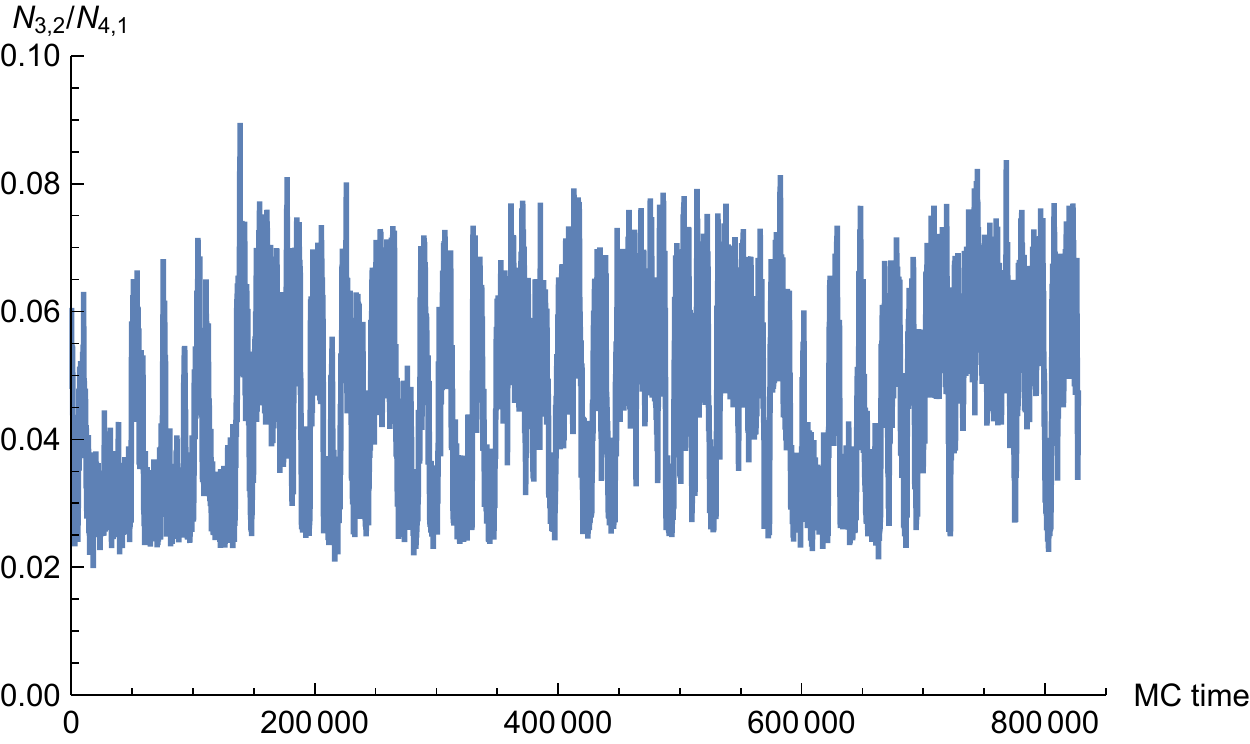}
\caption{\small Monte Carlo time history of $OP_2\equiv N_{3,2}/N_{4,1}$ at $\kappa_{0}=4.710$, $\Delta=0.6$ for fixed $N_{4,1}=115.2k$ and with $T=80$ time slices.}
\label{N41-1}
\end{figure}

\begin{figure}[H]
\centering
\includegraphics[width=0.6\linewidth]{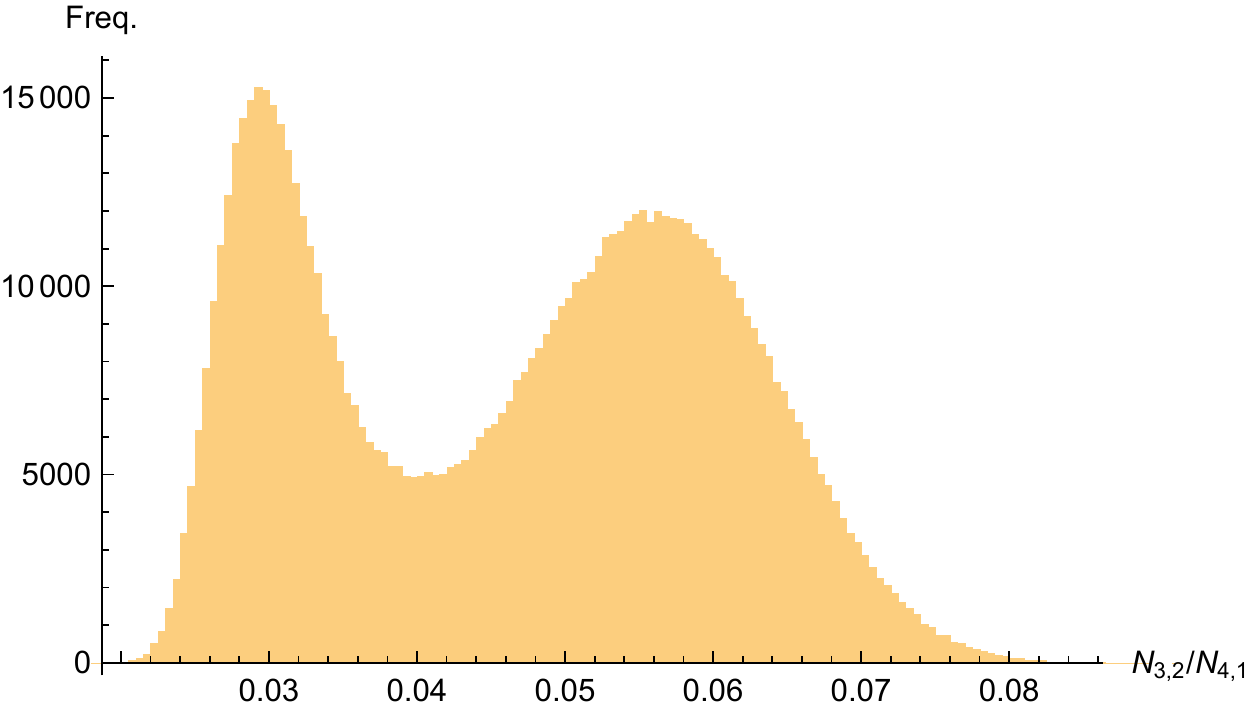}
\caption{\small Histogram of $OP_2\equiv N_{3,2}/N_{4,1}$ values at $\kappa_{0}=4.710$, $\Delta=0.6$ for fixed $N_{4,1}=115.2k$  and with $T$=80 time slices.}
\label{N41-2}
\end{figure}

\end{subsubsection}
\end{subsection}


\begin{subsection}{Impact of the time slicing}\label{tt-1}

As demonstrated in subsection~\ref{t80-1}, the presence of a double 
peak structure in the histogram of $OP_2\equiv N_{3,2}/N_{4,1}$ at the transition point does not seem to depend on the particular volume fixing method. However, might it depend on the number of time slices used in the numerical simulations? In particular, limiting the number of time slices below a certain limit  changes the semiclassical spatial volume profile of phase $C$ from the  de Sitter solution of a $\cos^3$ curve to a flat profile as, due to the limited number of time slices, the blob-like solution cannot form. Consequently, the phase transition no longer ``breaks'' the time-translation symmetry (in the sense described in footnote \ref{foot1}) when moving from  phase $A$  to phase $C$. This in principle may have an impact on the nature of the phase transition. To 
investigate the impact of limiting the CDT proper-time period we repeat the analysis of subsection~\ref{t80-1} using $T=4$ time slices.

\begin{subsubsection}{Fixed $N_{4}=120k$ ($T=4$)}\label{JGS111}

Due to the different number of time slices one may expect a shift in the finite size effects, and so the transition point is expected to shift
. Therefore, we must again locate the pseudo-critical $\kappa_{0}^{crit}$ value by finding the peak of the susceptibility $\chi_{OP_2}$ as a function of $\kappa_{0}$. 
{As shown in Fig.~\ref{N4T4-1}, the peak of susceptibility of $OP_2$ is not pronounced but,  as described in section \ref{OrderOfTrans},  one can observe a clear peak in susceptibility of $\sqrt{OP_2}$ for $\kappa_0\approx4.825$. We then analyse the Monte Carlo time history of $OP_2\equiv N_{3,2}/N_{4,1}$ in the form of histograms in the vicinity of this $\kappa_{0}$ value, as shown in Fig. \ref{N4T4-3}.  No double peak structure in the histogram of $OP_2$ is observed up to a resolution of three decimal places in $\kappa_{0}$. However, at $\kappa_{0}=4.825$ the value of $OP_2$ evolves entirely within one metastable state and at $\kappa_{0}=4.826$ within a distinctly different state (see Fig.~\ref{N4T4-3} for the histograms of this data), and so presumably somewhere within the range $\kappa_{0}=4.825-4.826$ lies the true pseudo-critical $\kappa_{0}^{crit}$ value, although this is yet to be established.}

\begin{figure}[H]
\centering
\includegraphics[width=0.45\linewidth]{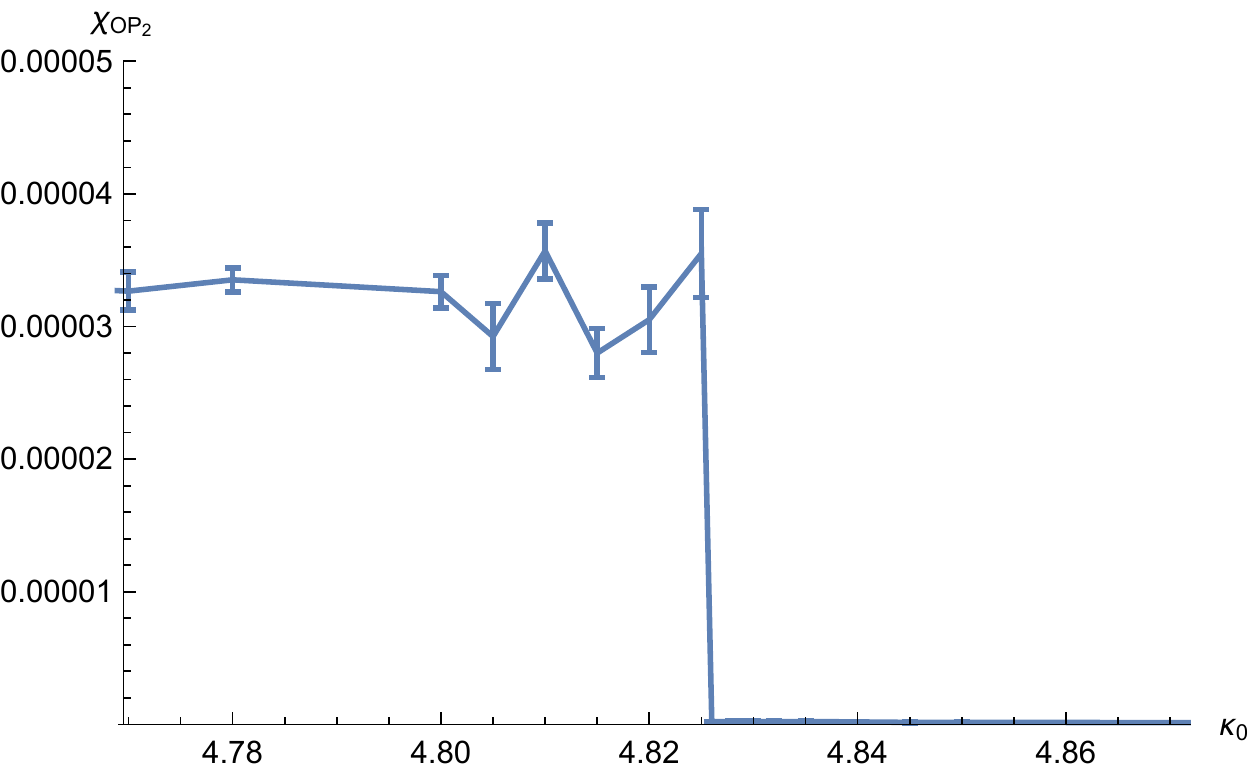}
\includegraphics[width=0.45\linewidth]{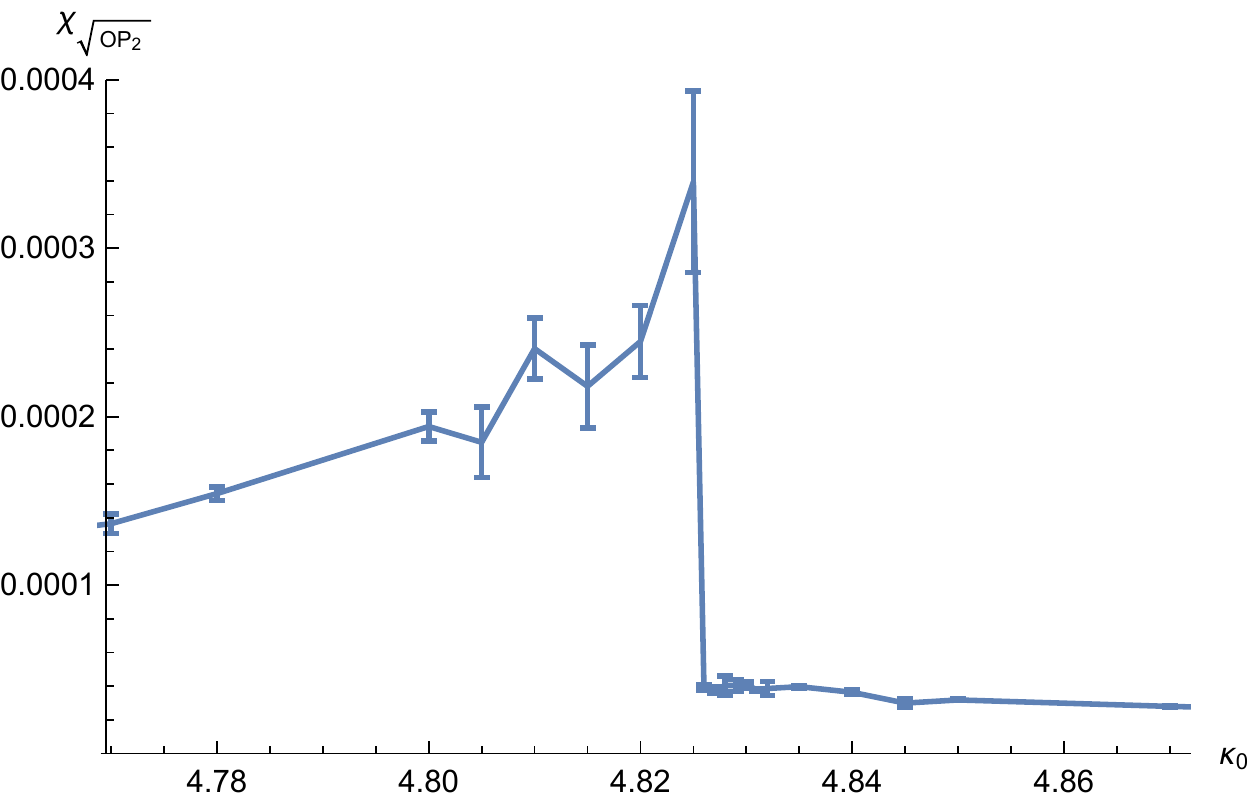}
\caption{\small The susceptibility $\chi_{OP_2}$ as defined by Eq.~\rf{susc} {(left chart) and the susceptibility $\chi_{\sqrt{OP_2}}$ (right chart)} as a function of $\kappa_{0}$ for fixed $N_{4}=120k$ with $T=4$ time slices.}
\label{N4T4-1}
\end{figure}



\begin{figure}[H]
\centering
\includegraphics[width=0.45\linewidth]{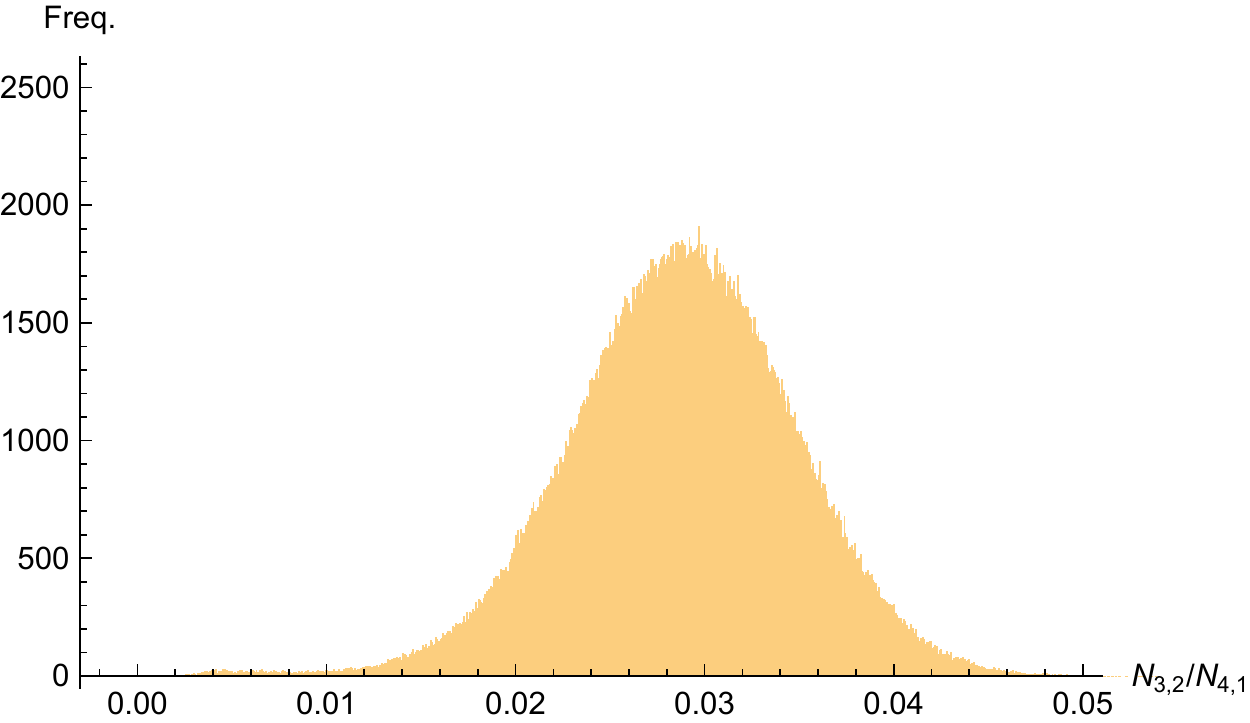}
\includegraphics[width=0.45\linewidth]{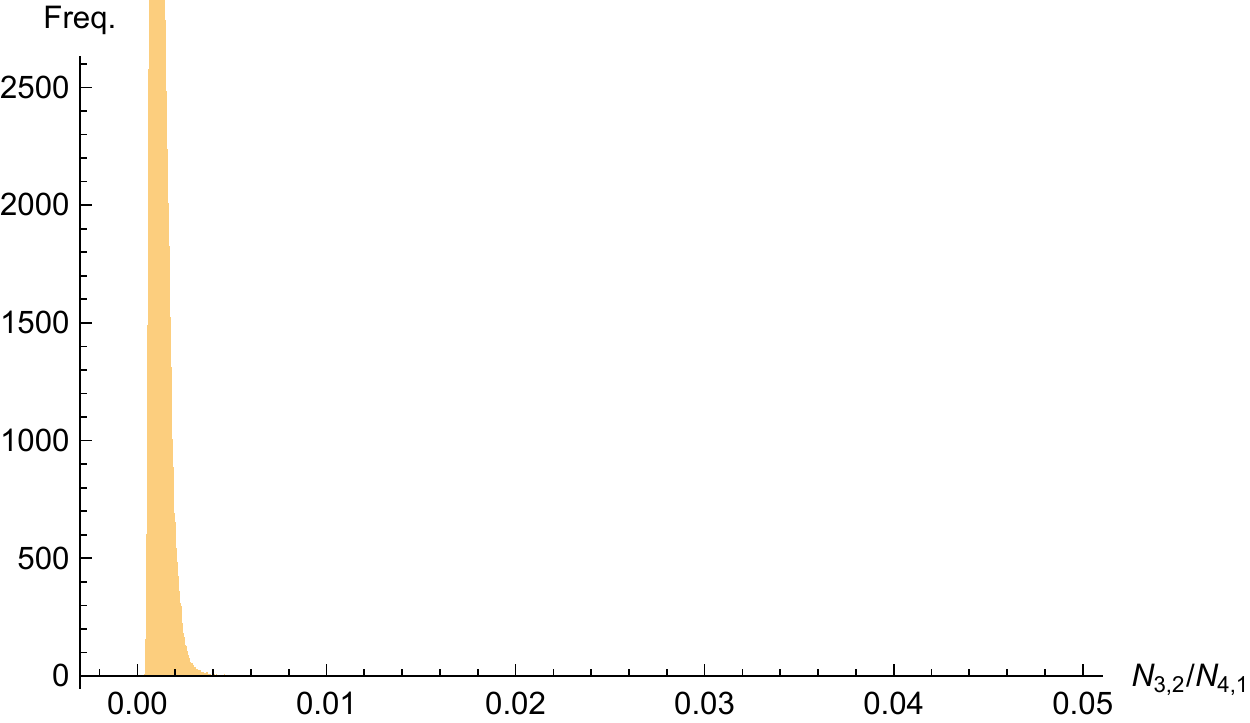}
\caption{\small Histogram of $OP_2\equiv N_{3,2}/N_{4,1}$ values at $\kappa_{0}=4.825$, $\Delta=0.6$ (left) and $\kappa_{0}=4.826$, $\Delta=0.6$ (right)  for fixed $N_{4}=120k$ and with $T=4$ time slices.}
\label{N4T4-3}
\end{figure}

\end{subsubsection}


\begin{subsubsection}{Fixed $N_{4,1}=115.2k$ ($T=4$)}\label{t4v115}

The pseudo-critical $\kappa_{0}$ value is again determined by locating the peak value of the susceptibility $\chi_{OP_2}$ { or (like in section \ref{JGS111}) $\chi_{\sqrt{OP_2}}$ }as a function of $\kappa_{0}$, as shown in Fig.~\ref{N41T4-1}. {Again,} {such an analysis indicates the $\kappa_{0}^{crit}$ value is likely in the range $\kappa_{0}^{crit}=4.828-4.829$, however no double peak structure in the histogram of the Monte Carlo time history of $OP_2$ is observed up to a resolution of three decimal places in $\kappa_{0}$. However, at $\kappa_{0}=4.828$ the value of $OP_2$ evolves entirely within one metastable state and at $\kappa_{0}=4.829$ within a distinctly different state (see Fig.~\ref{N41T4-2} for the histograms of this data), and so presumably somewhere within the range $\kappa_{0}=4.828-4.829$ lies the true pseudo-critical $\kappa_{0}$ value, although this is yet to be established.}



\begin{figure}[H]
\centering
\includegraphics[width=0.45\linewidth]{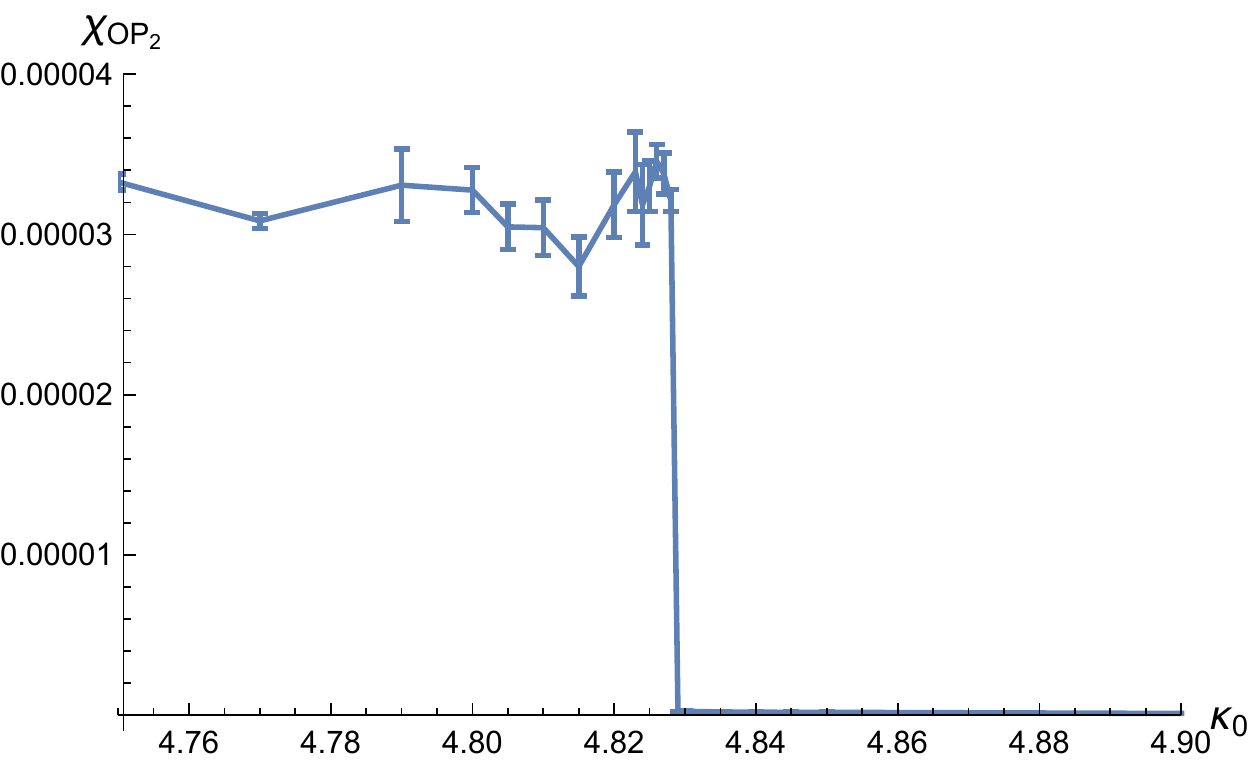}
\includegraphics[width=0.45\linewidth]{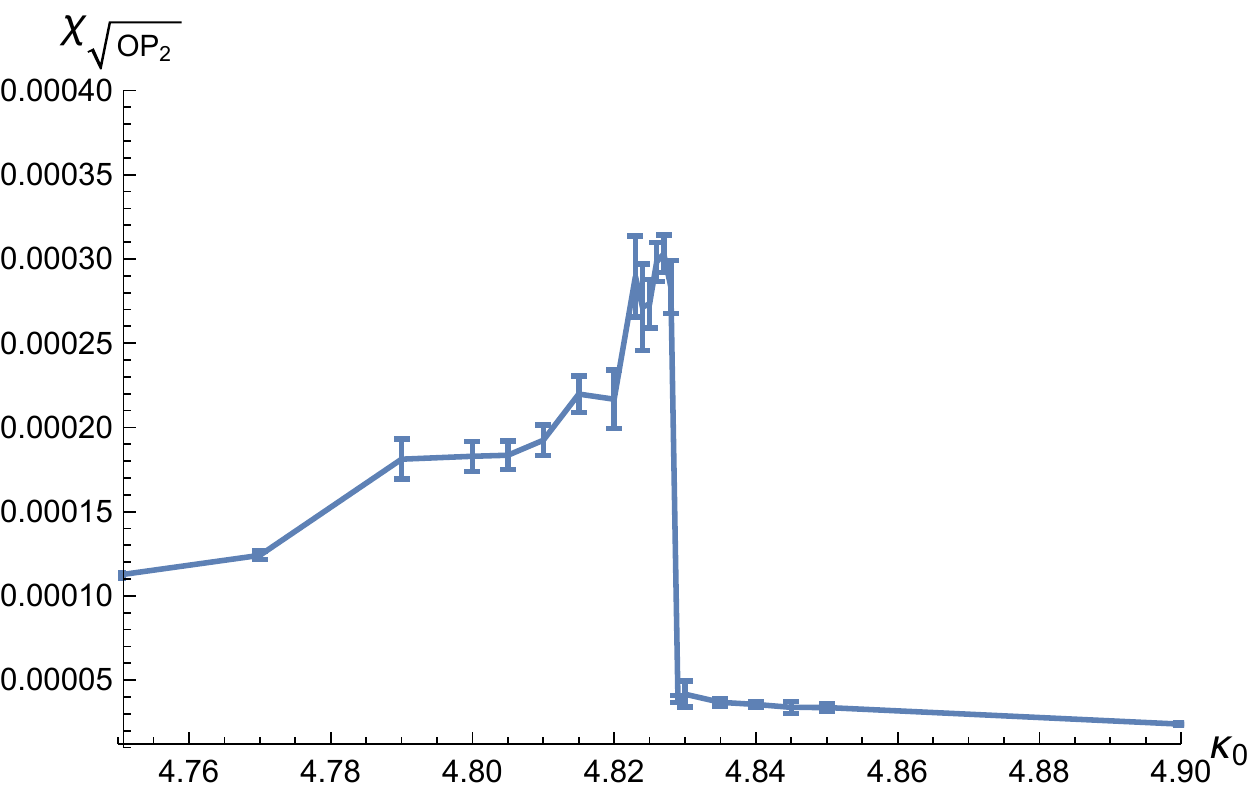}
\caption{\small The susceptibility $\chi_{OP_2}$ as defined by Eq.~(\ref{susc}) {(left chart) and the susceptibility $\chi_{\sqrt{OP_2}}$ (right chart)} as a function of $\kappa_{0}$ for fixed $N_{4,1}=115.2k$ with $T=4$ time slices.  }
\label{N41T4-1}
\end{figure}

\begin{figure}[H]
  \centering
  \scalebox{.6}{\includegraphics{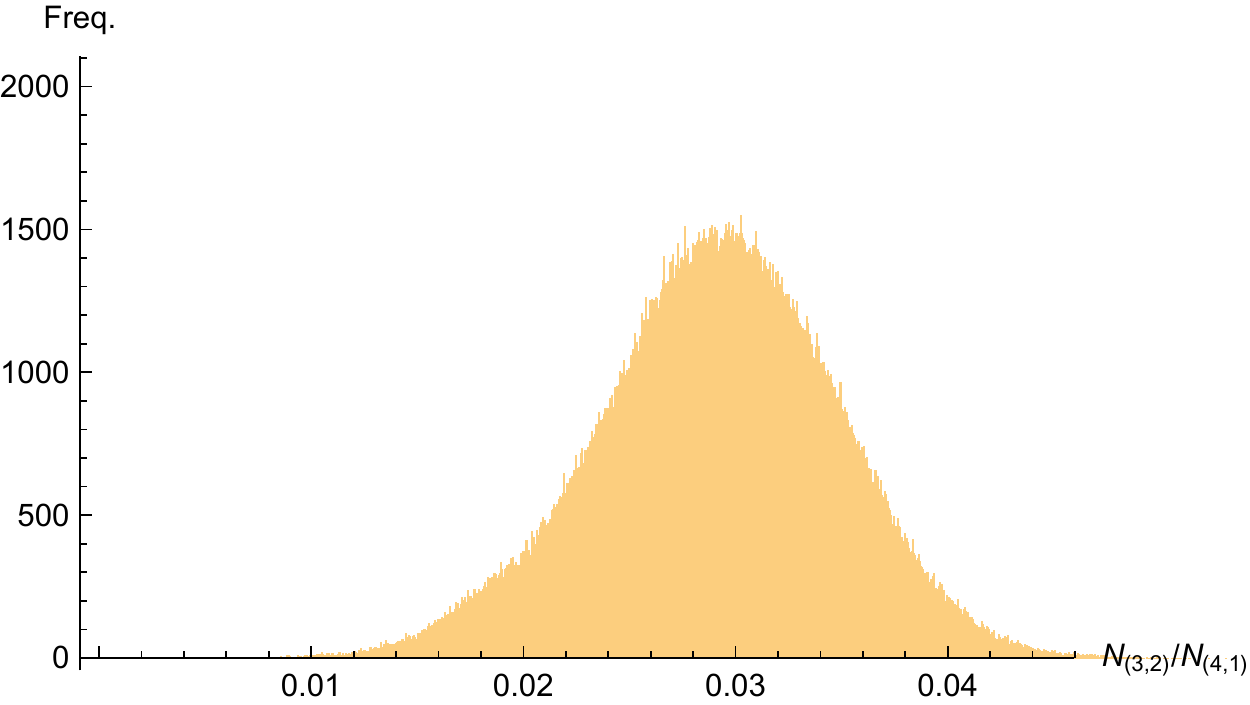}}
  \scalebox{.6}{\includegraphics{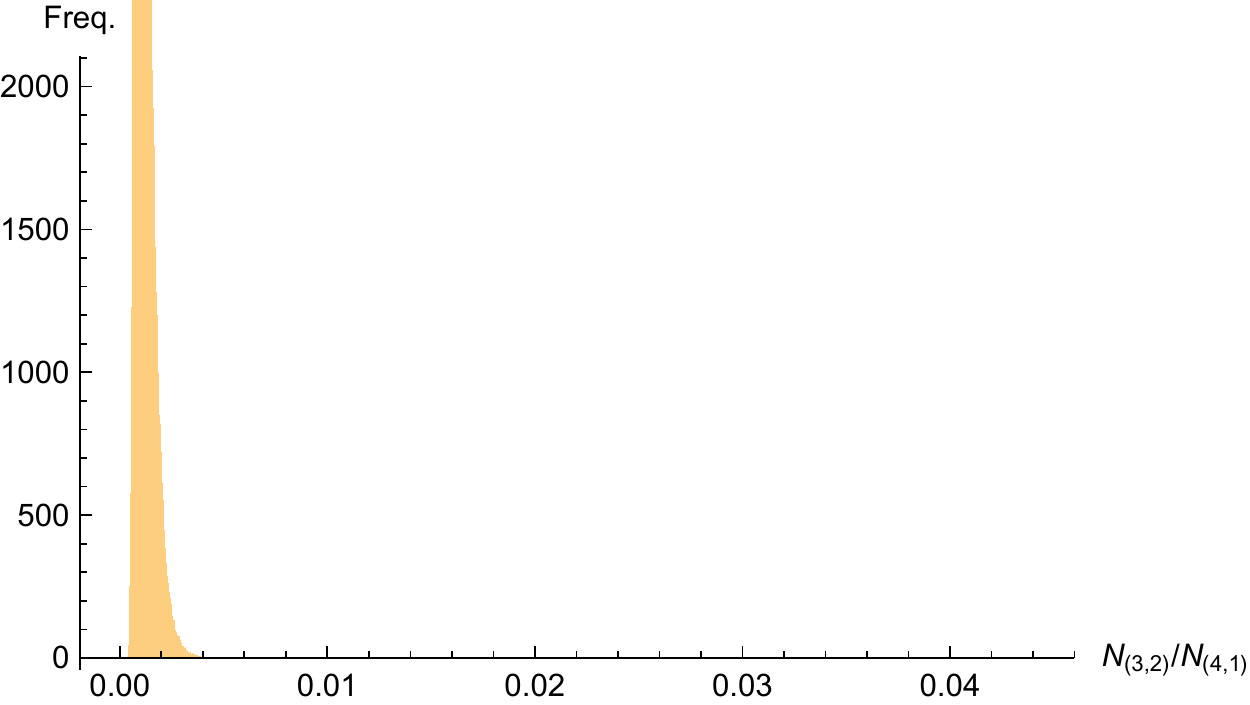}}
\caption{\small {Histogram of $OP_2\equiv N_{3,2}/N_{4,1}$ values at $\kappa_{0}=4.828$, $\Delta=0.6$ (left) and $\kappa_{0}=4.829$, $\Delta=0.6$ (right) for fixed $N_{4,1}=115.2k$ and with $T=4$ time slices.} }
\label{N41T4-2}
\end{figure}



\end{subsubsection}

\end{subsection}

\begin{subsection}{Summary for the spherical topology case.}
Summing up this part, using the $OP_2\equiv N_{3,2} / N_{4,1}$ order parameter we have analysed in detail the $A$-$C$ transition for a system with spherical spatial topology, $T=80$ time slices and $N_{4,1}$ volume fixing. We have confirmed all signatures of the first order transition, i.e. the  $OP_2$ histograms showing double peak structure at the transition points, the scaling exponent of $\gamma\approx 1$ and the divergence of $B_{OP_2}^{min}$ from zero when the lattice volume is increased, as earlier reported in Ref.~\cite{Ambjorn:2012ij} for the system with $N_4$ volume fixed and  the $N_0$ order parameter. We then checked the impact of the volume fixing method and the number of time slices on our results. We  have focused on the  existence  of the double peak structure in the histograms of $OP_2$ measured at the transition points and we have skipped the detailed discussion of the transition point position and the Binder cumulant scaling with lattice volume. The reason is two-fold. Firstly, phase transition studies are very time consuming and thus we were not able to repeat the numerical simulations for all 8 system sizes described in section \ref{sphere1} for all the cases discussed herein, but we focused on gaining large statistics in only  a few (big size) systems. Secondly, the $A$-$C$ transition with the $N_4$ volume fixed was already analysed in detail in Ref.~\cite{Ambjorn:2012ij}, albeit for a different order parameter $N_0$, but we believe that all the previous results concerning finite size scaling become valid also for the $OP_2$ when the $N_4$ volume is fixed. This is corroborated by our measurements of the histogram of the $OP_2$ which behaves exactly the same way as the $N_0$ parameter described in Ref. \cite{Ambjorn:2012ij}. 
{All the results for the $T=80$ time slices show a double peak structure in the histogram of $OP_2$ measured at the transition points, which provides strong evidence that the transition is first order independent of the volume fixing method. {The  ambiguous results come from measurements 
with $T=4$ time slices (see section~\ref{tt-1}). In this case, we can not confirm a double peak structure in the histograms up to a resolution of three decimal places in $\kappa_0$. However, an analysis of the Monte Carlo time evolution either side of the putative transition suggests a double peak structure is likely to emerge for a greater resolution of $\kappa_{0}$.} This is probably a consequence of finite size effects being different for the $T=4$ systems compared to the  $T=80$ triangulations, where the effective volume per slice is much lower. As a result, the separation of the metastable states is much more pronounced in the $T=4$ systems making the transition much sharper. Such a behaviour corroborates  the first order nature of the $A-C$ transition.} 

\end{subsection}

\end{section}

\begin{section}{Toroidal topology}\label{toroidal}

\begin{subsection}{The order of the $A$-$C$ transition}\label{J1}
We will now study the impact of topology change  on the properties of the $A$-$C$ transition. We will focus on systems with toroidal  topology of spatial slices, which we started to investigate some time ago (for details see Refs. \cite{Ambjorn:2016fbd,Ambjorn:2017ogo,kevin,Ambjorn:2018qbf}). We start our discussion from a system with $T=4$ time slices and with $N_{4,1}$ volume fixing. The reason for this choice is two-fold. Firstly, the systems with such parameters were earlier used to explore the toroidal CDT phase  diagram presented in Ref. \cite{Ambjorn:2018qbf}, 
where it was noted that the order parameters change smoothly between the $A$ and $C$  phases and one
does not observe any separation of states, nor any double peak structure on the $OP$ histograms at the transition points. Such a behaviour may suggest that the $A$-$C$ transition is now higher order. Therefore we want to check this in detail by performing a proper finite-size scaling analysis. Secondly, as noted in Ref. \cite{Ambjorn:2016fbd}, it seems that due to a much larger minimal triangulation the finite size effects are bigger in toroidal CDT compared to spherical CDT  and  thus in the former case one should use systems with a much larger spatial volume for single time slices, which is obtained for small $T$. We will then investigate the impact that both time slicing and volume fixing has on our results.
 
 We  again focus on the $OP_2$ parameter as defined by Eq.\rf{op2}. 
 Measurements presented in this section were  made for the same choice of $\Delta=0.6$ as for the spherical topology case, discussed in section \ref{fields}. We analysed systems with  $T=4$ time slices  for 11 different lattice volumes\footnote{The results for the Binder cumulant were obtained only for 8 lattice volumes: $N_{4,1} = 20k, 40k, 60k, 100k, 200k, 300k, 400k$ and $500k$.}: $N_{4,1} = 20k, 40k, 60k, 80k, 100k, 120k, 160k, 200k, 300k, 400k$ and $500k$. In order to  ensure a proper thermalization of our data, and to estimate the accuracy of our measurements, as well as to check possible hysteresis effects we performed two independent runs for each Monte Carlo simulation, one initiated with a  configuration from phase $A$ and another one with a configuration from phase $C$. After thermalizing both runs should converge and give (statistically) similar results. Data collected during the thermalization period is not included in our final measurements. 

 As already discussed in Section \ref{OrderOfTrans}, for toroidal topology  one cannot observe the susceptibility peaks when using $OP_2$, but they can be observed when using  $\sqrt{OP_2}$, thus in the following we will focus on this latter function. We again start by locating the pseudo critical $\kappa_0^{crit}$ values by finding maxima in the susceptibility $\chi_{\sqrt{OP_2}}\approx \frac{\chi_{OP_2}}{\langle OP_2\rangle}$ as a function of $\kappa_0$ (see Fig.~\ref{FigT1}) separately for each lattice volume. 
 
 \begin{figure}[H]
  \centering
  \scalebox{.6}{\includegraphics{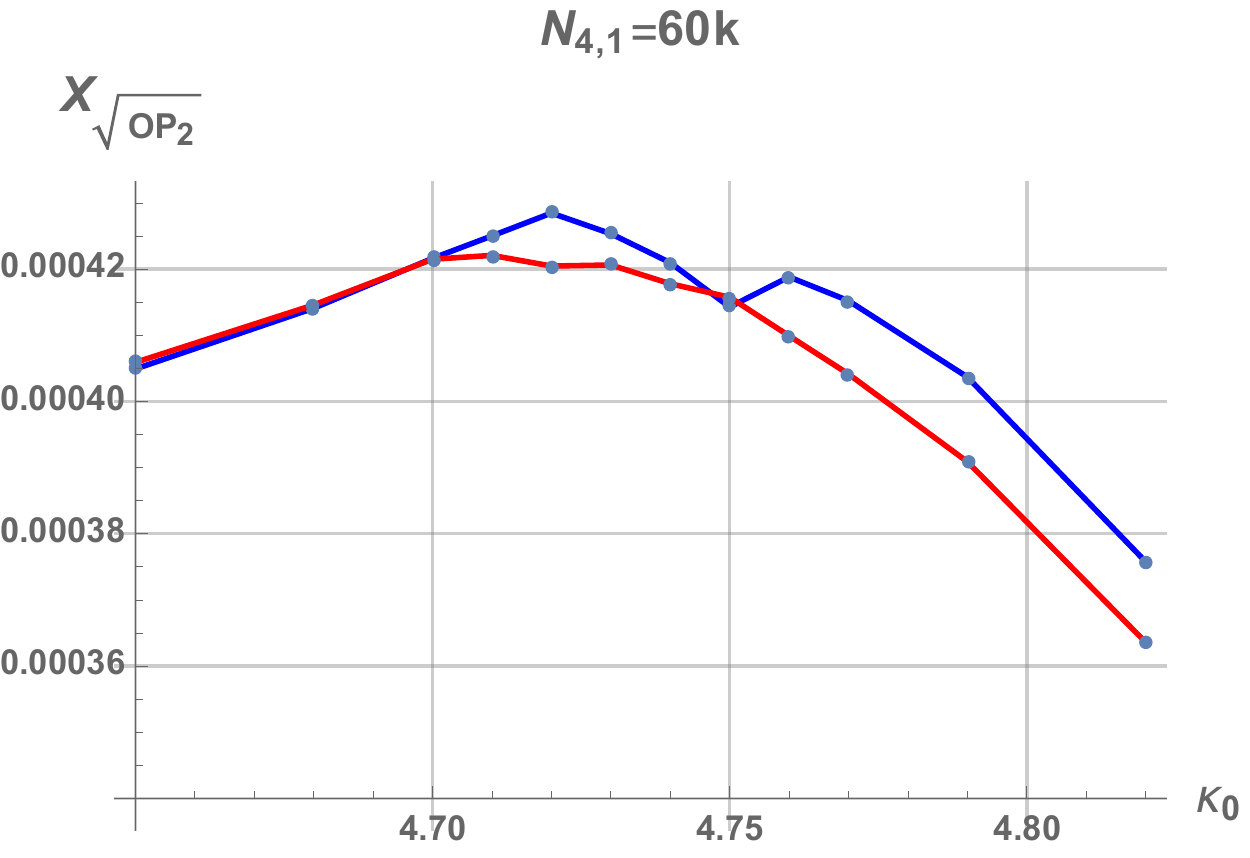}}
  \scalebox{.6}{\includegraphics{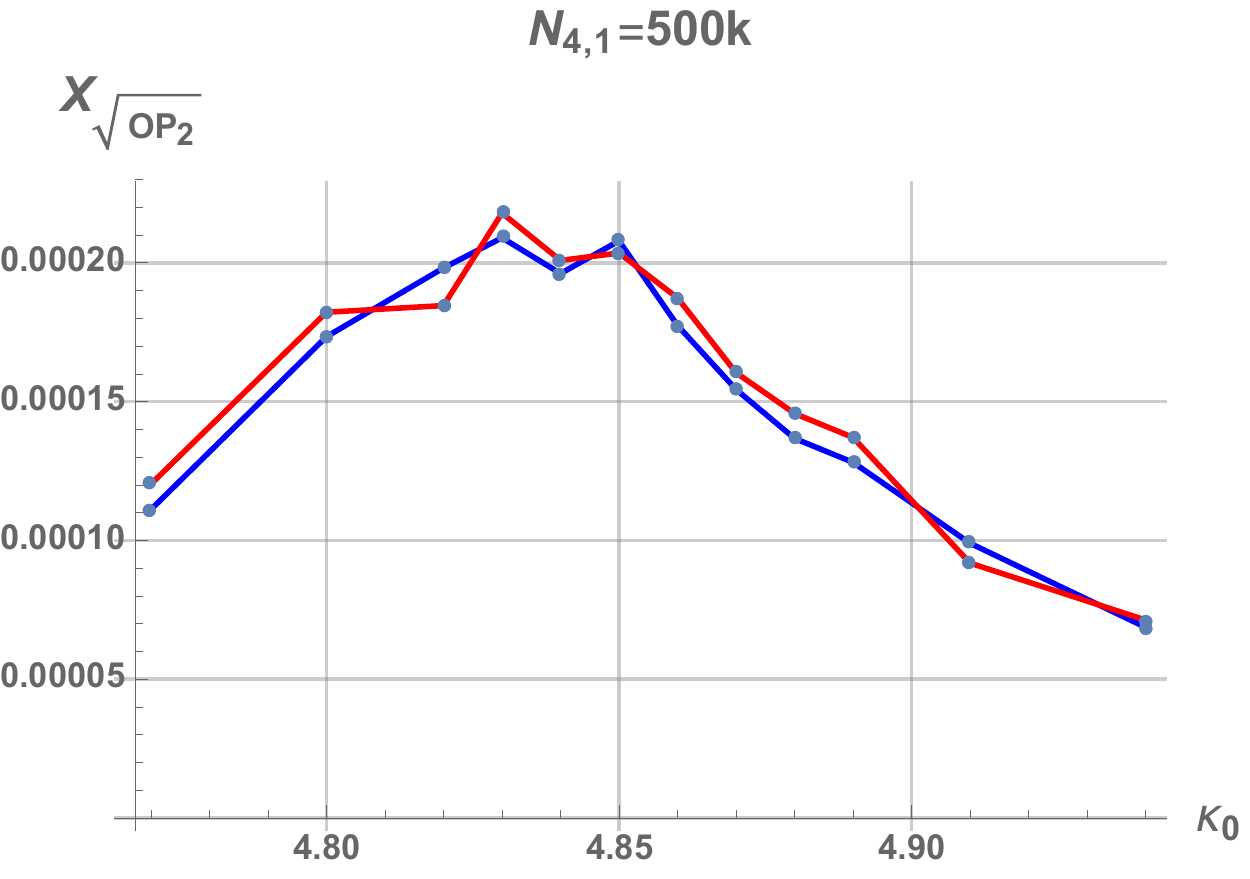}}
\caption{\small The susceptibility $\chi_{\sqrt{OP_2}}\approx \frac{\chi_{OP_2}}{\langle OP_2\rangle} $ as defined by Eqs.~(\ref{susc}) and \rf{suscsqrt}
as a function of $\kappa_{0}$ for $N_{4,1}=60k$ (left) and $N_{4,1}=500k$ (right). The blue line corresponds to simulations initiated deep in phase $A$ and the red line to simulations initiated deep in phase $C$.
}
\label{FigT1}
\end{figure}

 One can see that the peaks in Fig.~\ref{FigT1} are relatively flat and despite the long simulation time\footnote{The Monte Carlo simulations described in this section had around $5 \times 10^{12}$ attempted MC moves each and they took almost half a year to complete.} there is still a small discrepancy between data measured using the two alternative starting configurations, yielding  relatively large error bars for the position of the pseudo-critical $\kappa_0^{crit}$ values, as shown in Figure \ref{FigT2}. The position of the susceptibility peaks and the resulting position of the pseudo-critical points at which the transition occurs shows a strong volume dependence. One can again use the finite-size scaling relation of equation \rf{powerlaw} to fit the critical exponent $\gamma$ to the measured $\kappa_0^{crit}(N_{4,1})$ data (see Fig. \ref{FigT2}). The best fit yields a shift exponent of $\gamma = 1.22 \pm 0.08$ which is slightly higher than the $\gamma$ value measured for the spherical case (see section \ref{sphere1}), but still consistent with $\gamma=1$ of a first order transition. 

\begin{figure}[H]
  \centering
  \scalebox{.75}{\includegraphics{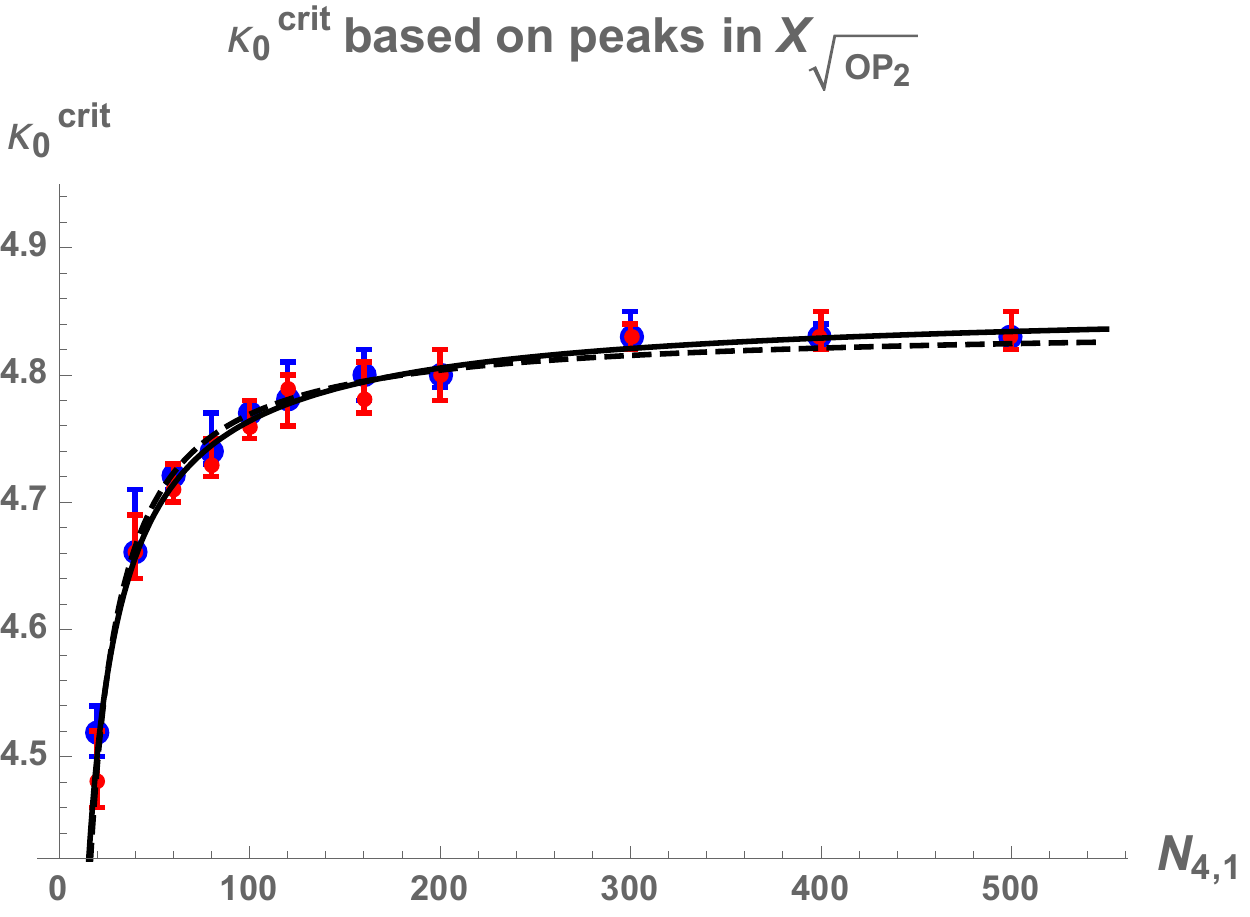}}
  \caption{\small Lattice volume dependence of pseudo-critical points $\kappa_0^{crit}$ based on the position of susceptibility $\chi_{\sqrt{OP_2}}$ peaks measured  for 11 different lattice volumes $N_{4,1} = 20k, 40k, 60k, 80k, 100k, 120k, 160k, 200k, 300k, 400k, 500k$ together with a fit to Eq. \rf{powerlaw} (solid line) for which $\gamma = 1.22 \pm 0.08$ and the same fit with a forced value of $\gamma = 1$ (dashed line). The blue data are for simulations initiated deep in phase $A$ and the red data for simulations initiated deep in phase $C$.
}
\label{FigT2}
\end{figure}

 \begin{figure}[H]
  \centering
  \scalebox{.6}{\includegraphics{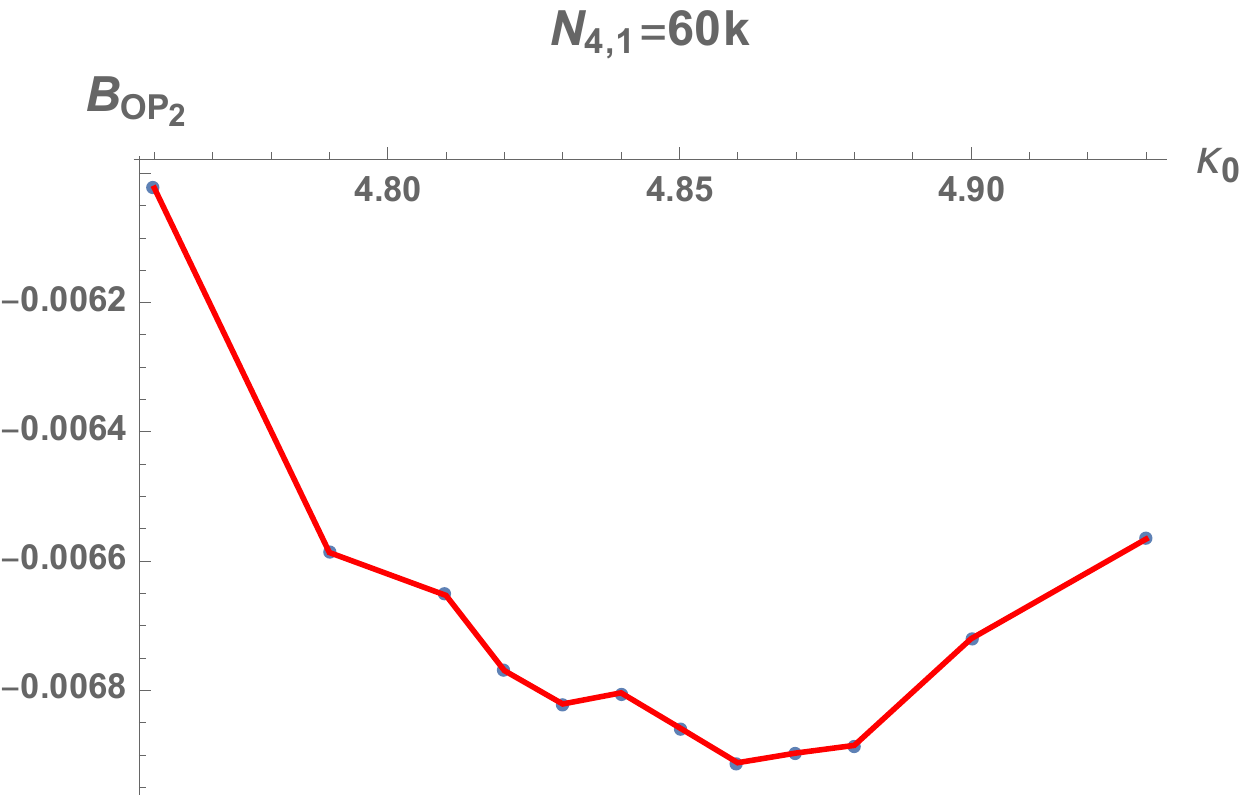}}
  \scalebox{.6}{\includegraphics{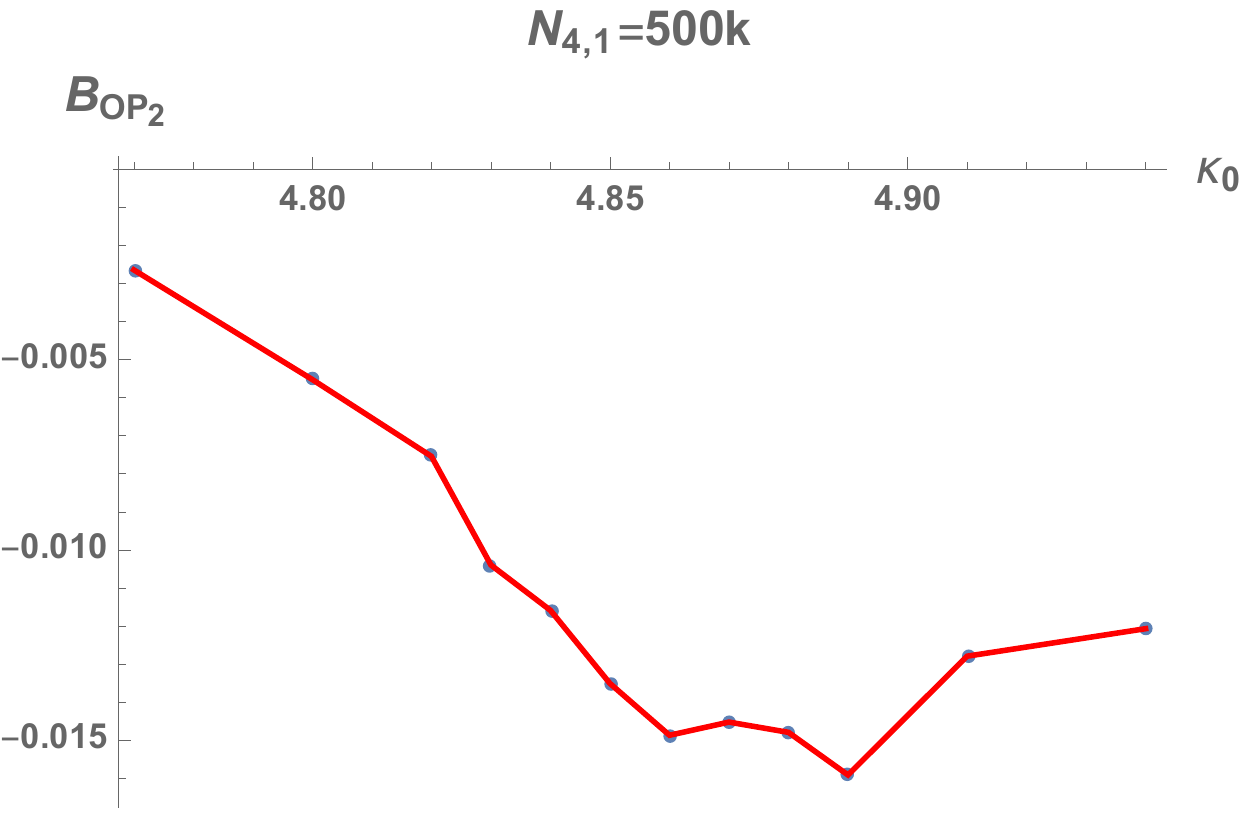}}
\caption{\small The Binder cumulant $B_{OP_2}\approx B_{\sqrt{OP_2}}\approx {-\chi_{\ln OP_2}} $ as defined by Eqs.~(\ref{binder}) (and Eqs. \rf{relations}, \rf{relations3})
as a function of $\kappa_{0}$ for {$N_{4,1}=60k$} (left) and $N_{4,1}=500k$ (right). 
}
\label{FigT3}
\end{figure}

One can repeat the above finite size analysis, now looking at  minima (see Fig. \ref{FigT3}) of  the Binder cumulant of ${OP_2}$ defined in Eq. \rf{binder}.\footnote{As discussed in section \ref{OrderOfTrans}, it is also consistent with the Binder cumulant of $\sqrt{OP_2}$ and with the  susceptibility of $\ln{OP_2}$ (see Eqs. \rf{relations} and \rf{relations3}): $B_{OP_2}\approx B_{\sqrt{OP_2}}\approx -\chi_{\ln OP_2}$.} The Binder cumulant (as a function of $\kappa_0$) was measured for 8 different lattice volumes: $N_{4,1} = 20k, 40k, 60k, 100k, 200k, 300k, 400k$ and $500k$. The pseudo-critical $\kappa_0^{crit}$ values located this way are shifted towards a higher $\kappa_0$  from the values measured by the location of susceptibility peaks, discussed above. This kind of discrepancy is possible as positions of the Binder cumulant minima are expected to show smaller finite size dependence than the positions of the susceptibility maxima. This is indeed the case as illustrated in Fig. \ref{FigT4}, where we plot the  lattice volume dependence of $\kappa_0^{crit}$ values based on the Binder cumulant minima. 

\begin{figure}[H]
  \centering
  \scalebox{.75}{\includegraphics{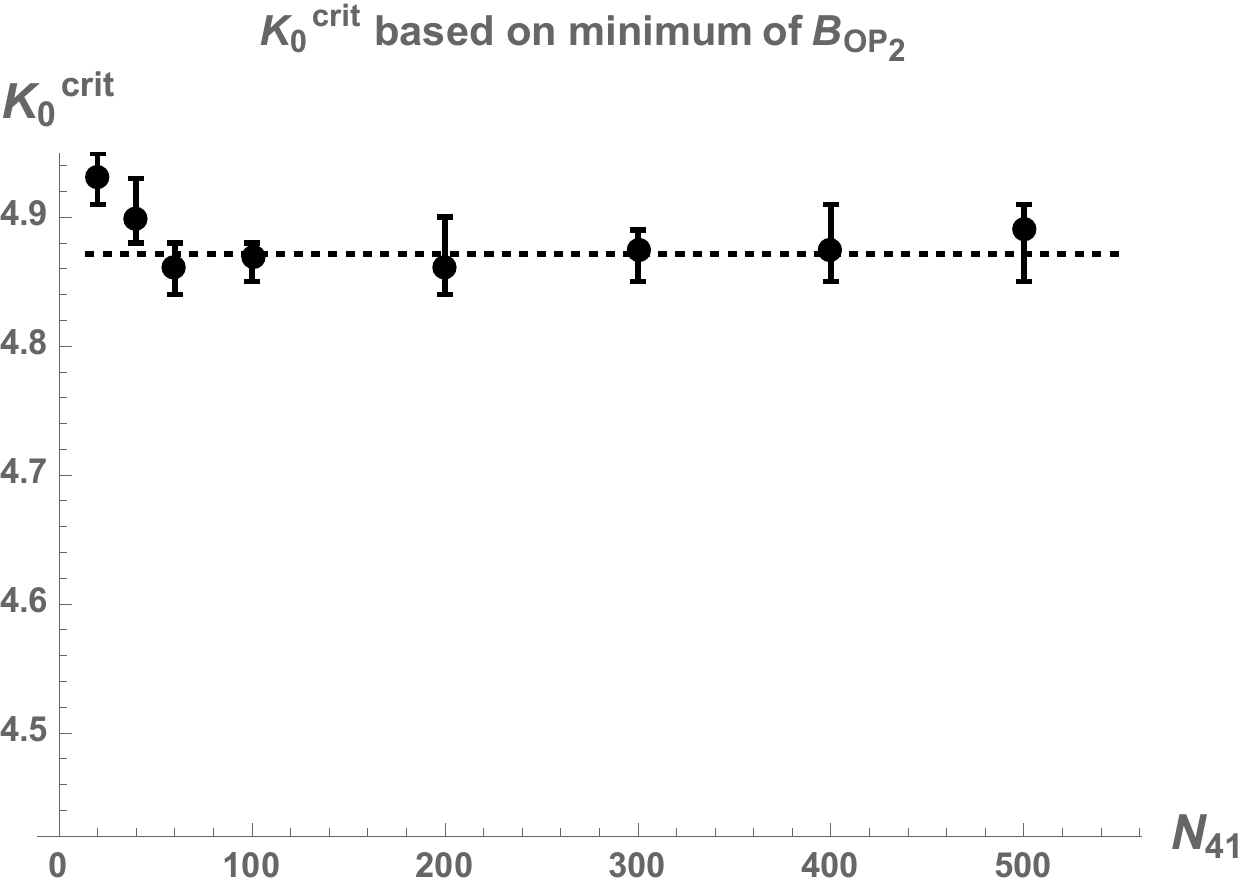}}
  \caption{\small Lattice volume dependence of pseudo-critical points $\kappa_0^{crit}$ based on the position of the Binder cumulant $B_{{OP_2}}$ minima measured  for 8 different lattice volumes $N_{4,1} = 20k, 40k, 60k, 100k, 200k, 300k, 400k, 500k$ together with an estimate of the true critical value of $\kappa_0^{crit}(\infty)= 4.87 \pm 0.01$ (dotted line). 
}
\label{FigT4}
\end{figure}

In Fig.\ref{FigT4} we keep the same range of the vertical axis as in Fig. \ref{FigT2} so that it is clearly visible that the pseudo-critical value $\kappa_0^{crit}$ is now (almost) volume independent. This may suggest that, due to smaller finite size effects when using the Binder cumulants, one has reached the system size at which the measured pseudo-critical $\kappa_0^{crit}(N_{4,1})$ values are very close to the true critical $\kappa_0^{crit}(\infty)$ value in the infinite volume limit. By taking the mean value of   $\kappa_0^{crit}(N_{4,1})$ measured for $N_{4,1} \geq 60k$~\footnote{We have excluded the first two data points, for $N_{4,1}=20k$ and $40k$, from the mean as they still show a small volume dependency - see Fig.~\ref{FigT4}.} one can estimate the true critical value to be $\kappa_0^{crit}(\infty)= 4.87 \pm 0.01$.
One can then use this estimate to refit the pseudo-critical $\kappa_0^{crit}(N_{4,1})$ values, measured using the susceptibility peaks method, to the finite-size scaling relation \rf{powerlaw}, now with forced value of $\kappa_0^{crit}(\infty)= 4.87$. This way one obtains a slightly corrected estimate of the critical exponent of $\gamma = 1.31 \pm 0.03$. {One should note that the above accuracy of the  $\gamma$ exponent is just the fit error and it does not take into account statistical errors of the data points. The result is now less consistent with $\gamma=1$ expected from a first order transition, but, as presented in Figure \ref{FigT5},  the critical exponent of $\gamma=1$ cannot be excluded within the error bars of the measured $\kappa_0^{crit}$ data points.}  


\begin{figure}[H]
  \centering
  \scalebox{.75}{\includegraphics{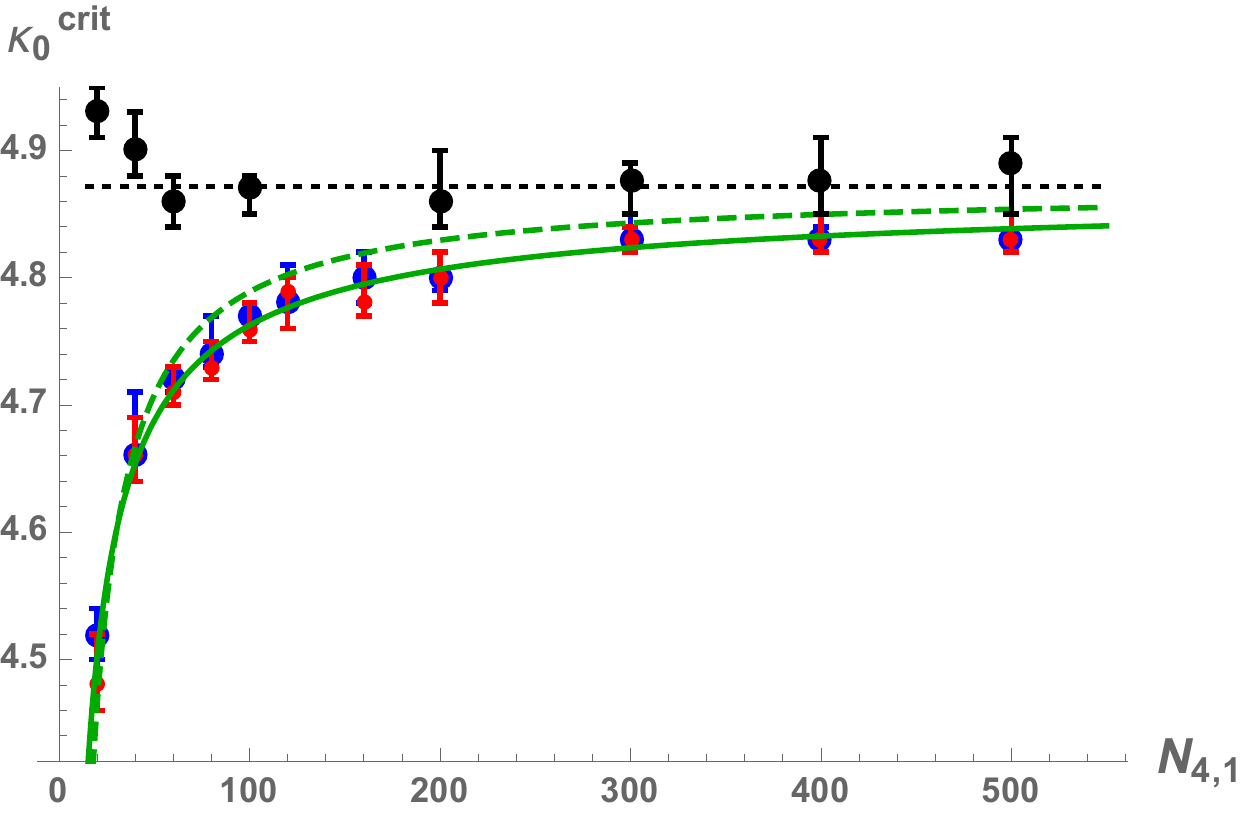}}
  \caption{\small Lattice volume dependence of pseudo-critical points $\kappa_0^{crit}$ based on the susceptibility maxima (blue and red data points) and on the Binder cumulant minima (black data points). The estimate of the true critical value of $\kappa_0^{crit}(\infty)= 4.87 \pm 0.01$ is presented as a black dotted line.  The fit  to Eq. \rf{powerlaw} with forced $\kappa_0^{crit}(\infty)= 4.87$ for which $\gamma = 1.31 \pm 0.03$  is drawn as a green solid line and the same fit with a forced value of $\gamma = 1$ as a green dashed line.
}
\label{FigT5}
\end{figure}

The Binder cumulant data of the $OP_2$ parameter, discussed above, can also be used to check how the minimal (critical) value $B_{OP_2}^{min}$, defined by Eq. \rf{Bmin}, depends on $N_{4,1}$. As shown in Fig.~\ref{FigT6}, the value of $B_{OP_2}^{min}$ moves away from zero when the lattice volume is increased. Although the scaling of  $B_{OP_2}^{min}$ as a function of  $N_{4,1}$ seems to be power-like, it is quite unlikely that this kind of behaviour can persist in the infinite volume limit. Anyway, the observed divergence from zero with increased lattice volume is expected for a first order transition and it is usually attributed to the  existence of two metastable states  at the transition point.

\begin{figure}[H]
  \centering
  \scalebox{.6}{\includegraphics{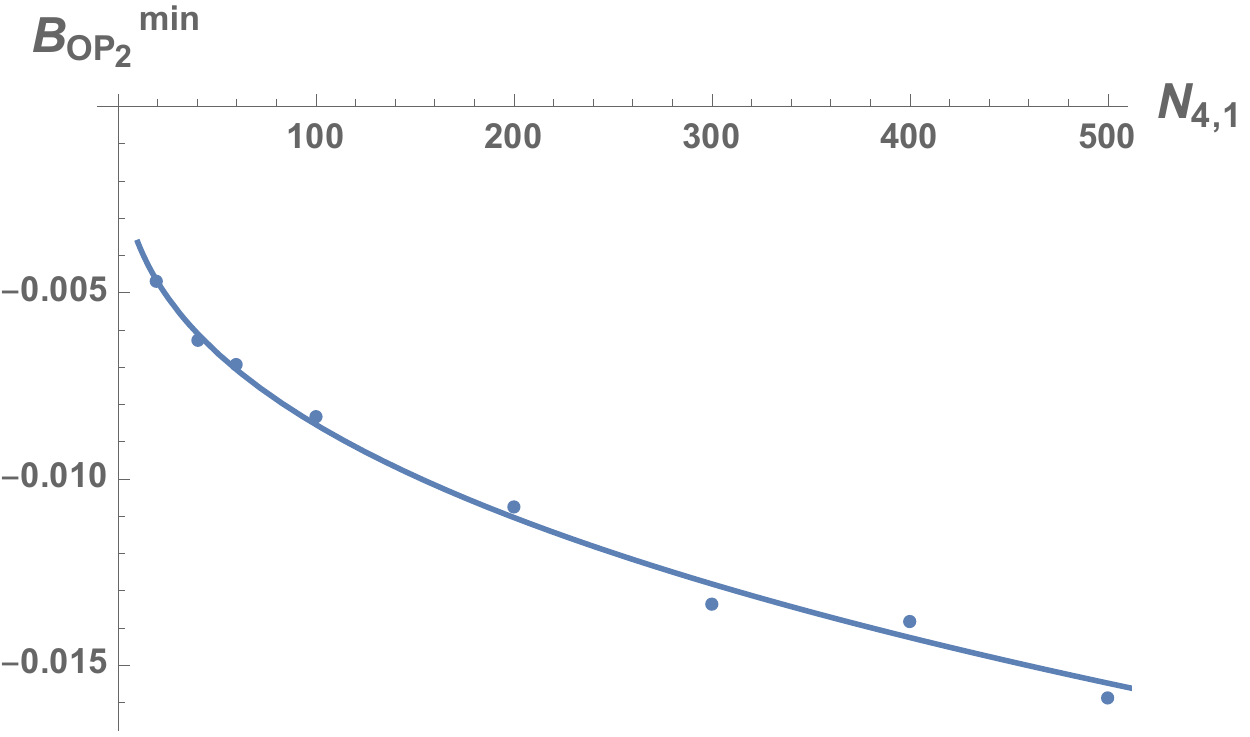}}
  \scalebox{.6}{\includegraphics{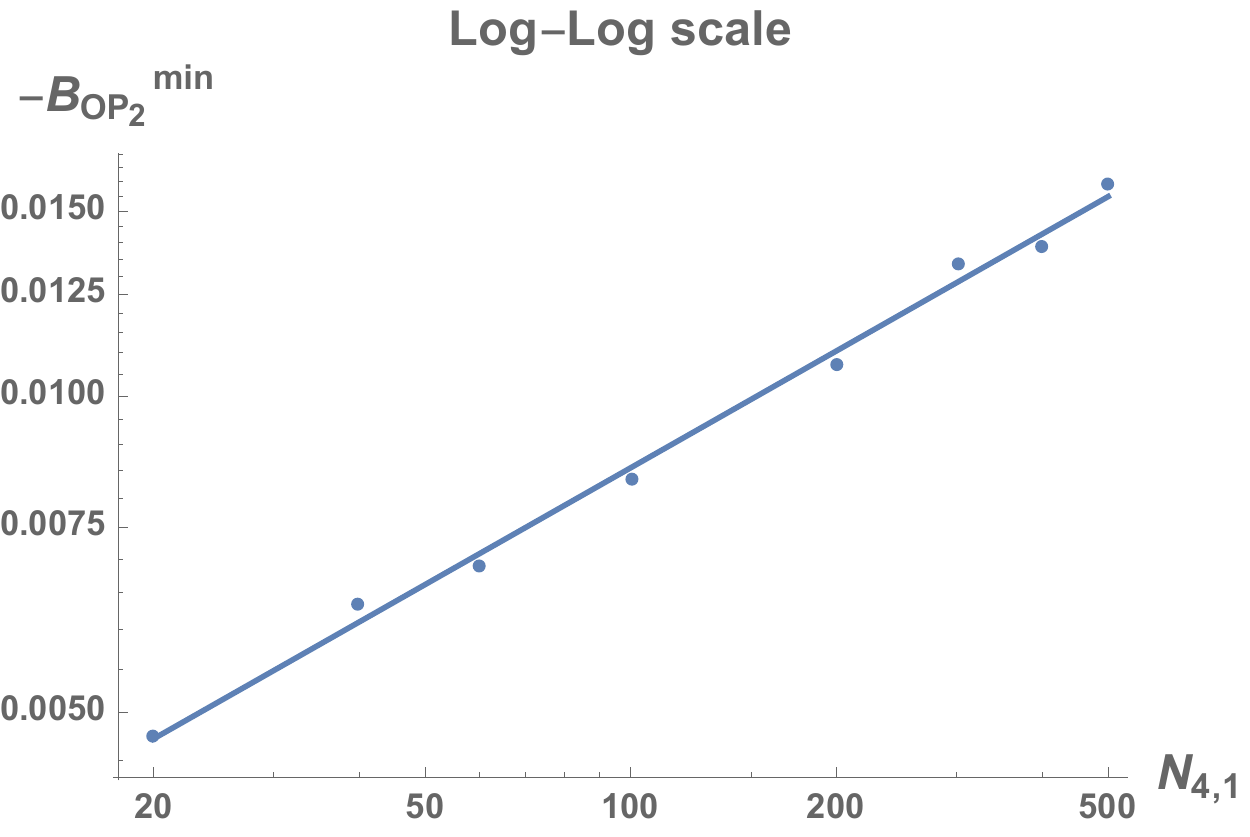}}

  \caption{\small The dependence of (the minimum of) the Binder cumulant $B_{OP_2}^{min}$ as defined by Eqs.~\rf{binder} and \rf{Bmin} on the system size $N_{4,1}$ at the $A$-$C$ transition for $\Delta=0.6$ . The log-log plot on the right shows a clear powerlaw behaviour. 
}
\label{FigT6}
\end{figure}

Therefore, we have finally performed a very careful Monte Carlo time history analysis for all our data  in search of the double peak structure  in the measured $OP_2$ histograms. In
agreement with preliminary findings of Ref.~\cite{Ambjorn:2018qbf}, we could not observe neither metastable state jumping nor double peaks of the $OP_2$  parameter nor its functions\footnote{As we have located the pseudo-critical $\kappa_0^{crit}$ values by using either the susceptibility  $\chi_{\sqrt{OP_2}}$ or the Binder cumulant $B_{OP_2} \approx -\chi_{\ln OP_2} $ we have also checked the histograms of $\sqrt{OP_2}$ in the former and of $\ln{OP_2}$ in the later case, respectively.} at any of the transition points - all measured histograms have a Gaussian-like shape with only one peak, as presented in Fig.~\ref{FigT7}. As the position of the pseudo-critical points $\kappa_0^{crit}$ is different for the susceptibility  and for the Binder cumulant observables we have also analysed the histograms for all other  measured $\kappa_0$ data points, but in each case a single Gaussian peak was present. 

\begin{figure}[H]
  \centering
  \scalebox{.65}{\includegraphics{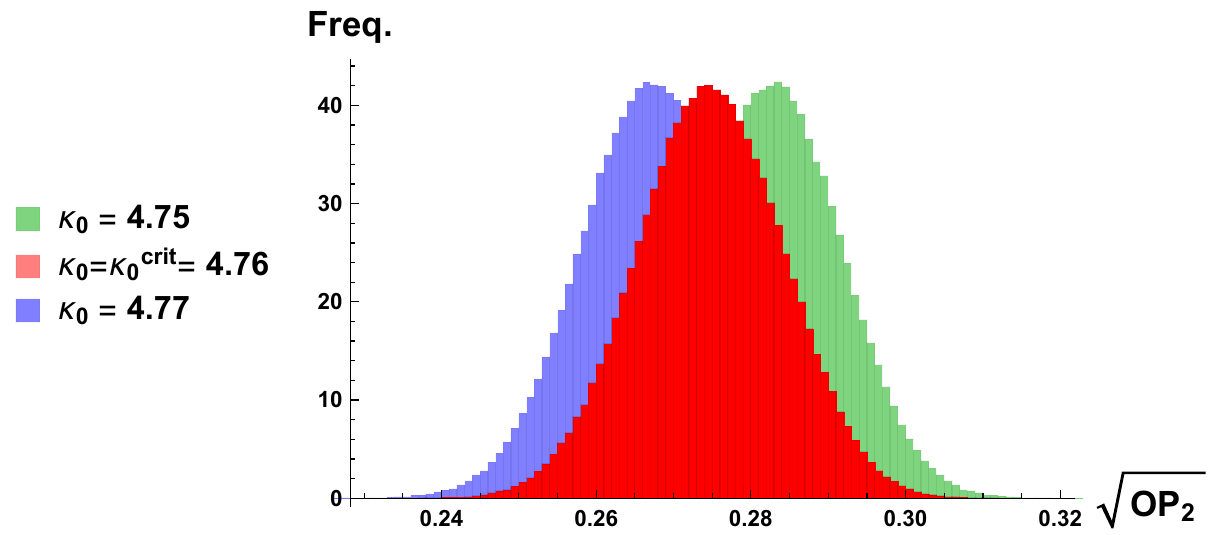}}
  \scalebox{.65}{\includegraphics{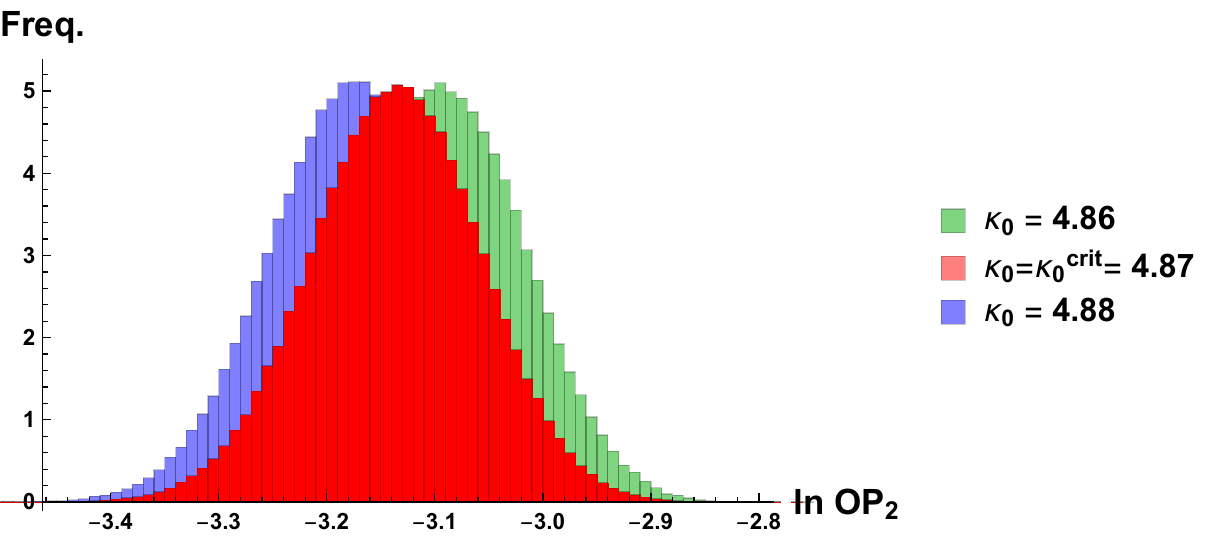}}
  \caption{\small Exemplary histograms of the (functions of) order parameter $OP_2\equiv {N_{3,2}/N_{4,1}}$ at pseudo-critical $\kappa_{0}^{crit}$ values for $\Delta=0.6$ with fixed lattice volumes $N_{4,1}=100k$. The left chart presents the histograms of $\sqrt{OP_2}$ measured in the vicinity of  $\kappa_{0}^{crit}= 4.76 $ based on the position of the susceptibility $\chi_{\sqrt{OP_2}}$ peak, and the right chart shows the histograms of $\ln{OP_2}$ measured in the vicinity of  $\kappa_{0}^{crit} = 4.87 $ based on the position of the Binder cumulant $B_{{OP_2}}$ minimum. The  histograms plotted in red are exactly at the transition points, while the green / blue data are for histograms measured for slightly lower / higher value of $\kappa_0$ than the critical value. The red, green and blue histograms overlap showing that the transition is  smooth, allowing no space for the existence of any more than one state at the transition point.}
\label{FigT7}
\end{figure}

To conclude this part, {although we did not observe any metastable state jumping of the order parameter (all the $OP_2$ histograms had just a single peak with a Gaussian-like shape), both the value of the critical shift exponent  $\gamma$ (which is close to one) and  scaling of the Binder cumulant $B_{OP_2}^{min}$  with the lattice volume ($B_{OP_2}^{min}$ diverges from zero) suggest that the $A$-$C$ transition remains first order in the case of toroidal topology. The transition is not as sharp as in the spherical case, which may be attributed to much larger finite size effects in toroidal CDT compared with spherical CDT.}


\end{subsection} 

\begin{subsection}{Impact of the time slicing}\label{J3}
In this section we will investigate what impact, if any, the choice of the number of time slices used in the MC numerical simulations  has on critical phenomena at the $A$-$C$ transition, now in the toroidal CDT case. We keep the volume fixing method of Section \ref{J1}, i.e. the total number of $(4,1)$ simplices is fixed at $N_{4,1}=200k$ and we change the number of time slices to $T=10, 20, 40$. 

We start by investigating the scaling properties of  the  order parameter $OP_2\equiv N_{4,1}/N_{3,2}$ measured at the vicinity of the $A$-$C$ transition. In particular, we want to check the scaling of the mean value $\langle OP_2 \rangle$ and the susceptibility $\chi_{OP_2}$ with $T$ when the average spatial volume $\langle n_t \rangle =N_{4,1} / T$ is kept fixed.  By using approximations \rf{suscsqrt}, \rf{susclog}  \rf{relations} and \rf{relations3}  these scaling relations will automatically translate into scaling of $\chi_{\sqrt{OP_2}}\approx \frac{\chi_{OP_2}}{\langle OP_2 \rangle}$ and $B_{OP_2}\approx -\frac{\chi_{OP_2}}{\langle OP_2 \rangle^2}$.

In the toroidal CDT case, as opposed to the spherical case, the average volume profile does not change between phase $A$ and phase $C$ where in both phases $\langle n_t \rangle = N_{4,1} / T = const $. We therefore suspect that the order parameter and its fluctuations depend on  $\langle n_t \rangle$. As demonstrated in Fig. \ref{FigT401}, the mean $\langle OP_2 \rangle$ does not depend on $T$  when $\langle n_t \rangle$ is kept fixed, however it is different for each value of $\langle n_t \rangle=const$.  The $OP_2$ fluctuations, measured by the susceptibility $\chi_{OP_2}\equiv \langle OP_2^2 \rangle - \langle OP_2 \rangle$, scale (approximately) as $1/T$ when $\langle n_t \rangle$ is kept fixed. This is shown
in Fig. \ref{FigT402}.
This kind of scaling may suggest that,  in the vicinity of the $A$-$C$ transition, the $OP_2$ can be modelled by $T$ statistically independent identical random variables indexed by  $t=1,..., T$. In such a case one has $OP_2= \frac{1}{T}\sum_t  \widetilde{OP}_2(n_t)$, where $\widetilde{OP}_2(n_t)$ denotes the value of the order parameter in each time slice, which depends solely on $ n_t $. Consequently
\beql{meanOP2}
\langle OP_2 \rangle = \Big\langle  \frac{1}{T}\sum_t  \widetilde{OP}_2(n_t) \Big\rangle=\Big\langle \widetilde{OP}_2\left({N_{4,1}}/{T}\right) \Big\rangle \, ,
\eeq
\beql{VarOP2}
\chi_{OP_2}  = \chi_ {\frac{1}{T}\sum_t  \widetilde{OP}_2(n_t)} =\frac{\chi_{\widetilde{OP}_2({N_{4,1}}/{T})}}{T},
\eeq
in accordance with the scaling observed  for various choices of $N_{4,1}$ and $T$ such that  $ \langle n_t \rangle={N_{4,1}}/{T}$ is kept fixed, {see Fig. \ref{FigT401} and \ref{FigT402}.}

\begin{figure}[H]
  \centering
  \scalebox{.6}{\includegraphics{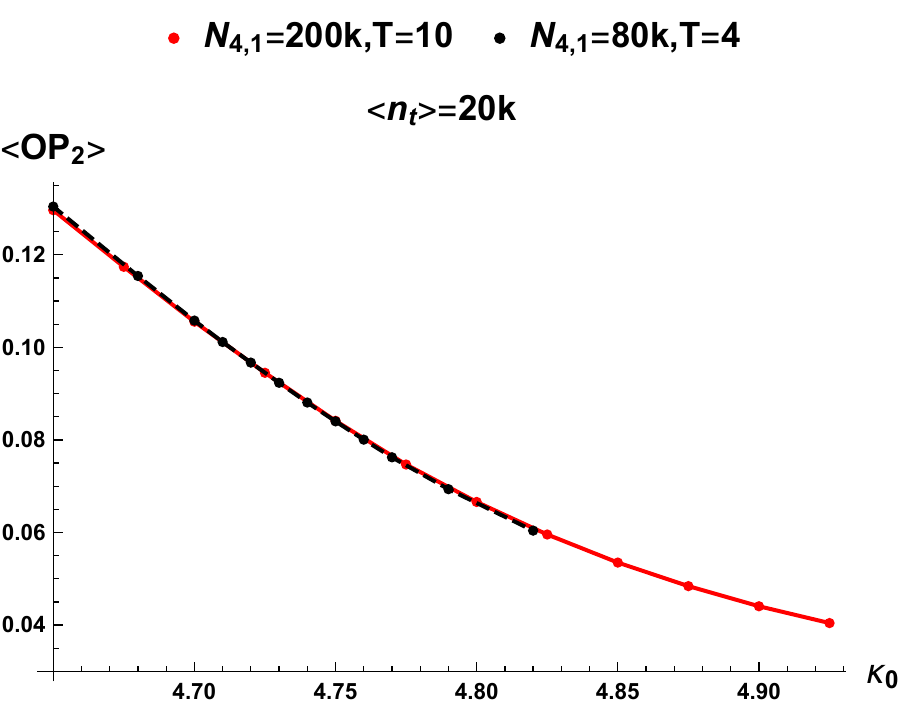}}
  \scalebox{.6}{\includegraphics{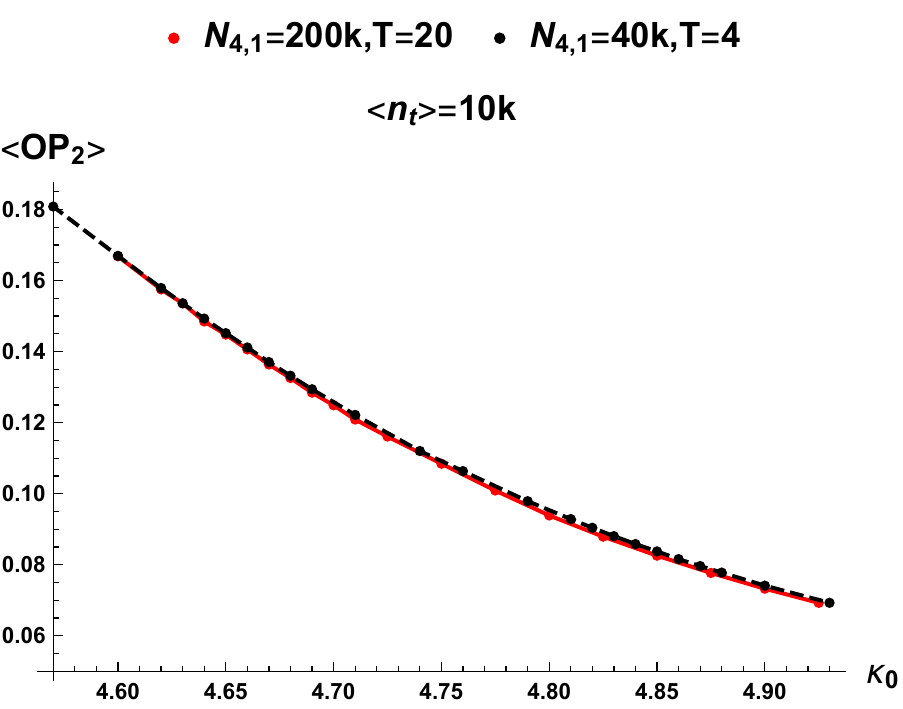}}
    \scalebox{.6}{\includegraphics{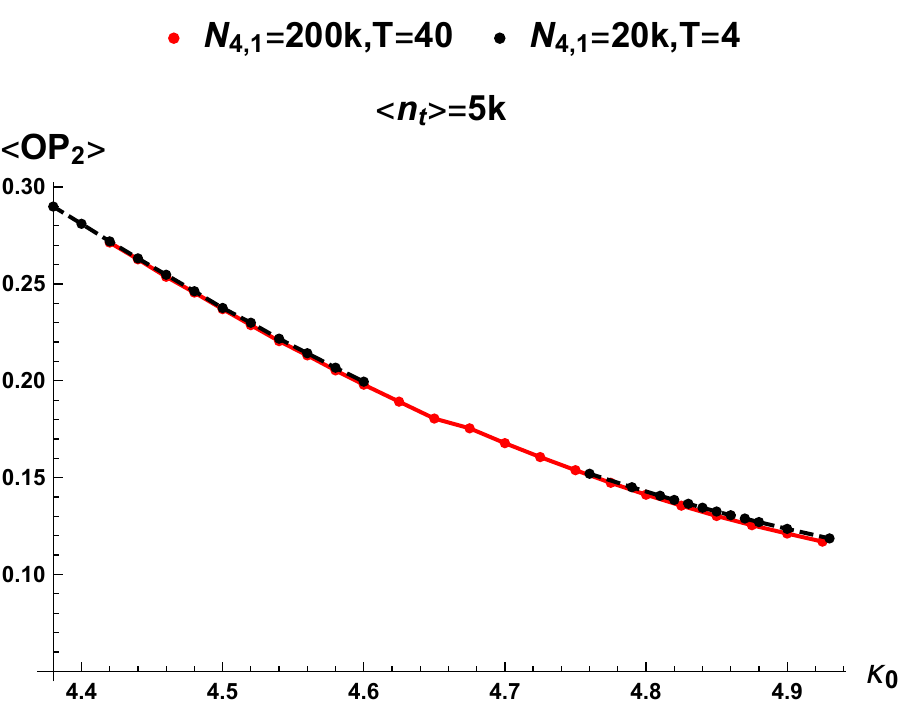}}
\caption{\small Mean value of the order parameter $\langle{{OP_2}}\rangle$ as a function of $\kappa_{0}$ (red line)  measured for fixed $N_{4,1}=200k$ and $T=10$ (left),  $T=20$  (middle) and $T=40$ (right) compared to data measured for fixed $T=4$ time slices (black-dashed line) and $N_{4,1}=80k$ (left), $N_{4,1}=40k$ (middle) and $N_{4,1}=20k$ (right), respectively. 
}
\label{FigT401}
\end{figure}

\begin{figure}[H]
  \centering
  \scalebox{.6}{\includegraphics{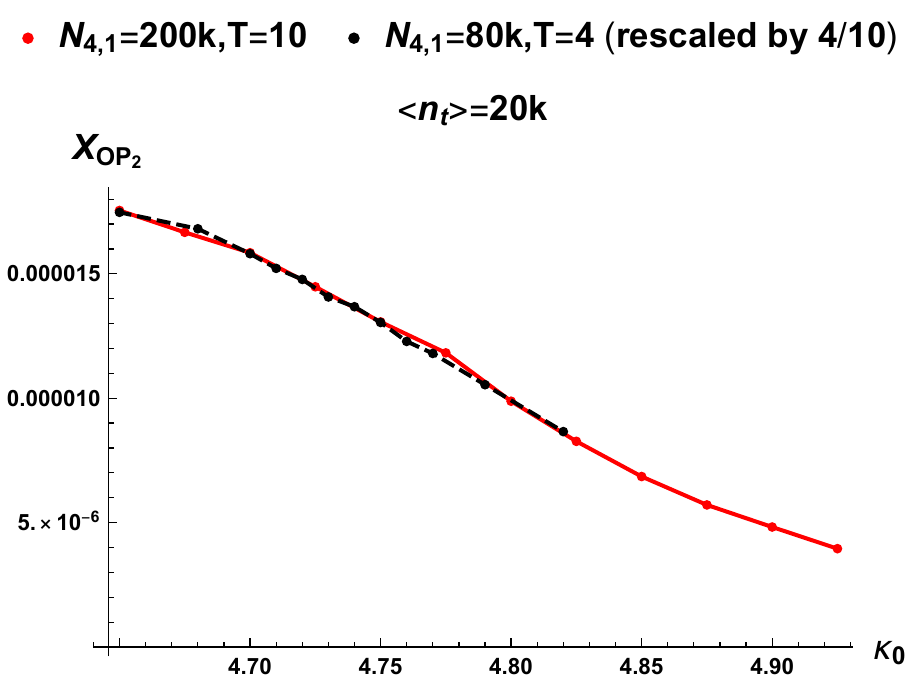}}
  \scalebox{.6}{\includegraphics{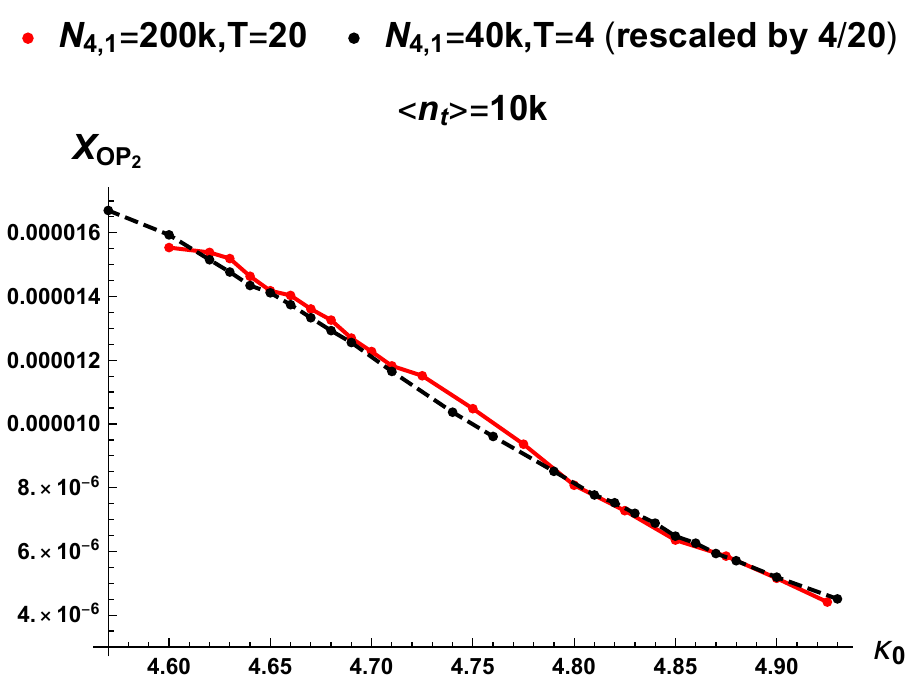}}
    \scalebox{.6}{\includegraphics{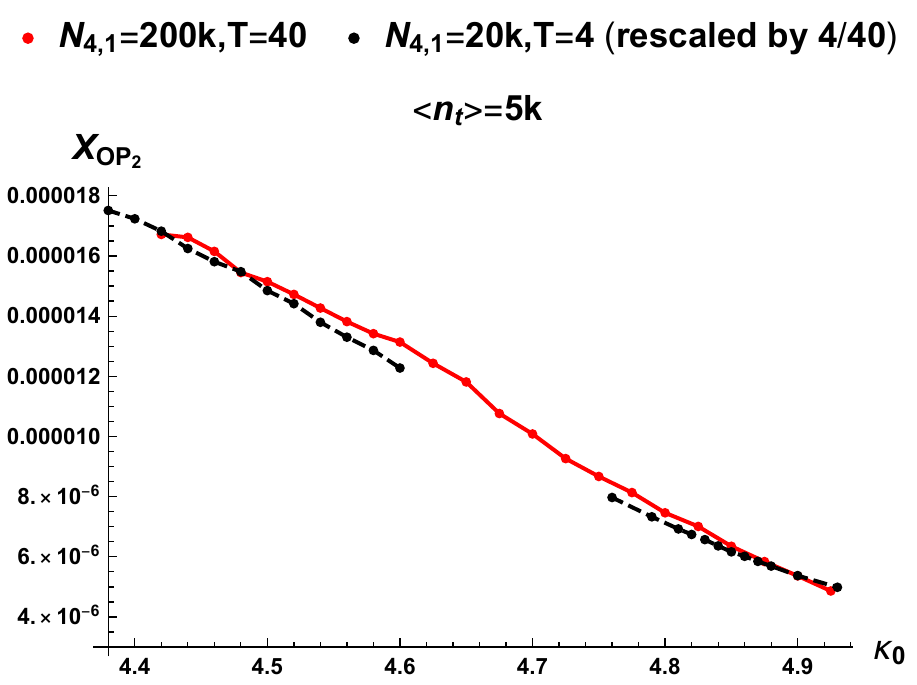}}
\caption{\small The susceptibility $\chi_{{OP_2}}$ as defined by Eq.~(\ref{susc}) as a function of $\kappa_{0}$ (red line)  measured for fixed $N_{4,1}=200k$ and $T=10$ (left),  $T=20$  (middle) and $T=40$ (right) compared to rescaled data measured for fixed $T=4$ time slices (black-dashed line) and $N_{4,1}=80k$ (left), $N_{4,1}=40k$ (middle) and $N_{4,1}=20k$ (right), respectively. }
\label{FigT402}
\end{figure}

\begin{figure}[H]
  \centering
  \scalebox{.75}{\includegraphics{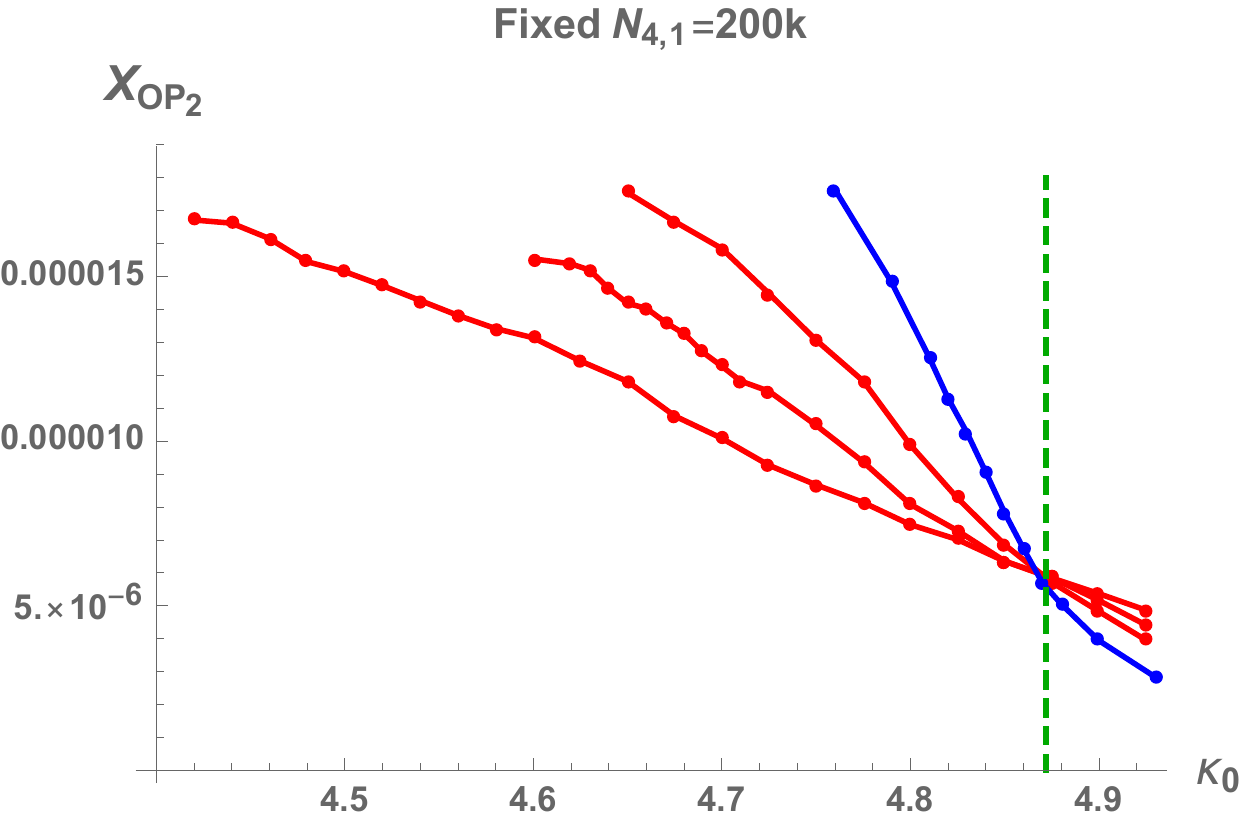}}
\caption{\small The susceptibility $\chi_{{OP_2}}$ as a function of $\kappa_{0}$ measured for fixed $N_{4,1}=200k$ and various $T=4$ (blue line) and $T=10, 20, 40$ (red lines) and thus various $\langle n_t \rangle = N_{4,1} / T = 50k, 20k, 10k, 5k$, respectively. The curves cross at the $\kappa_0^{crit}(\infty)=4.87$ point whose position is denoted by a  green-dashed line.}
\label{FigT403}
\end{figure}

What is more interesting is the following:  for fixed $N_{4,1}$ the $OP_2$ fluctuations  
expressed by $\chi_{OP_2}$  as a function of $\kappa_0$  depend on $\langle n_t \rangle=N_{4,1}/T$, see Fig. \ref{FigT403}, {\it but they seem to be universal at the true (infinite volume) transition point} $\kappa_0^{crit}(\infty)=4.87$ (see Section \ref{J1}). This is seen in Fig.\ \ref{FigT404} where the curves plotted for various $\langle n_t \rangle=N_{4,1}/T$ cross. Therefore, for fixed $N_{4,1}$,  the true critical susceptibility  $\chi^{crit}_{OP_2}\equiv \chi_{OP_2}(\kappa_0^{crit}(\infty))$ does not depend on $T$, which, in conjunction with relation \rf{VarOP2}, implies a universal critical scaling 
\beql{ChiScaling}
\chi^{crit}_{OP_2} = \frac{\chi^*}{N_{4,1}} \ , \ \chi^*=const.,
\eeq
valid for any  $N_{4,1}$ and $T$.
This is indeed the case, as illustrated in Fig. \ref{FigT404}, where we plot $\chi_{OP_2} \cdot  N_{4,1}$ as a function of $\kappa_0$ measured for various choices of both $N_{4,1}$ and $T$. In accordance with relation \rf{ChiScaling} 
all the curves cross at a single critical point $\kappa_0^{crit}(\infty)=4.87$ resulting in a 
universal value of $\chi^*\approx 1.2$.  

\begin{figure}[H]
  \centering
  \scalebox{.75}{\includegraphics{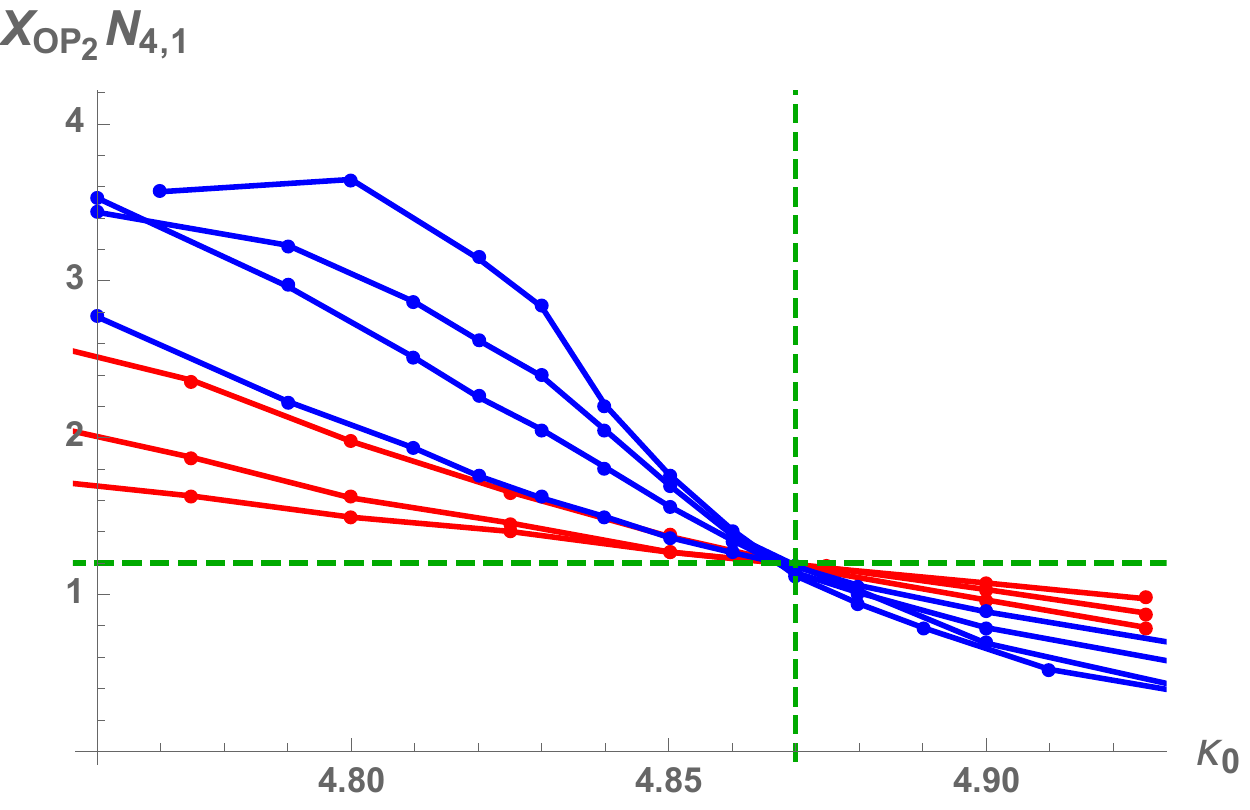}}
\caption{\small The (rescaled) susceptibility $\chi_{{OP_2}} \cdot N_{4,1}$  as a function of $\kappa_{0}$ measured for $N_{4,1}=200k$ and $T=10, 20, 40$ (red lines) and for $T=4$  and $N_{4,1}=100, 200, 300, 400$ (blue lines). All the curves cross at a single  $(\kappa_0^{crit}(\infty)\,;\,\chi^*)\approx(4.87\,;\,1.2)$ point whose position is denoted by green-dashed  lines.}
\label{FigT404}
\end{figure}

All the above results show that data measured in numerical MC simulations with the $N_{4,1}$ volume fixed  for various number of time slices $T$ and various lattice volumes $N_{4,1}$ can be  simply rescaled, {resulting in universal  critical behaviour of Eq. \rf{ChiScaling}, and  consistent with $\kappa_0^{crit}(\infty) = 4.87$ computed in Section \ref{J1}. The critical scaling \rf{ChiScaling} of susceptibility of $OP_2 \equiv N_{3,2} / N_{4,1}$, which is  an intensive parameter, automatically translates into the following scaling of susceptibility of the extensive   parameter $N_{3,2}$:
\beql{ChiN32Scaling}
\chi^{crit}_{N_{3,2}} = {\chi^*}\cdot {N_{4,1}} \ , \ \chi^*=const.\ ,
\eeq
which is typical for a first order transition. Thus the result strongly supports the first order nature of the $A-C$ transition in the toroidal topology.}
\newline

{We have also performed the}
histogram analysis of the $OP_2$ parameter  at the transition points measured for fixed $N_{4,1}=200k$ and various $T=10$ (Fig. \ref{FigT43}), $T=20$ (Fig. \ref{FigT53}) and $T=40$ (Fig. \ref{FigT63}). As in Section \ref{J1}, in each case we have located the pseudo-critical $\kappa_0^{crit}$ values by either looking at the susceptibility $\chi_{\sqrt{OP_2}}$ maxima or the Binder cumulant $B_{OP_2} $ minima (see Fig. \ref{FigT41}, Fig. \ref{FigT51} and Fig. \ref{FigT61}, respectively), and we have measured the $OP_2$ Monte Carlo history at these points  (see Fig. \ref{FigT42}, Fig. \ref{FigT52} and Fig. \ref{FigT62}, respectively). In each case, and also for any other $\kappa_0$ data point measured, the $OP_2$ order parameter (and its functions) shows Gaussian-like fluctuations and there are no signs of the metastable states jumping in any case. {Again, this behaviour is not typical for a first order transition, but most likely, due to much larger finite size effects in the toroidal topology, we have not yet reached the lattice size where two distinct  metastable states can be observed.}

\begin{subsubsection}{Fixed $N_{4,1}=200k$ ($T=10$)}

\begin{figure}[H]
  \centering
  \scalebox{.65}{\includegraphics{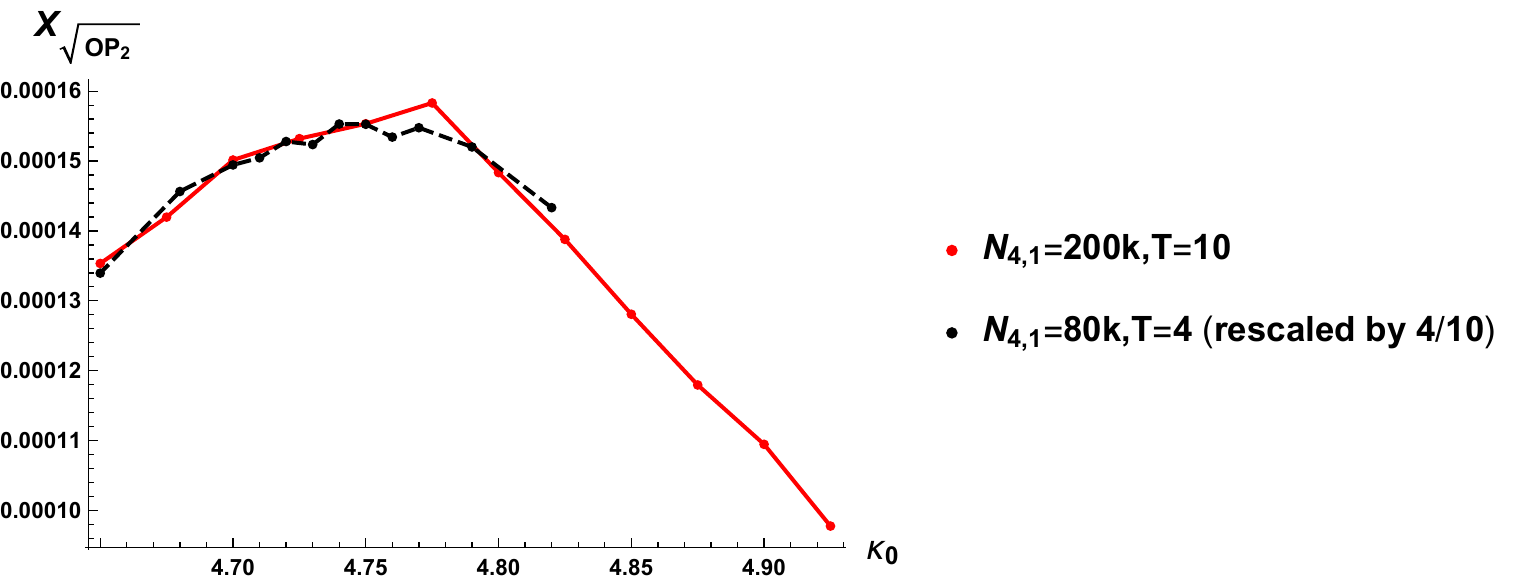}}
  \scalebox{.50}{\includegraphics{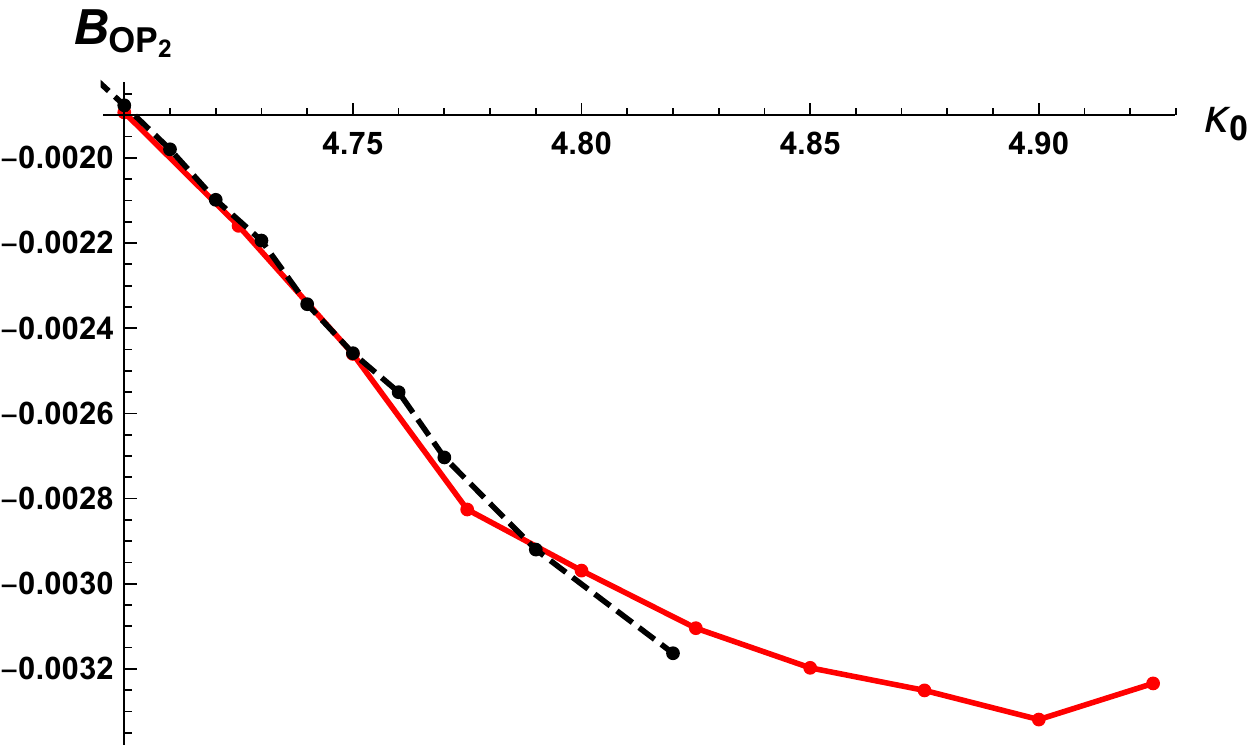}}
\caption{\small The susceptibility $\chi_{\sqrt{OP_2}}\approx\frac{\chi_{OP_2}}{\langle OP_2\rangle} $ as defined by Eq.~(\ref{susc})  (left chart) and the Binder cumulant $B_{OP_2}$ as defined by Eq.~(\ref{binder})  (right chart)
as a function of $\kappa_{0}$ measured for fixed $N_{4,1}=200k$ and $T=10$ time slices (red line) and rescaled data for fixed $N_{4,1}=80k$ and $T=4$ time slices (black-dashed line). 
}
\label{FigT41}
\end{figure}

\begin{figure}[H]
  \centering
  \scalebox{.6}{\includegraphics{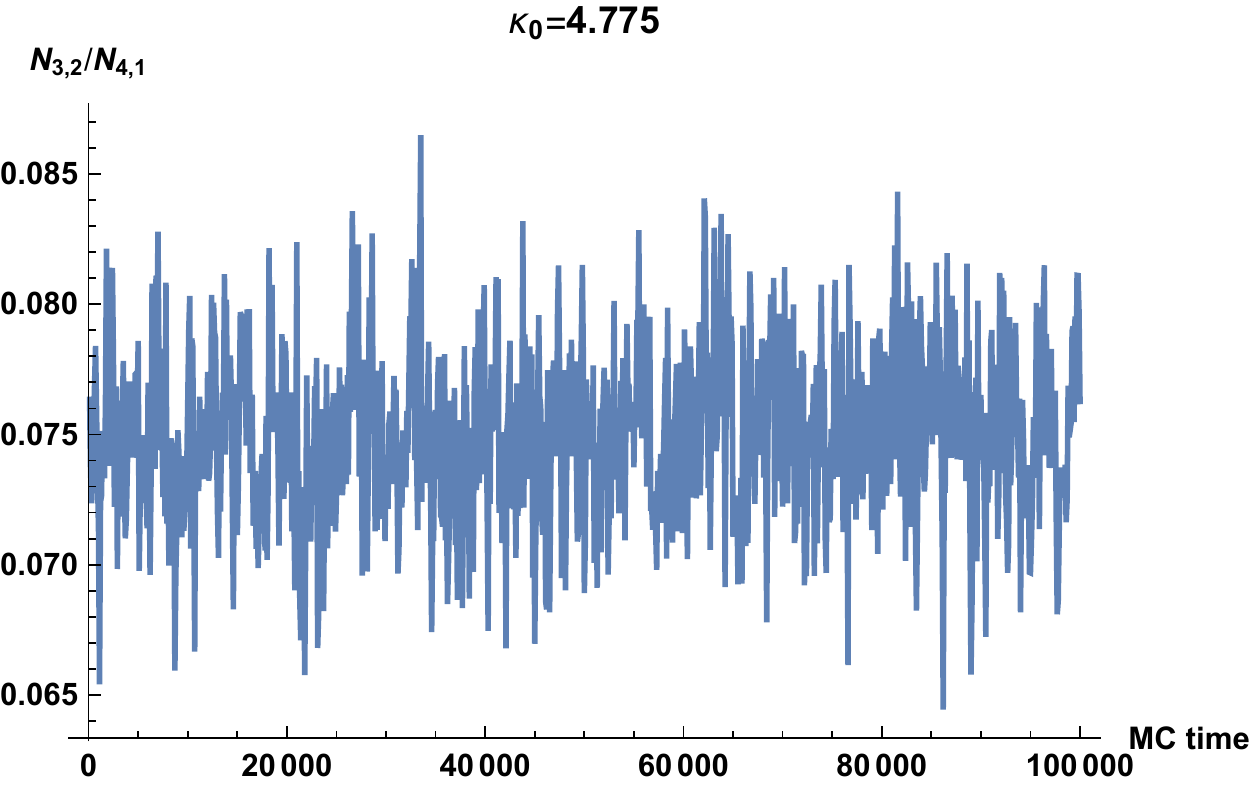}}
  \scalebox{.6}{\includegraphics{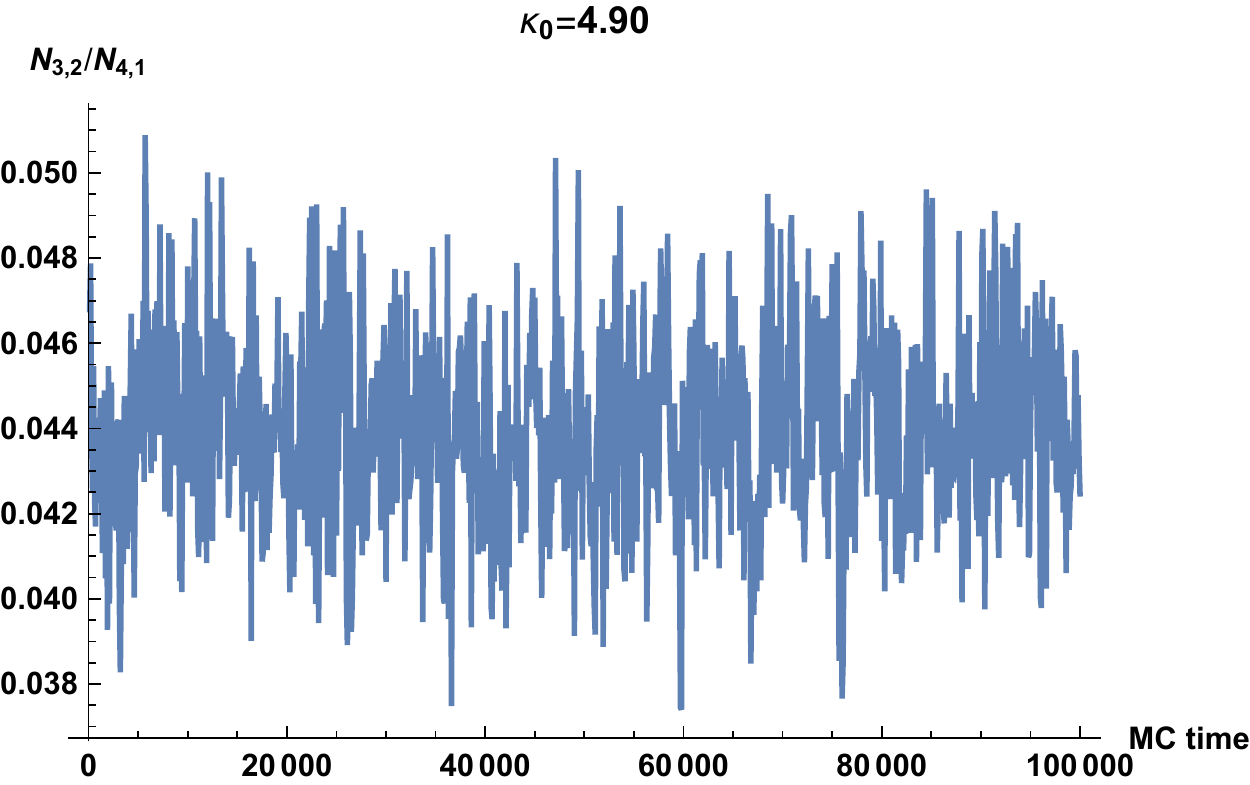}}
\caption{\small Monte Carlo time history of $OP_2\equiv N_{3,2}/N_{4,1}$ at pseudo-critical points measured using the susceptibility method: $\kappa_0 = 4.775$ (left chart) and using the Binder cumulant method: $\kappa_0 = 4.90$ (right chart) for fixed $N_{4,1}=200k$ and $T=10$ time slices. 
}
\label{FigT42}
\end{figure}

\begin{figure}[H]
  \centering
  \scalebox{.65}{\includegraphics{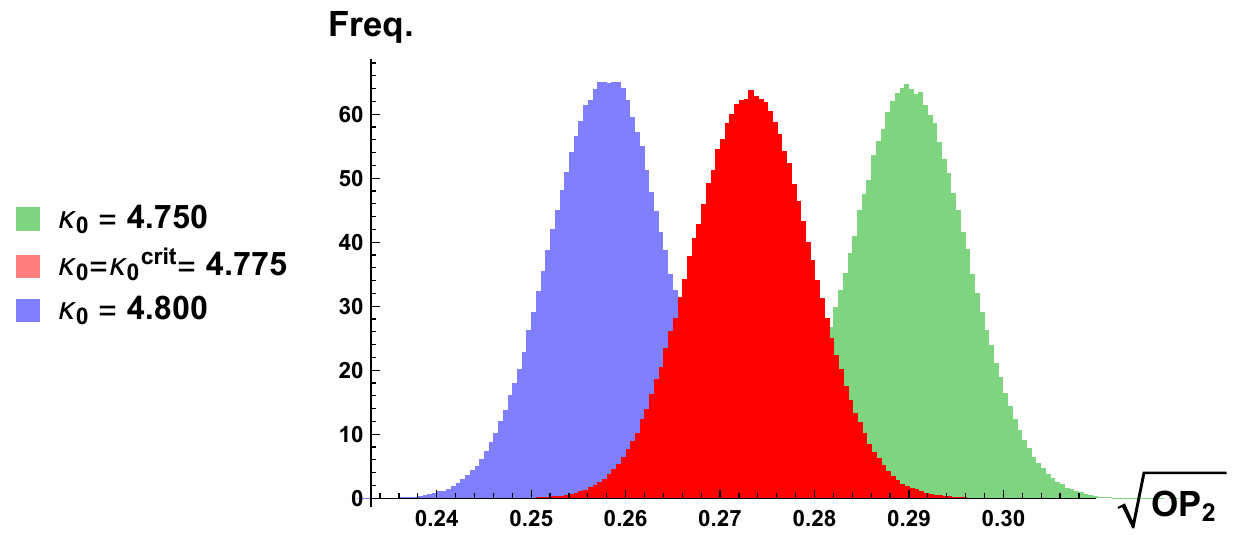}}
  \scalebox{.65}{\includegraphics{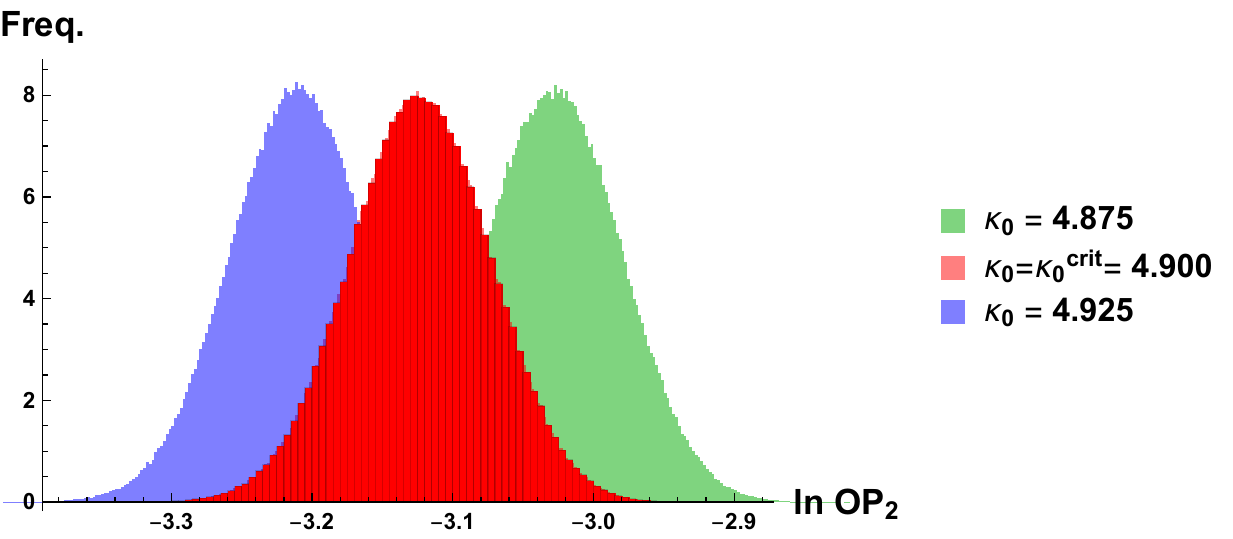}}
  \caption{\small Exemplary histograms of the (functions of) order parameter $OP_2\equiv {N_{3,2}/N_{4,1}}$ at pseudo-critical $\kappa_{0}^{crit}$ values  with fixed $N_{4,1}=200k$ and $T=10$ time slices . The left chart presents the histograms of $\sqrt{OP_2}$ measured in the vicinity of  $\kappa_{0}^{crit}= 4.775 $ based on the position of the susceptibility $\chi_{\sqrt{OP_2}}$ peak (Fig. \ref{FigT41}, left), and the right chart shows the histograms of $\ln{OP_2}$ measured in the vicinity of  $\kappa_{0}^{crit} = 4.90 $ based on the position of the Binder cumulant $B_{{OP_2}}$ minimum (Fig. \ref{FigT41}, right). The  histograms plotted in red are exactly at the transition points, while the green / blue data are for histograms measured for slightly lower / higher value of $\kappa_0$ than the critical value. The red, green and blue histograms overlap showing that the transition is  smooth, allowing no space for the existence of any more than one state at the transition point.}
\label{FigT43}
\end{figure}

\end{subsubsection}

\begin{subsubsection}{Fixed $N_{4,1}=200k$ ($T=20$)}

\begin{figure}[H]
  \centering
  \scalebox{.65}{\includegraphics{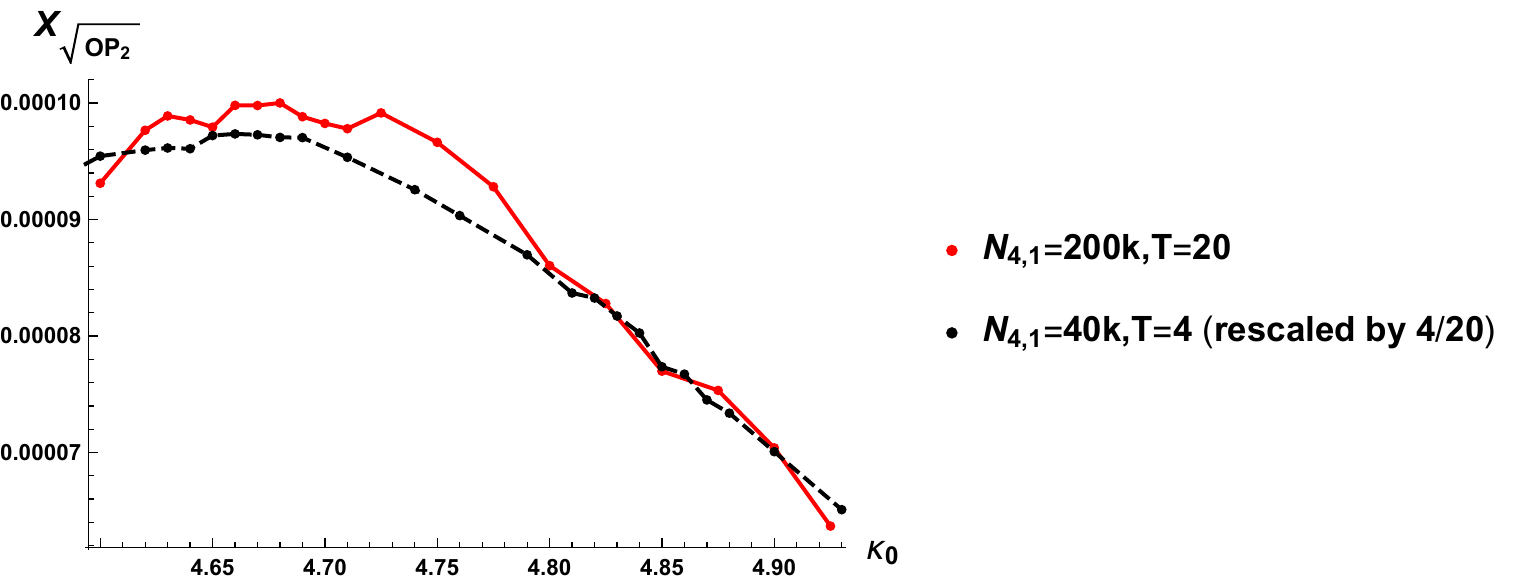}}
  \scalebox{.50}{\includegraphics{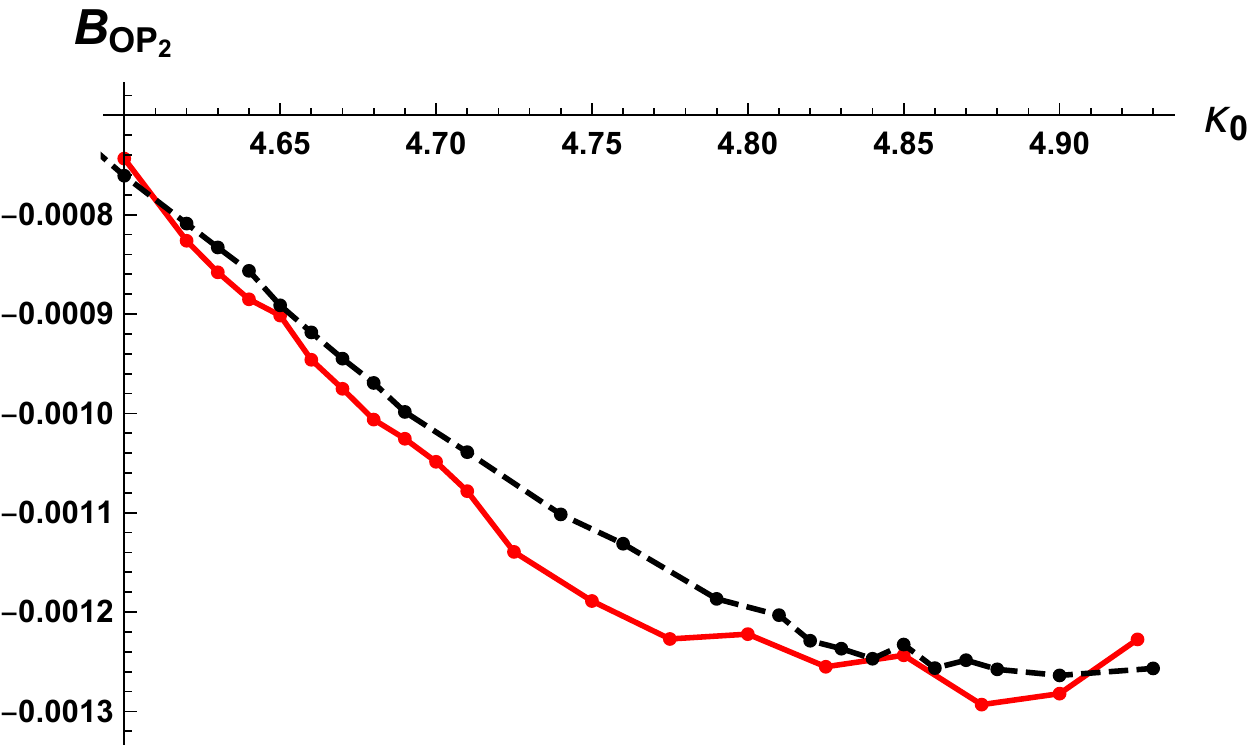}}
\caption{\small The susceptibility $\chi_{\sqrt{OP_2}}\approx\frac{\chi_{OP_2}}{\langle OP_2\rangle} $ as defined by Eq.~(\ref{susc})  (left chart) and the Binder cumulant $B_{OP_2}$ as defined by Eq.~(\ref{binder})  (right chart)
as a function of $\kappa_{0}$ measured for fixed $N_{4,1}=200k$ and $T=20$ time slices (red line) and rescaled data for fixed $N_{4,1}=40k$ and $T=4$ time slices (black-dashed line). 
}
\label{FigT51}
\end{figure}

\begin{figure}[H]
  \centering
  \scalebox{.6}{\includegraphics{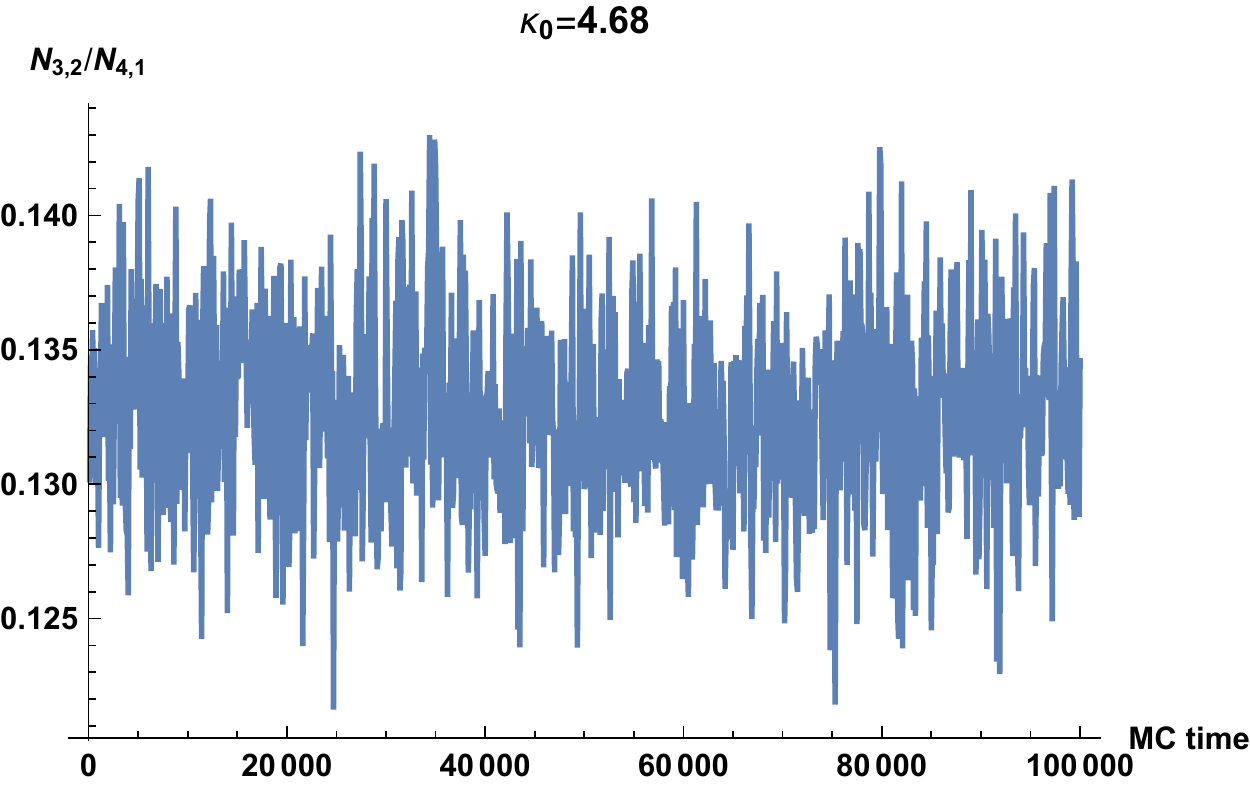}}
  \scalebox{.6}{\includegraphics{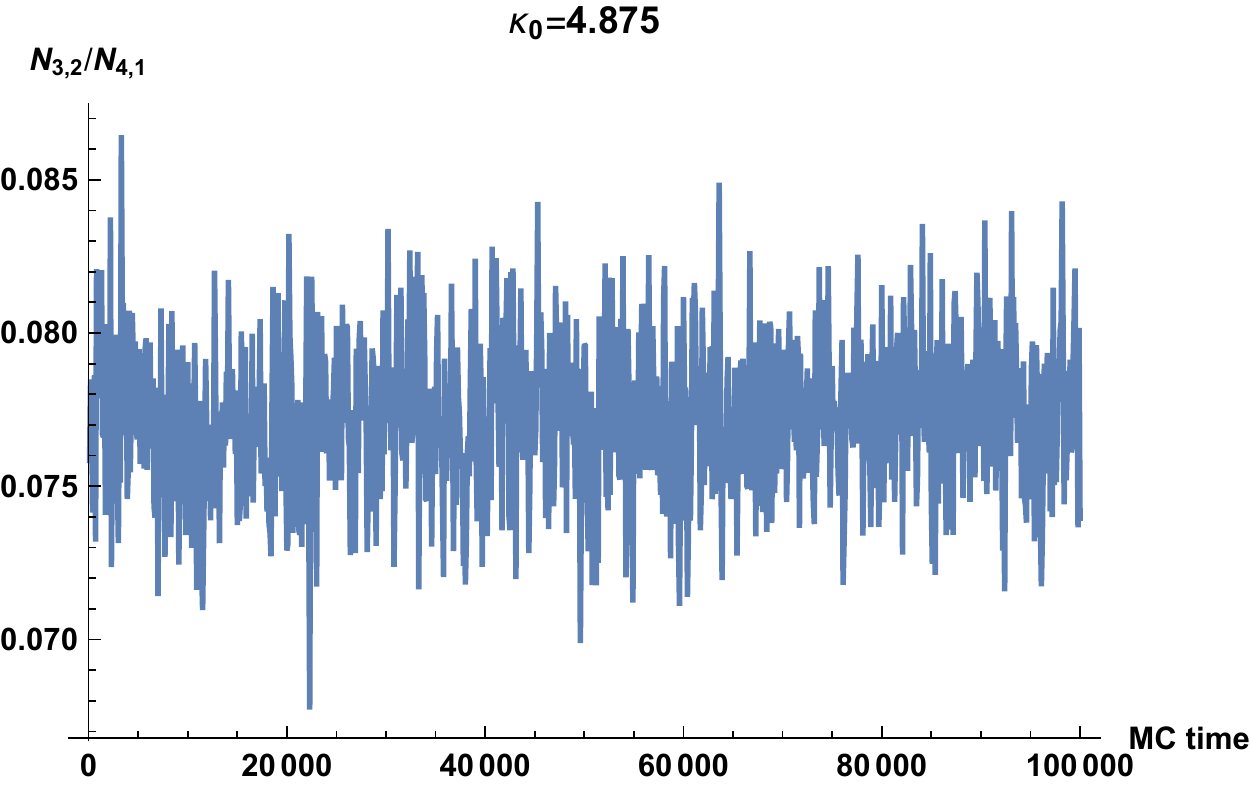}}
\caption{\small Monte Carlo time history of $OP_2\equiv N_{3,2}/N_{4,1}$ at pseudo-critical points measured using the susceptibility method: $\kappa_0 = 4.68$ (left chart) and using the Binder cumulant method: $\kappa_0 = 4.875$ (right chart) for fixed $N_{4,1}=200k$ and $T=20$ time slices. 
}
\label{FigT52}
\end{figure}

\begin{figure}[H]
  \centering
  \scalebox{.65}{\includegraphics{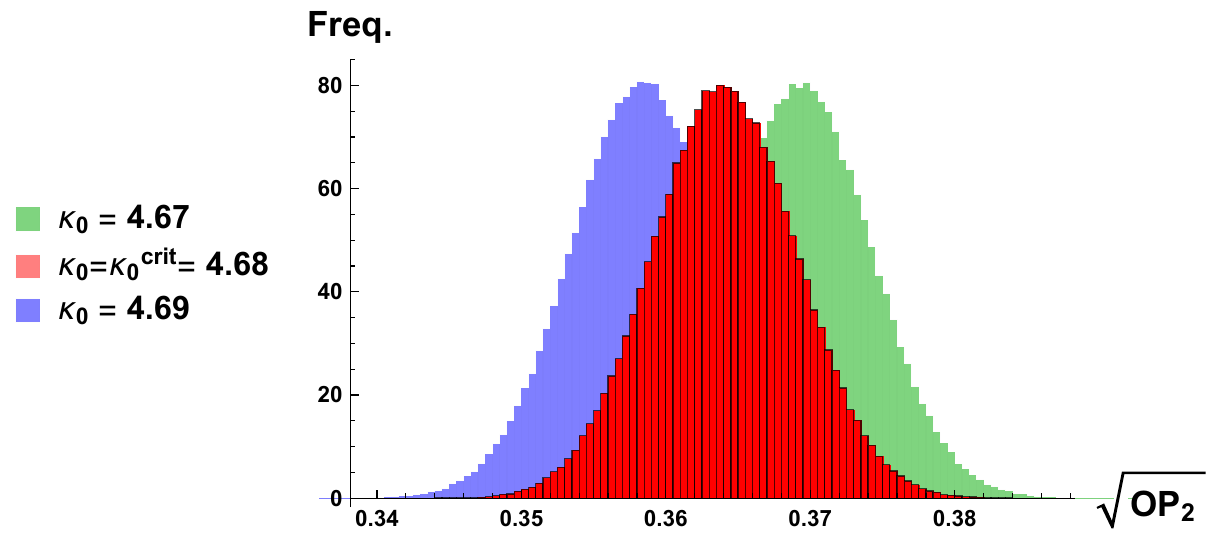}}
  \scalebox{.65}{\includegraphics{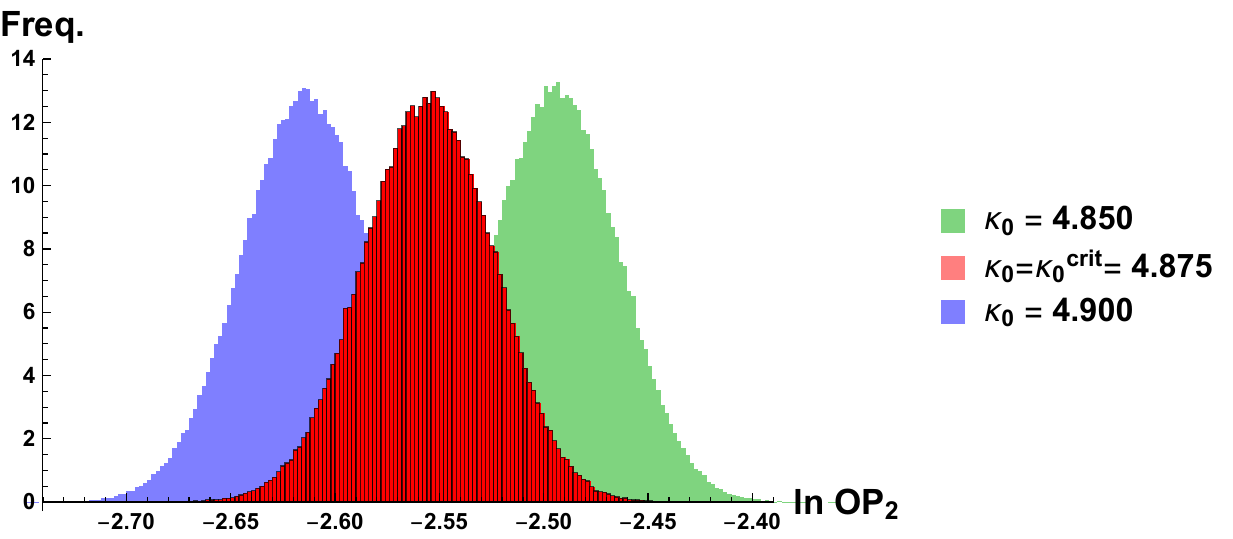}}
  \caption{\small Exemplary histograms of the (functions of) order parameter $OP_2\equiv {N_{3,2}/N_{4,1}}$ at pseudo-critical $\kappa_{0}^{crit}$ values  with fixed $N_{4,1}=200k$ and $T=20$ time slices. The left chart presents the histograms of $\sqrt{OP_2}$ measured in the vicinity of  $\kappa_{0}^{crit}= 4.68 $ based on the position of the susceptibility $\chi_{\sqrt{OP_2}}$ peak (Fig. \ref{FigT51}, left), and the right chart shows the histograms of $\ln{OP_2}$ measured in the vicinity of  $\kappa_{0}^{crit} = 4.875 $ based on the position of the Binder cumulant $B_{{OP_2}}$ minimum (Fig. \ref{FigT51}, right). The  histograms plotted in red are exactly at the transition points, while the green / blue data are for histograms measured for slightly lower / higher value of $\kappa_0$ than the critical value. The red, green and blue histograms overlap showing that the transition is  smooth, allowing no space for the existence of any more than one state at the transition point.}
\label{FigT53}
\end{figure}
\end{subsubsection}

\begin{subsubsection}{Fixed $N_{4,1}=200k$ ($T=40$)}

\begin{figure}[H]
  \centering
  \scalebox{.65}{\includegraphics{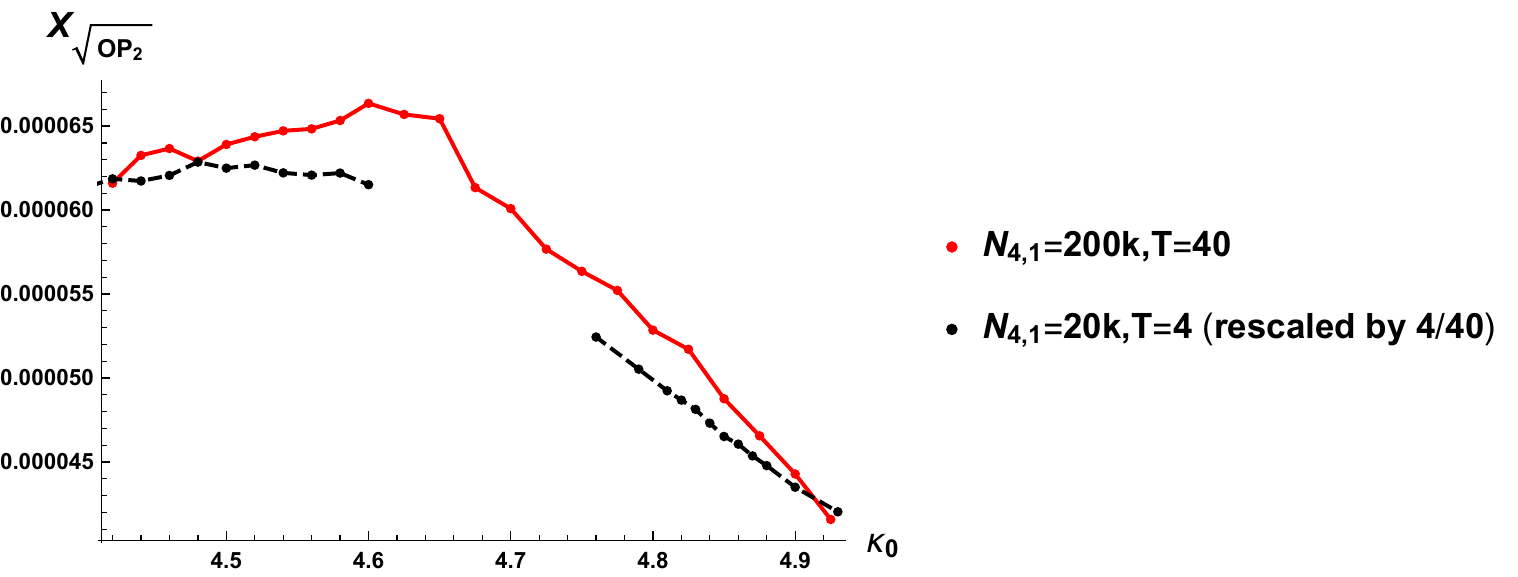}}
  \scalebox{.50}{\includegraphics{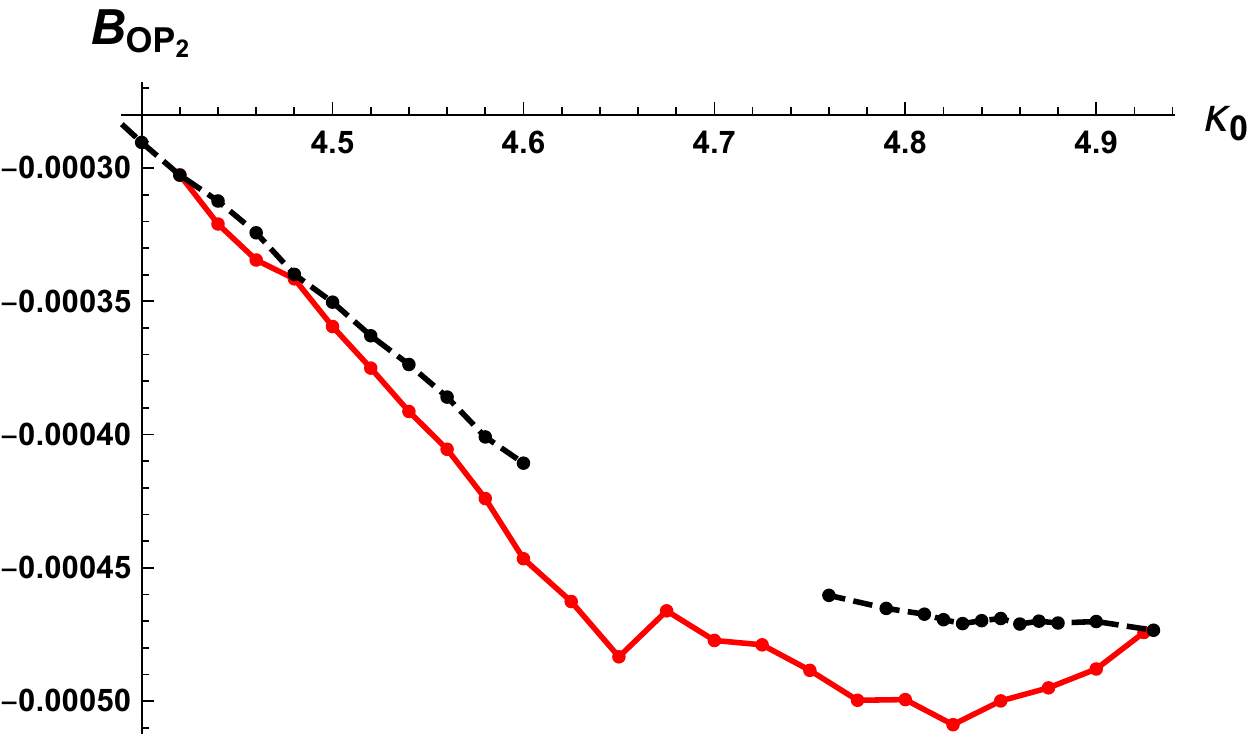}}
\caption{\small The susceptibility $\chi_{\sqrt{OP_2}}\approx\frac{\chi_{OP_2}}{\langle OP_2\rangle} $ as defined by Eq.~(\ref{susc})  (left chart) and the Binder cumulant $B_{OP_2}$ as defined by Eq.~(\ref{binder})  (right chart)
as a function of $\kappa_{0}$ measured for fixed $N_{4,1}=200k$ and $T=40$ time slices (red line) and rescaled data for fixed $N_{4,1}=20k$ and $T=4$ time slices (black-dashed lines). 
}
\label{FigT61}
\end{figure}

\begin{figure}[H]
  \centering
  \scalebox{.6}{\includegraphics{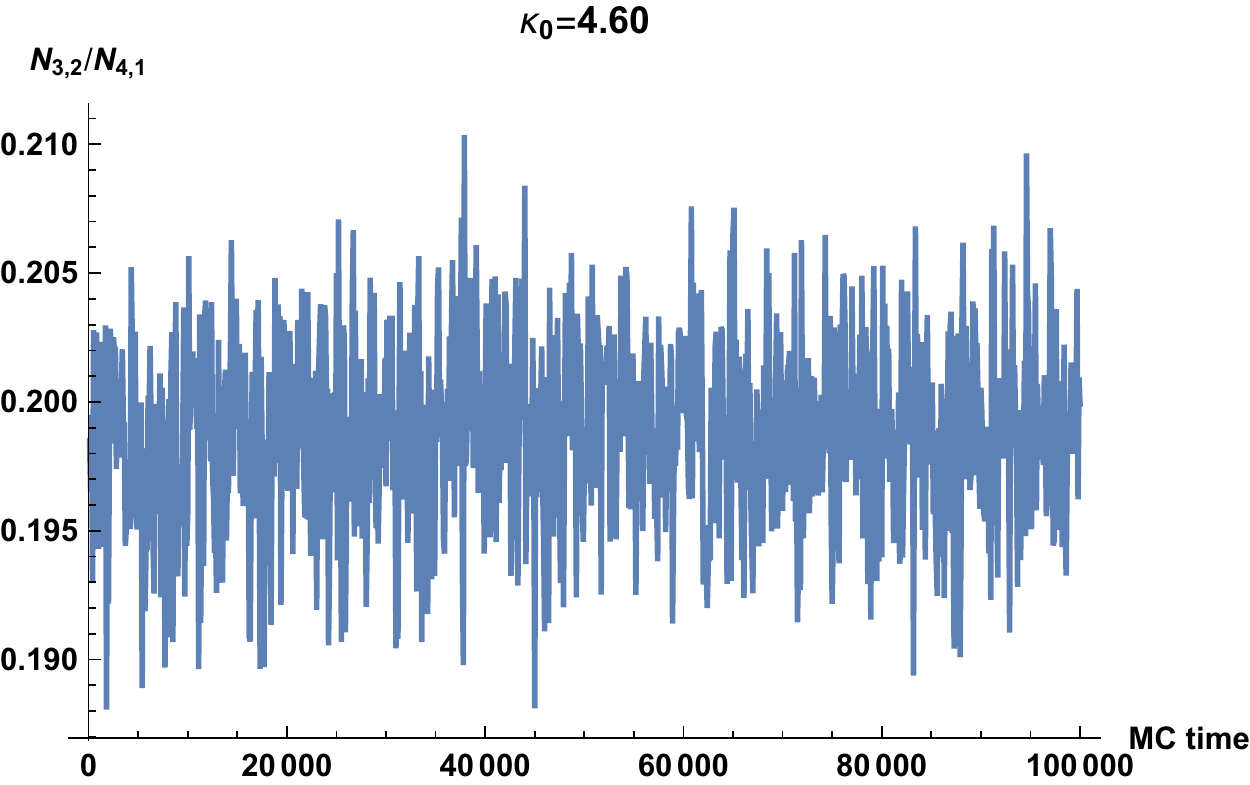}}
  \scalebox{.6}{\includegraphics{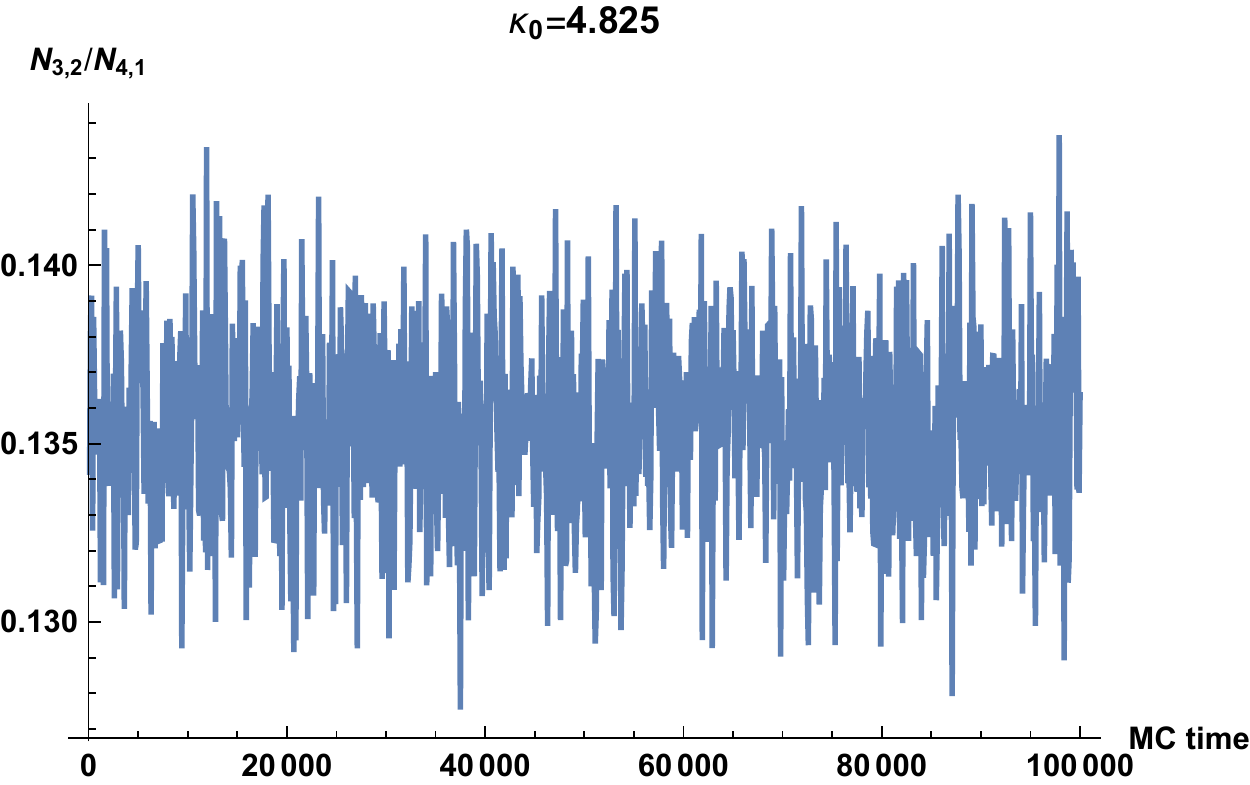}}
\caption{\small Monte Carlo time history of $OP_2\equiv N_{3,2}/N_{4,1}$ at pseudo-critical points measured using the susceptibility method: $\kappa_0 = 4.60$ (left chart) and using the Binder cumulant method: $\kappa_0 = 4.825$ (right chart) for fixed $N_{4,1}=200k$ and $T=40$ time slices. 
}
\label{FigT62}
\end{figure}

\begin{figure}[H]
  \centering
  \scalebox{.65}{\includegraphics{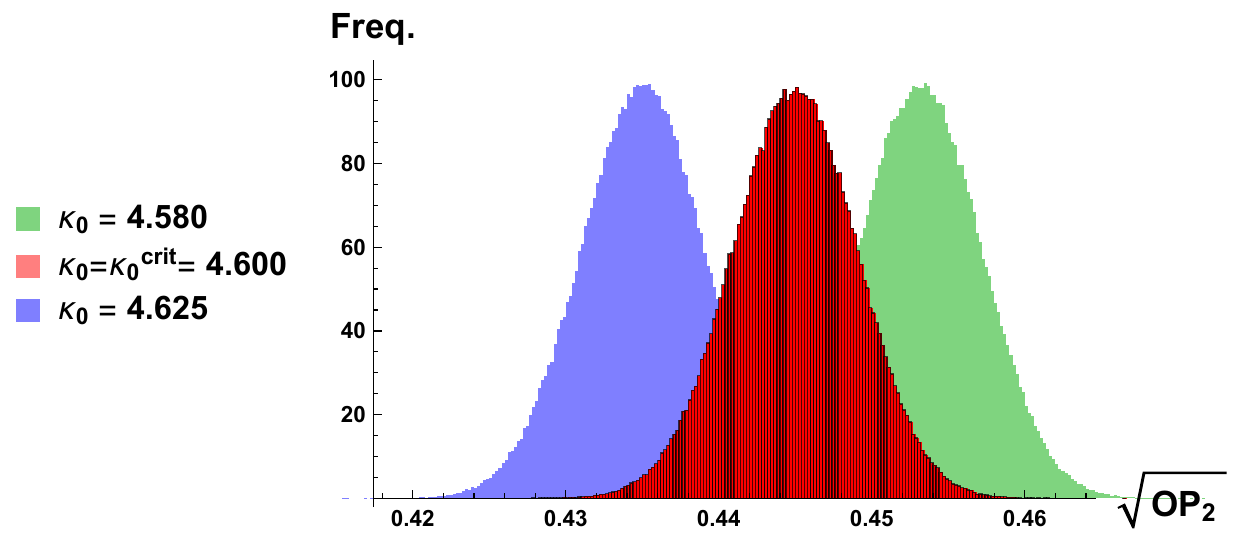}}
  \scalebox{.65}{\includegraphics{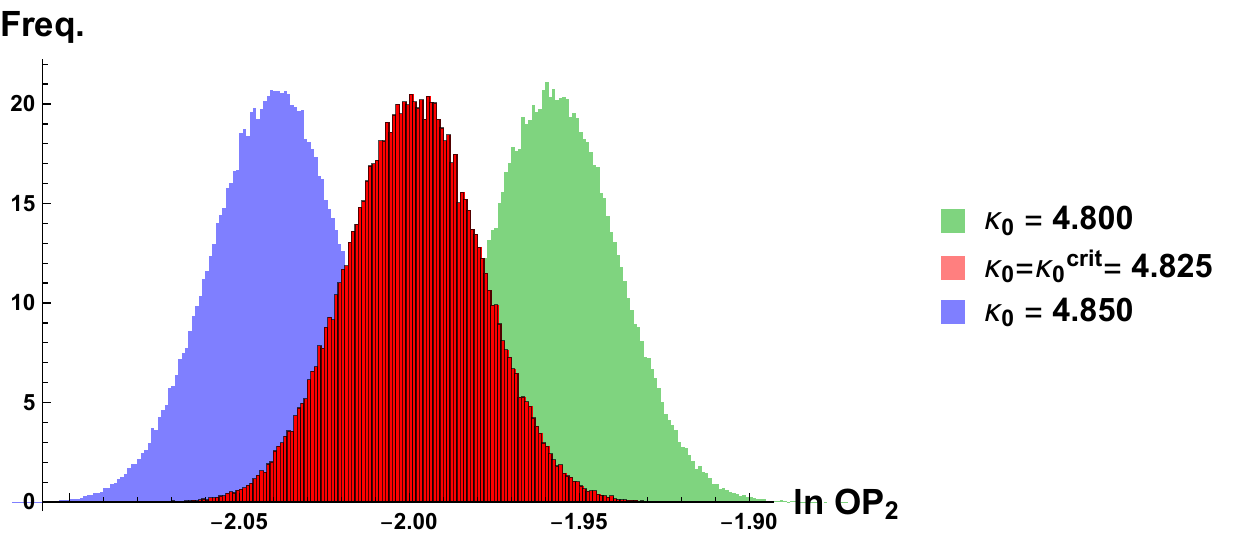}}
  \caption{\small Exemplary histograms of the (functions of) order parameter $OP_2\equiv {N_{3,2}/N_{4,1}}$ at pseudo-critical $\kappa_{0}^{crit}$ values  with fixed $N_{4}=200k$ and $T=40$ time slices. The left chart presents the histograms of $\sqrt{OP_2}$ measured in the vicinity of  $\kappa_{0}^{crit}= 4.60 $ based on the position of the susceptibility $\chi_{\sqrt{OP_2}}$ peak (Fig. \ref{FigT61}, left), and the right chart shows the histograms of $\ln{OP_2}$ measured in the vicinity of  $\kappa_{0}^{crit} = 4.825 $ based on the position of the Binder cumulant $B_{{OP_2}}$ minimum (Fig. \ref{FigT61}, right). The  histograms plotted in red are exactly at the transition points, while the green / blue data are for histograms measured for slightly lower / higher value of $\kappa_0$ than the critical value. The red, green and blue histograms overlap showing that the transition is  smooth, allowing no space for the existence of any more than one state at the transition point.}
\label{FigT63}
\end{figure}

\end{subsubsection}

\end{subsection}


\begin{subsection}{Impact of the volume fixing method}\label{J2}

In this section we will investigate what impact, if any, the choice of the volume fixing method has on the properties of the $A$-$C$ transition in toroidal CDT. As before, we will focus especially on the Monte Carlo time history analysis of histograms measured at the transition points. It is known from \cite{Ambjorn:2016mnn} that 
the choice of volume fixing can affect the occurrence of metastable states at the $B-C_b$ transition.
Fixing $N_4$ led to a Monte Carlo time history where configurations could jump between phase $B$ configurations
and phase $C_b$ configurations. If we fixed $N_{4,1}$ we would see no such jumps. Thus naively one could 
be led to the conclusion that a $N_4$ volume fixing was associated with a first order transition while a 
$N_{4,1}$ volume fixing was associated with a  second order transition. Closer analysis revealed that 
both volume fixings were associated with second order transitions, although it indeed is somewhat unusual
that one observes metastable states (which weakens for larger volumes) for a second order transition.
In the case of spherical spatial topology we have seen clear double peaks and other characteristics 
of a first order $A-C$ transition, as reported above, both in the case where $N_4$ was fixed and in the case
where $N_{4,1}$ was fixed.  Here we are now facing the opposite situation. In the case of the torus 
with $N_{4,1}$ fixed we have presented strong evidence above that the transition is still first order, but 
we saw no evidence of double peaks and metastable states at the transition point. Inspired by the 
situation at the $B-C_b$ transition we now want to fix $N_4$ instead of $N_{4,1}$ and see if that 
triggers the appearance of double peaks and metastable states.

\begin{subsubsection}{Changing global volume fixing: fixed $N_4=160k$ ($T=4$)}\label{N4fixedTorus}

We start by keeping the number of time slices fixed at $T=4$ and we change the volume fixing method to that of the total number of four-simplices fixed at  $N_4=160k$. We again  locate the pseudo-critical $\kappa_0^{crit}$ values by either looking at the susceptibility $\chi_{\sqrt{OP_2}}$ maximum or at the Binder cumulant $B_{OP_2}$ minimum, as shown in Fig.~\ref{FigT21}. The Monte Carlo time history of $OP_2\equiv N_{3,2}/N_{4,1}$ at the transition points located using both methods is shown in Fig.~\ref{FigT22} and the histograms measured at the transition  points $\kappa_0^{crit}$ and the neighbouring $\kappa_0$ values are presented in Fig.~\ref{FigT23}. We also analysed the histograms  at all other measured $\kappa_0$ data points. In each case a single Gaussian-like behaviour was observed and the evolution of the $OP_2$ parameter was smooth without exhibiting any metastable state switching. 

\begin{figure}[H]
  \centering
  \scalebox{.6}{\includegraphics{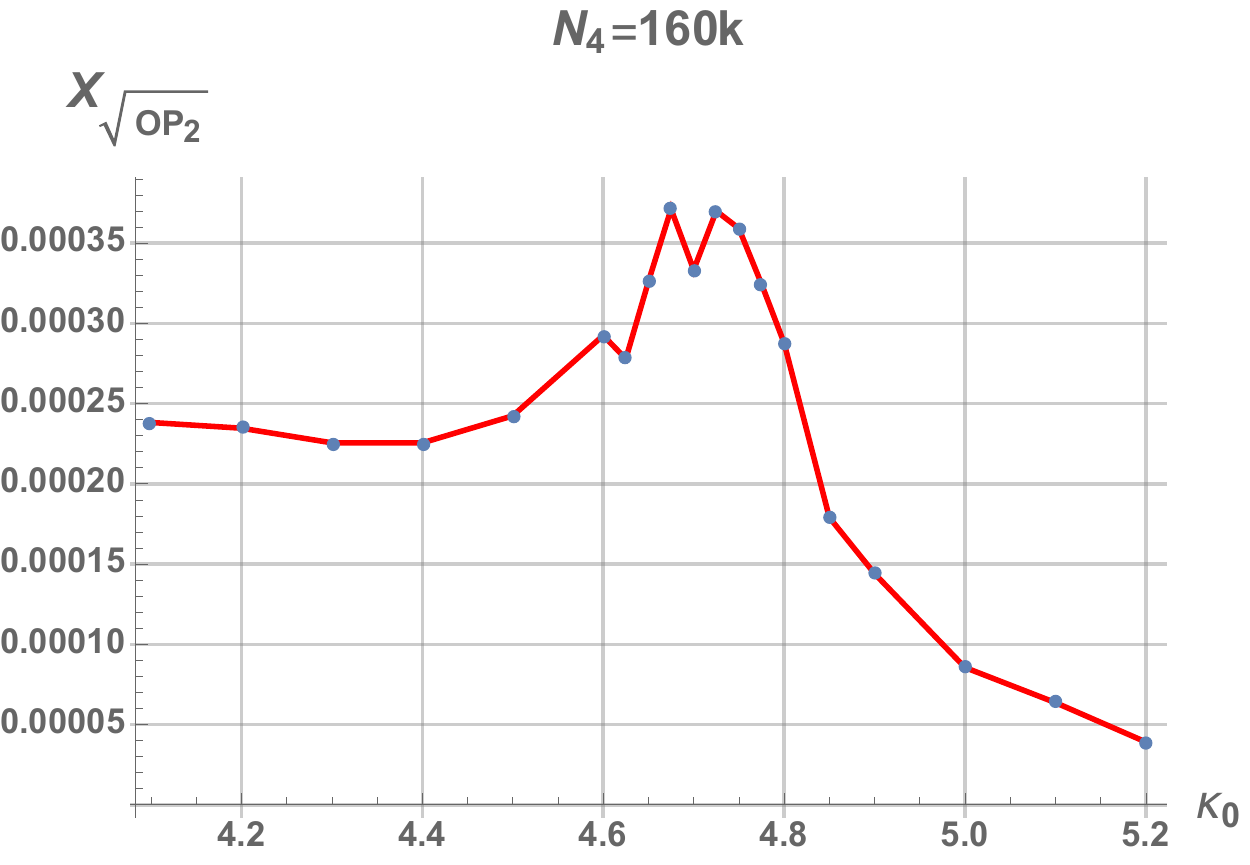}}
  \scalebox{.6}{\includegraphics{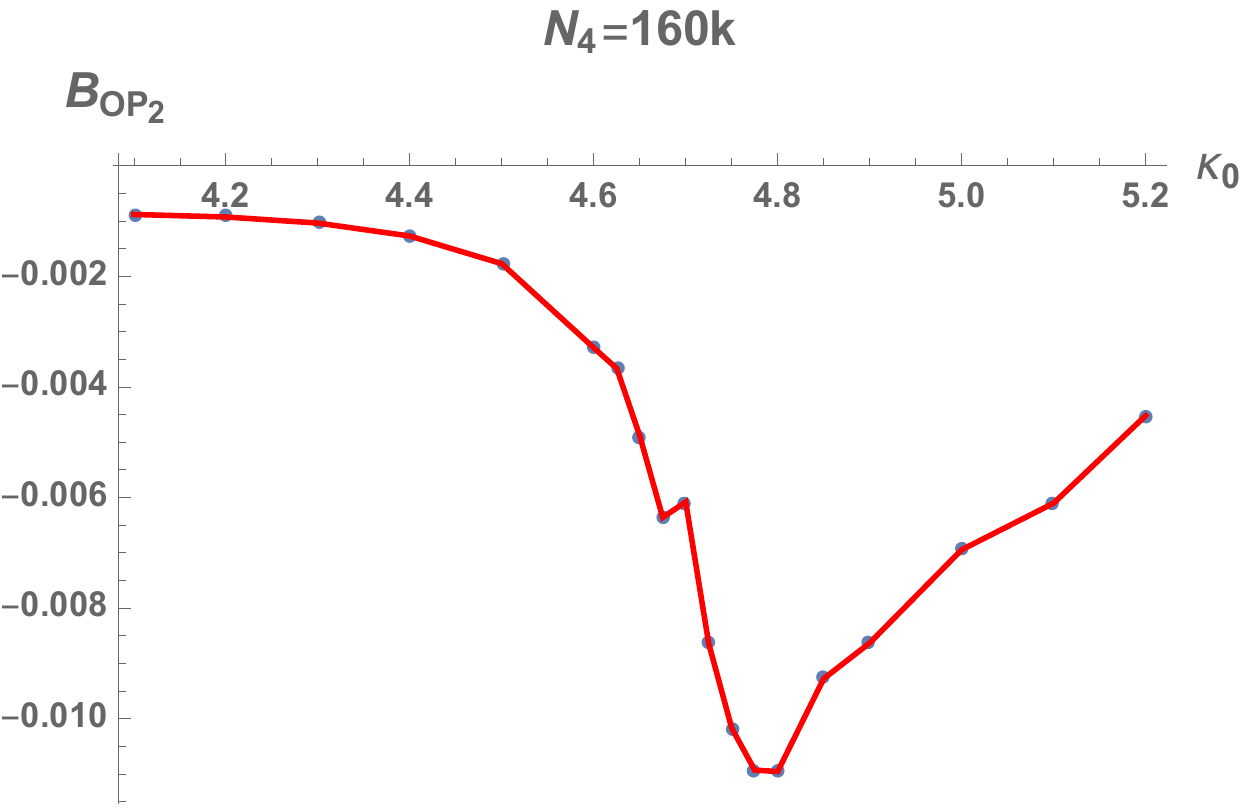}}
\caption{\small The susceptibility $\chi_{\sqrt{OP_2}}\approx\frac{\chi_{OP_2}}{\langle OP_2\rangle} $ as defined by Eq.~(\ref{susc})  (left chart) and the Binder cumulant $B_{OP_2}$ as defined by Eq.~(\ref{binder})  (right chart)
as a function of $\kappa_{0}$ for fixed $N_{4}=160k$ and $T=4$ time slices. 
}
\label{FigT21}
\end{figure}

\begin{figure}[H]
  \centering
  \scalebox{.6}{\includegraphics{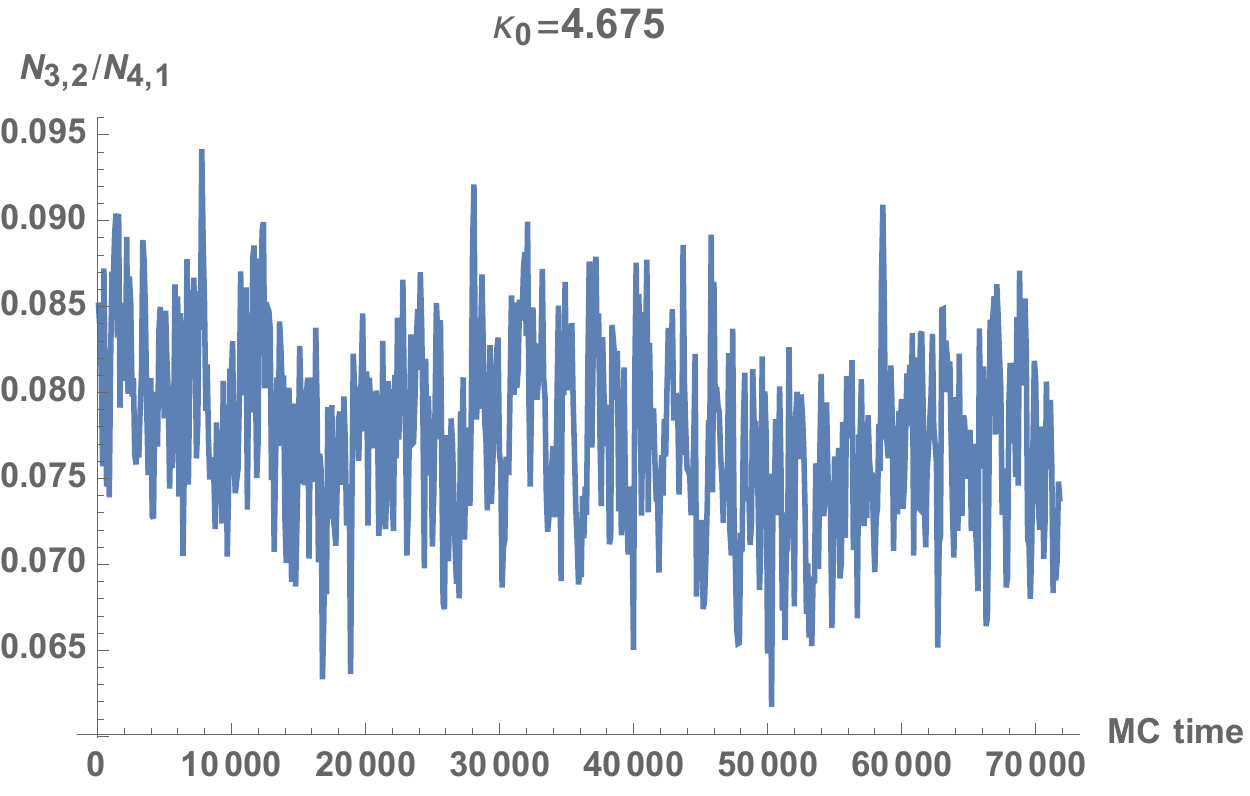}}
  \scalebox{.6}{\includegraphics{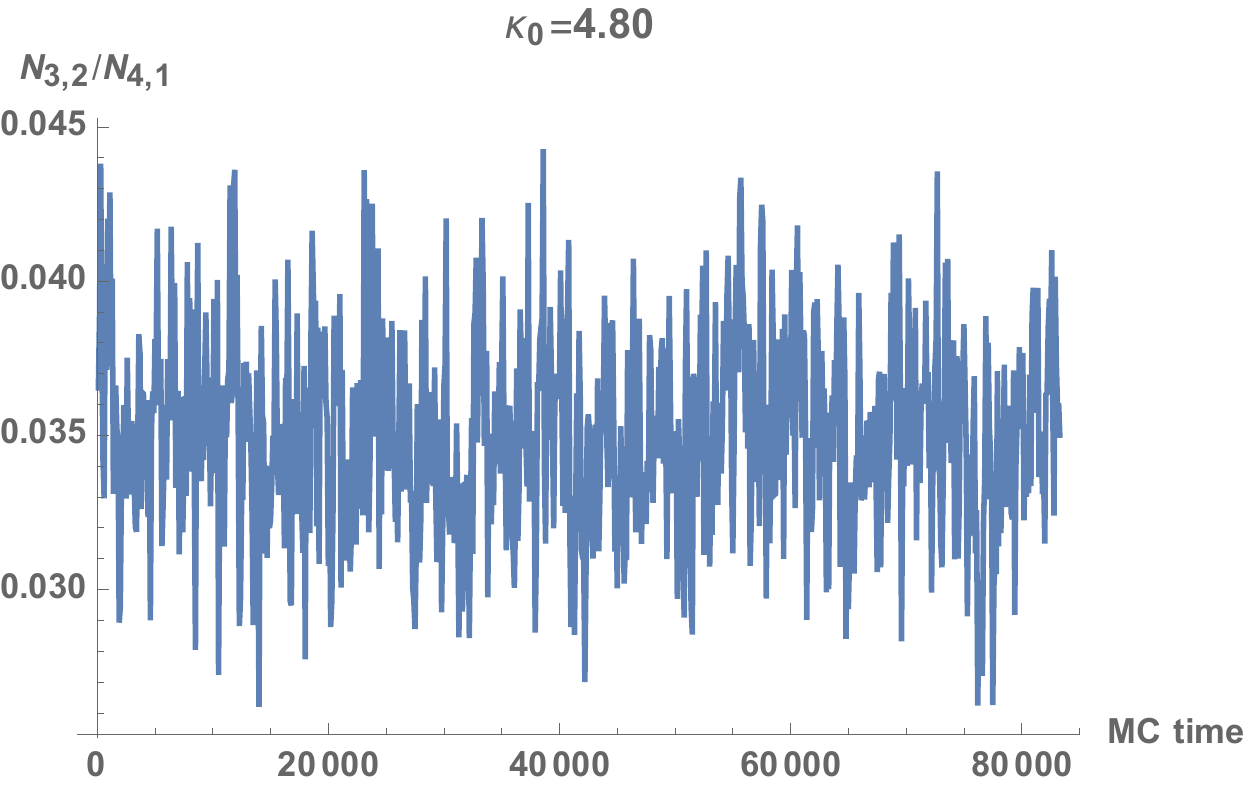}}
\caption{\small Monte Carlo time history of $OP_2\equiv N_{3,2}/N_{4,1}$ at pseudo-critical points measured using the susceptibility method: $\kappa_0 = 4.675$ (left chart) and using the Binder cumulant method: $\kappa_0 = 4.80$ (right chart) for fixed $N_{4}=160k$ and $T=4$ time slices. 
}
\label{FigT22}
\end{figure}

\begin{figure}[H]
  \centering
  \scalebox{.65}{\includegraphics{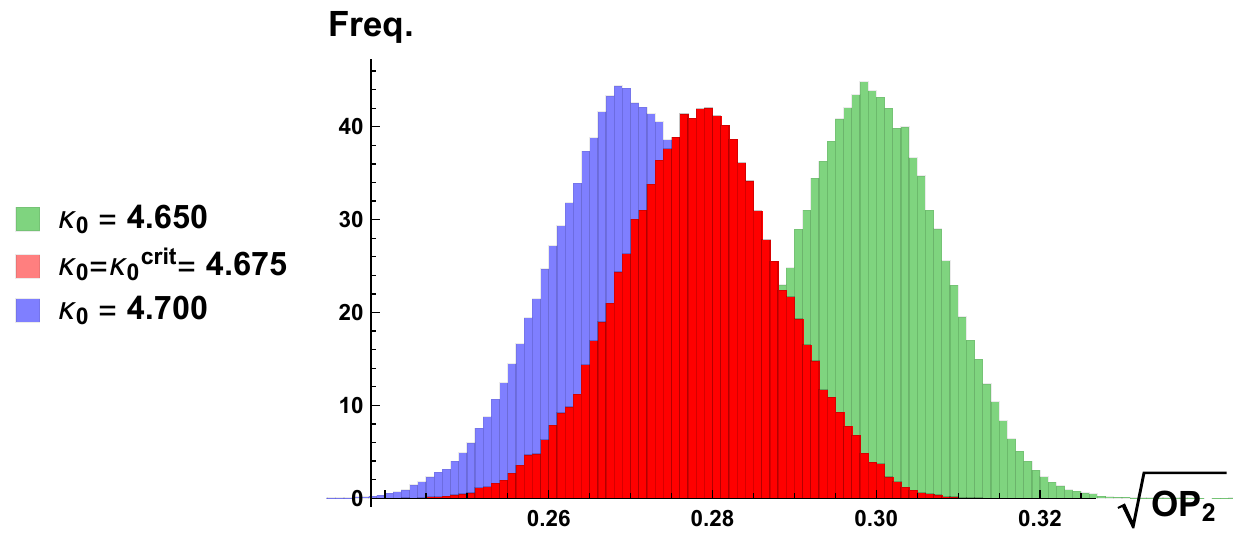}}
  \scalebox{.65}{\includegraphics{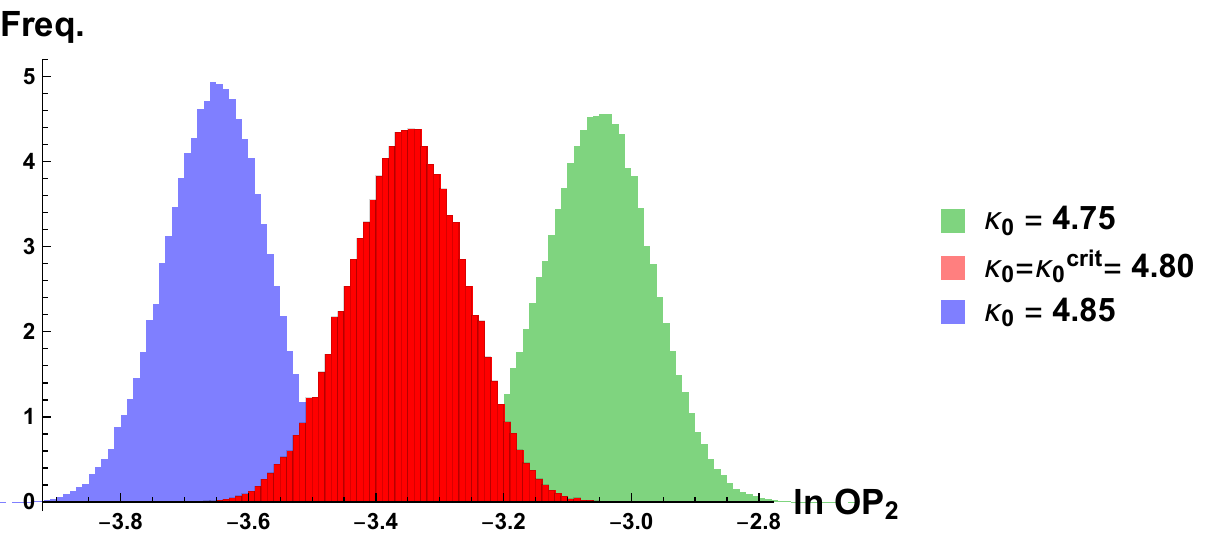}}
  \caption{\small Exemplary histograms of the (functions of) order parameter $OP_2\equiv {N_{3,2}/N_{4,1}}$ at pseudo-critical $\kappa_{0}^{crit}$ values  with fixed $N_{4}=160k$ and $T=4$ time slices. The left chart presents the histograms of $\sqrt{OP_2}$ measured in the vicinity of  $\kappa_{0}^{crit}= 4.675 $ based on the position of the susceptibility $\chi_{\sqrt{OP_2}}$ peak (Fig. \ref{FigT21}, left), and the right chart shows the histograms of $\ln{OP_2}$ measured in the vicinity of  $\kappa_{0}^{crit} = 4.8 $ based on the position of the Binder cumulant $B_{{OP_2}}$ minimum (Fig. \ref{FigT21}, right). The  histograms plotted in red are exactly at the transition points, while the green / blue data are for histograms measured for slightly lower / higher value of $\kappa_0$ than the critical value. The red, green and blue histograms overlap showing that the transition is  smooth, allowing no space for the existence of any more than one state at the transition point.}
\label{FigT23}
\end{figure}

\end{subsubsection}

\begin{subsubsection}{Adding a local volume fixing term ($T=40$)}

{So far we were not able to observe any jumping between metastable states for the  $A-C$ transition in 
the case of toroidal spatial topology, while such jumping is clearly visible in the spherical topology case.}
The  obvious difference between the two topologies lies in the shape of the CDT spatial volume profile in phase $C$, which is non-trivial, i.e. $\langle n_t \rangle \propto \cos^3(t/t_0)$, in the spherical case and is trivial, i.e. $\langle n_t \rangle=const.$, in the toroidal case (see Fig. \ref{FigT300}). One reason that we have not been able to observe 
distinctly different metastable states at the transition point could be that the dominant  ``semiclassical'' configuration
in phase $C$ in the case of toroidal spatial topology looks too similar to configurations in phase $A$. So, in order 
to try to force the configurations in phase $C$ to look more like the non-trivial, dominant configuration
in the case of spherical topology, we simply impose a non-trivial spatial volume $n_t$ dependence in the toroidal case. 
As discussed in detail in Ref. \cite{Ambjorn:2017ogo} this can be done by performing numerical MC simulations with a nontrivial local volume-fixing term:
\beql{localVolFixing}
\delta V^{local} = \epsilon \left[ (n_1 - \hat n_1)^2 + (n_h - \hat n_h)^2 \right] \ ,
\eeq 
which makes the spatial volume of two chosen slices, namely $n_1$ and $n_h$, fluctuate around $ \hat n_1$ and $\hat n_h$, respectively.  Due to the imposed  time-periodic boundary conditions  it is convenient to choose $h=T/2+1$. In the following we will describe the results 
obtained for fixed $N_{4,1}=160k$, $ \hat n_1 = 1k$, $\hat n_h=7k$ and $T=40$ time slices, resulting in $h=T/2+1=21$.
We can say that if the toroidal system does not at all have a tendency to form a non-trivial spatial distribution, we impose 
a ``boundary'' condition which should encourage it to form such a distribution (much like using a boundary condition
for a spin system which encourage the spin to point up rather than down).
However, as seen in  Fig. \ref{FigT30}, it only takes a few steps in the time direction for the configurations
to get back to the constant configuration shown in Fig.\ \ref{FigT300}.

\begin{figure}[H]
  \centering
  \scalebox{.55}{\includegraphics{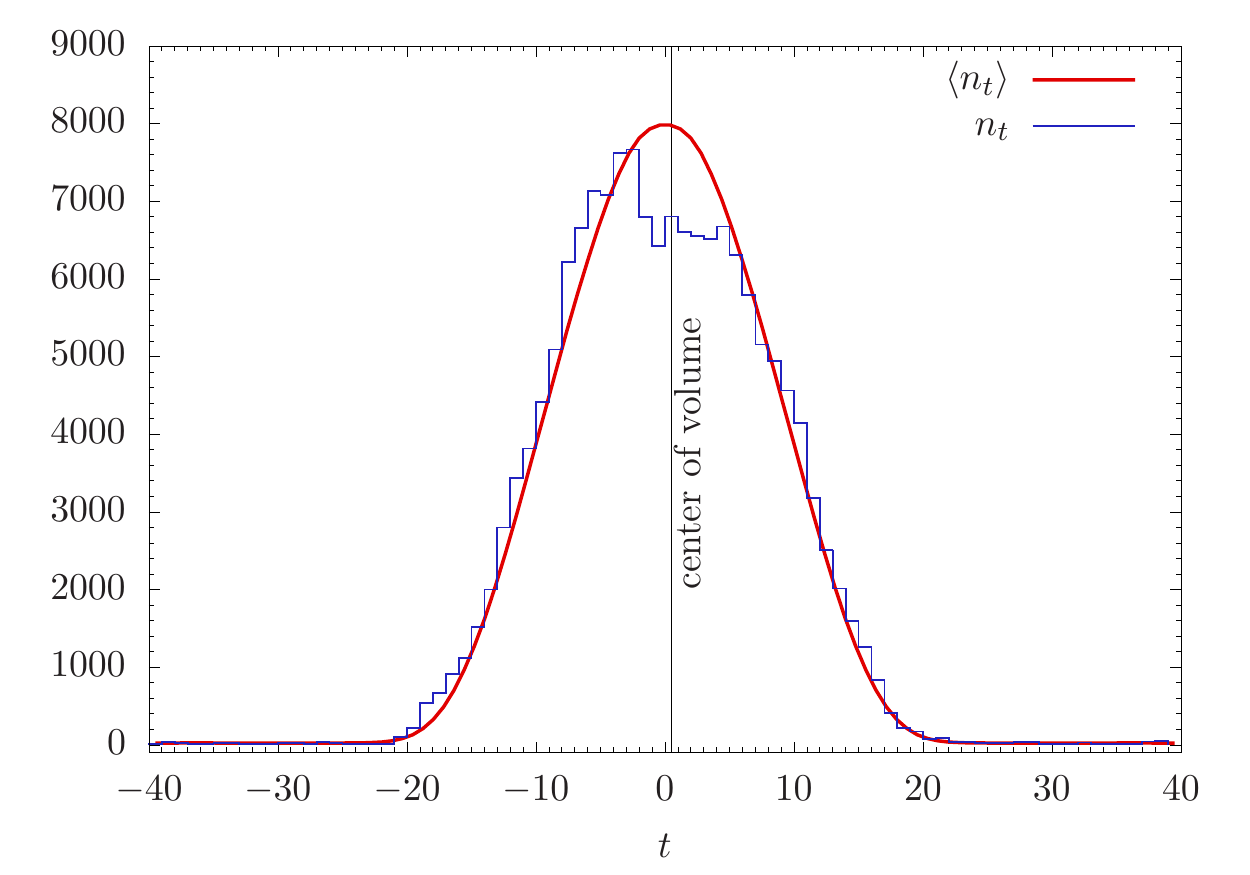}}
  \scalebox{.55}{\includegraphics{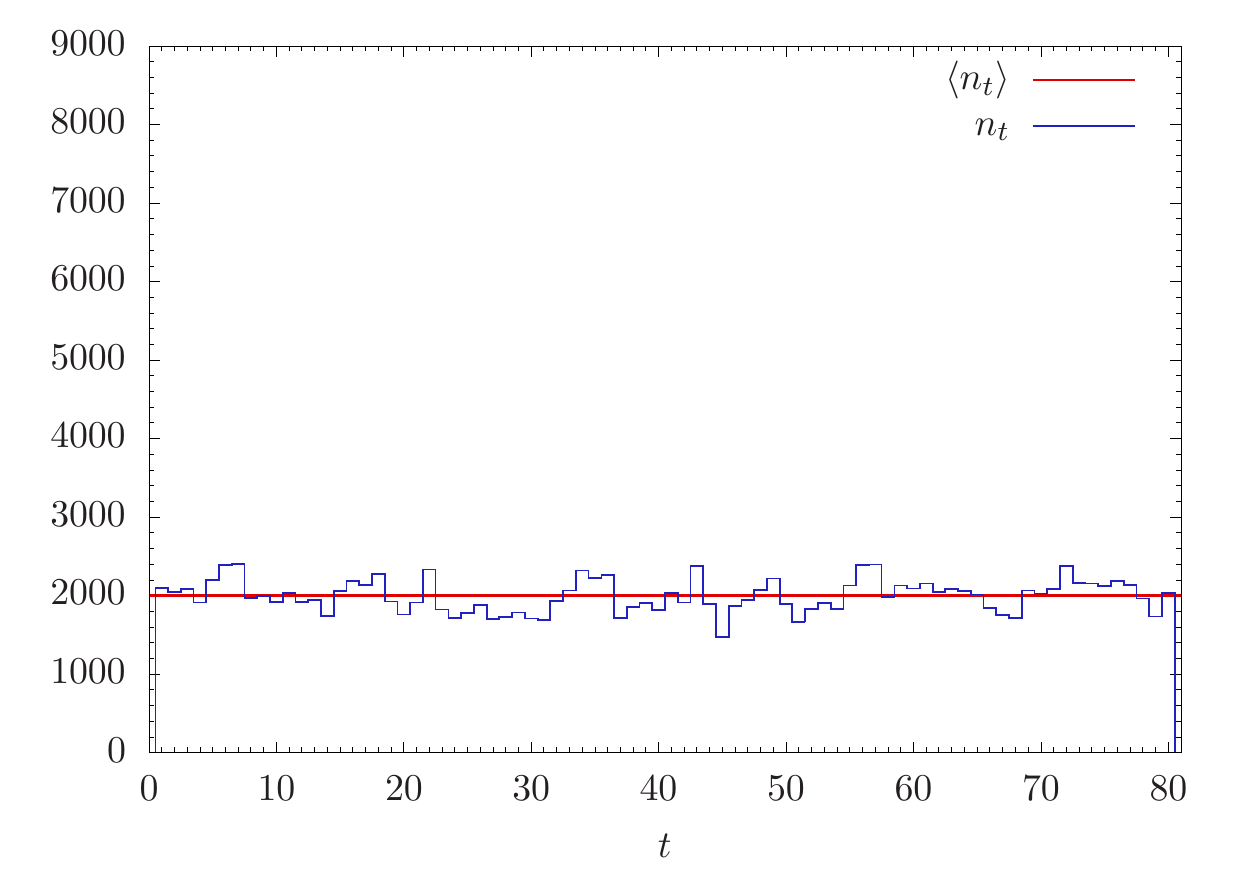}}
  \caption{\small The average (red) and typical (blue) volume profiles for the spherical spatial topology (left chart) and for the toroidal spatial topology (right chart) deep inside the phase $C$ ($\kappa_0=2.2$, $\Delta=0.6$) in CDT with $N_{4,1}=160k$ volume fixed and $T=80$ time slices.
  }
\label{FigT300}
\end{figure}

\begin{figure}[H]
  \centering
  \scalebox{.7}{\includegraphics{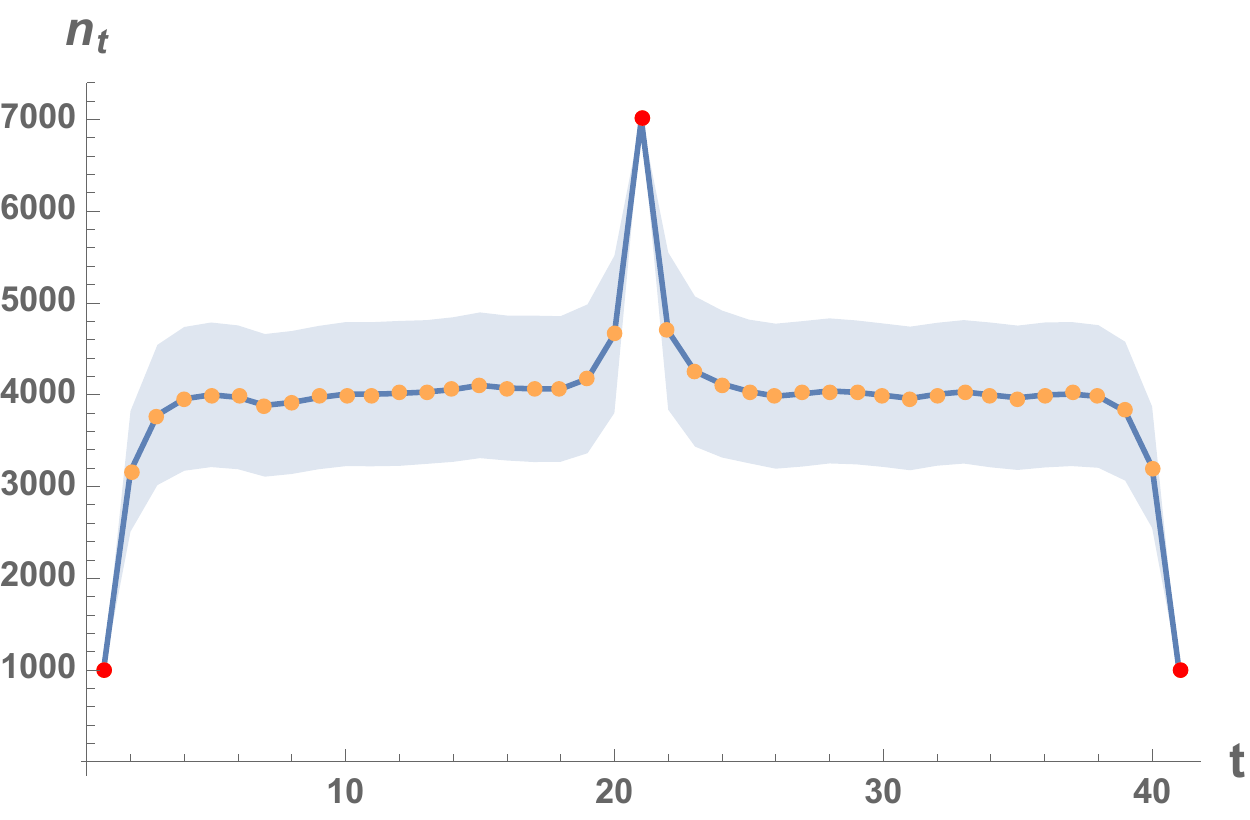}}
  \caption{\small The spatial volume profile measured for bare coupling constants $\kappa_0=4.54$, $\Delta=0.6$ and time period $T=40$ in toroidal CDT with fixed $N_{4,1}=160k$ and also local volume fixing \rf{localVolFixing}. Red dots denote time slices with fixed local volume $\hat n_1=1k$ and $\hat n_{h}=7k$, $h=T/2+1=21$. The average spatial volume $\langle n_t \rangle$ is plotted as orange points linked by a blue solid line, the amplitude of fluctuations $\langle n_t \rangle\pm \Delta n_t$,  $\Delta n_t = \sqrt {\langle \left( n_t -  \langle n_t \rangle\right)^2 \rangle }$ is shown as the shaded region.}
\label{FigT30}
\end{figure}

We again  repeat the analysis of Subsection \ref{N4fixedTorus}  by locating the pseudo-critical $\kappa_0^{crit}$ values looking for susceptibility $\chi_{\sqrt{OP_2}}$ peaks (see Fig.~\ref{FigT31}). The Monte Carlo time history of $OP_2\equiv N_{3,2}/N_{4,1}$ at the transition point  is shown in Fig.~\ref{FigT32} and the $\sqrt{OP_2}$ histogram is presented in Fig.~\ref{FigT33}. At the transition point, and also for all other measured $\kappa_0$ data points, one observes a single Gaussian-like behaviour of the $OP_2$ parameter. In view of the short range distortions of the volume profile 
shown in Fig.\ \ref{FigT30} this is not surprising. However, we also have to conclude that in the case of toroidal spatial
topology we have no natural geometric interpretation of the first order transition. It might be that the systems
we consider are simply too small to be able to observe $OP_2$ histograms with two distinct peaks, as we 
have already discussed\footnote{{Unfortunately, we were not able to observe the metastable state switching even for our biggest systems consisting of $N_{4,1}= 500k$ simplices, and, due to the exponentially growing thermalization time, we cannot go much further as regards the simulated system sizes. }  }    
, or it might be that this phase transition is atypical for a 
first order transition (like the $B-C_b$ transition was atypical for a second order transition).

\begin{figure}[H]
  \centering
  \scalebox{.7}{\includegraphics{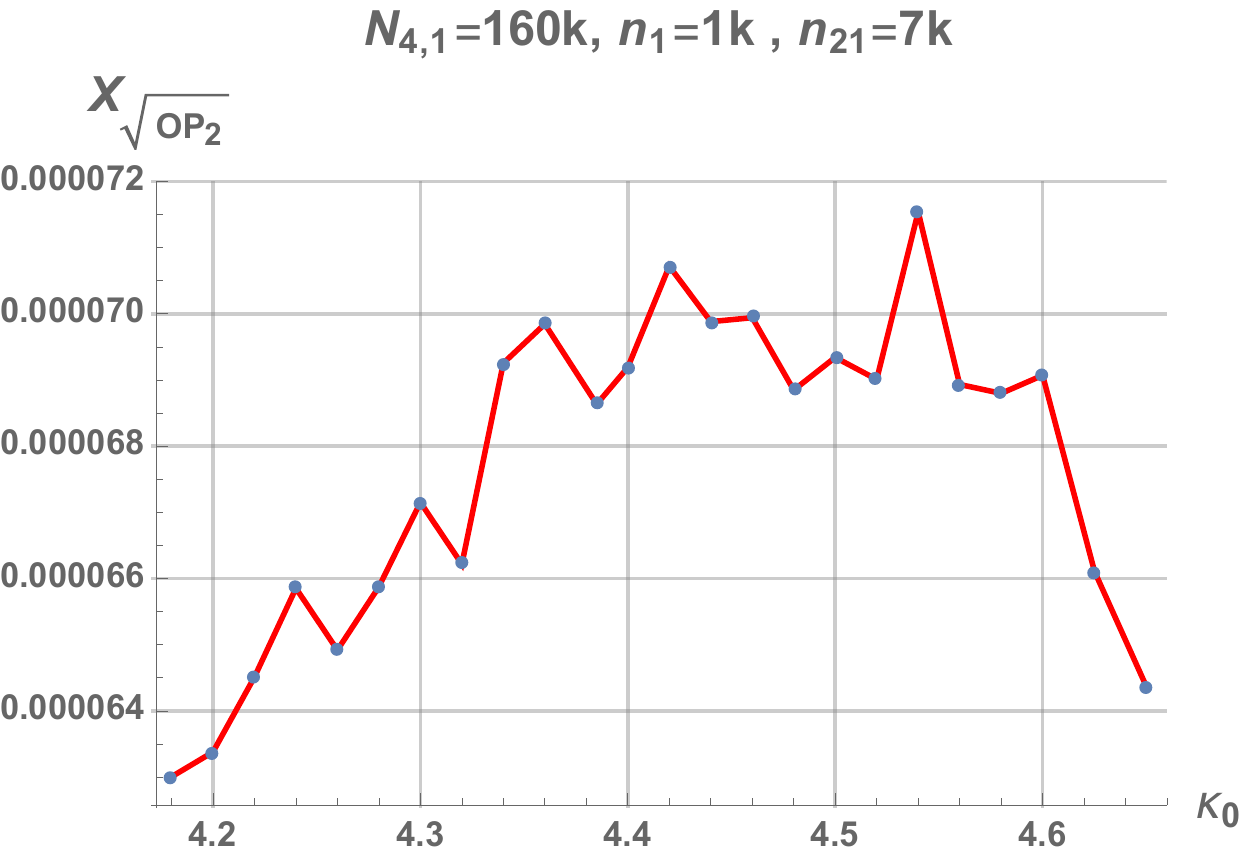}}
\caption{\small The susceptibility $\chi_{\sqrt{OP_2}}\approx\frac{\chi_{OP_2}}{\langle OP_2\rangle} $ as defined by Eq.~(\ref{susc}) 
as a function of $\kappa_{0}$ in toroidal CDT with $T=40$ time slices and with fixed $N_{4,1}=160k$ and also local volume fixing $n_1=1k$ and $n_{21}=7k$.
}
\label{FigT31}
\end{figure}

\begin{figure}[H]
  \centering
  \scalebox{.7}{\includegraphics{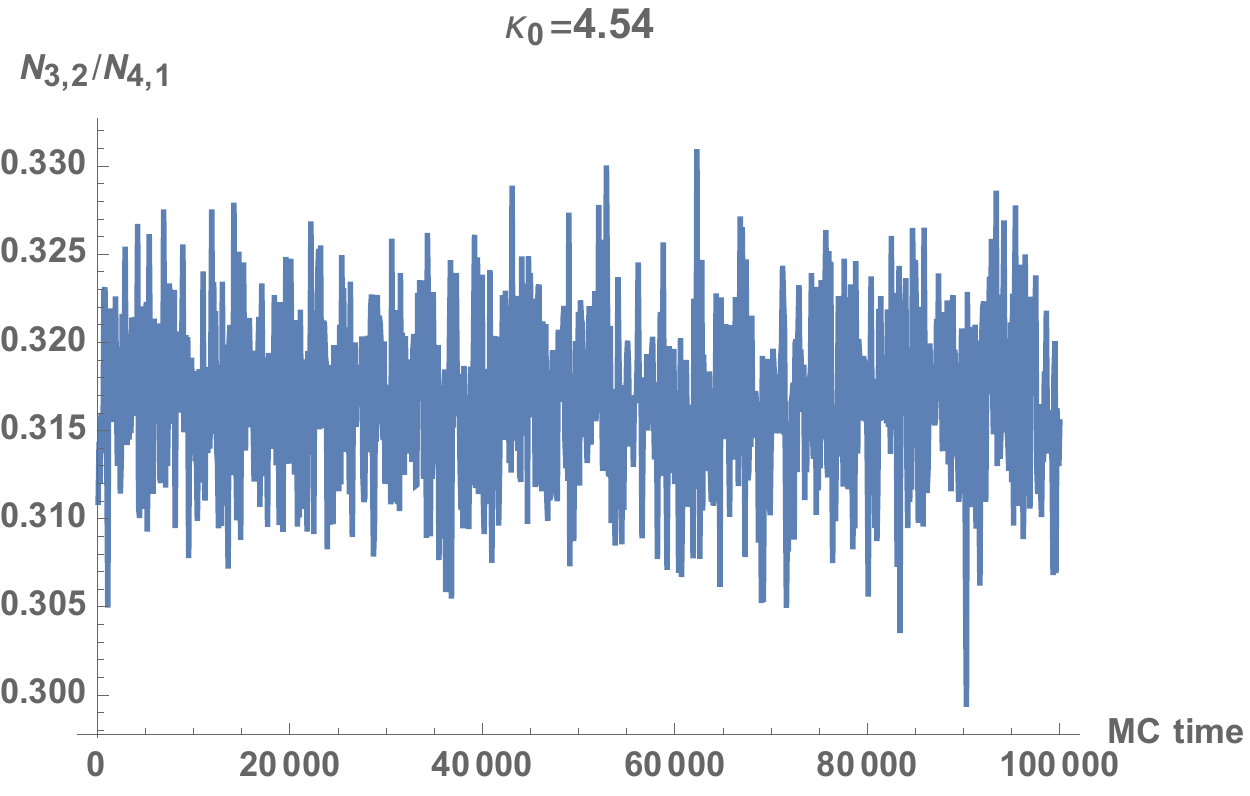}}
\caption{\small Monte Carlo time history of $OP_2\equiv N_{3,2}/N_{4,1}$ at pseudo-critical point $\kappa_0^{crit} = 4.54$ measured using the susceptibility peak method (see Fig. \ref{FigT31}) 
in toroidal CDT with $T=40$ time slices and with fixed $N_{4,1}=160k$ and also local volume fixing $n_1=1k$ and $n_{21}=7k$.
}
\label{FigT32}
\end{figure}

\begin{figure}[H]
  \centering
  \scalebox{.9}{\includegraphics{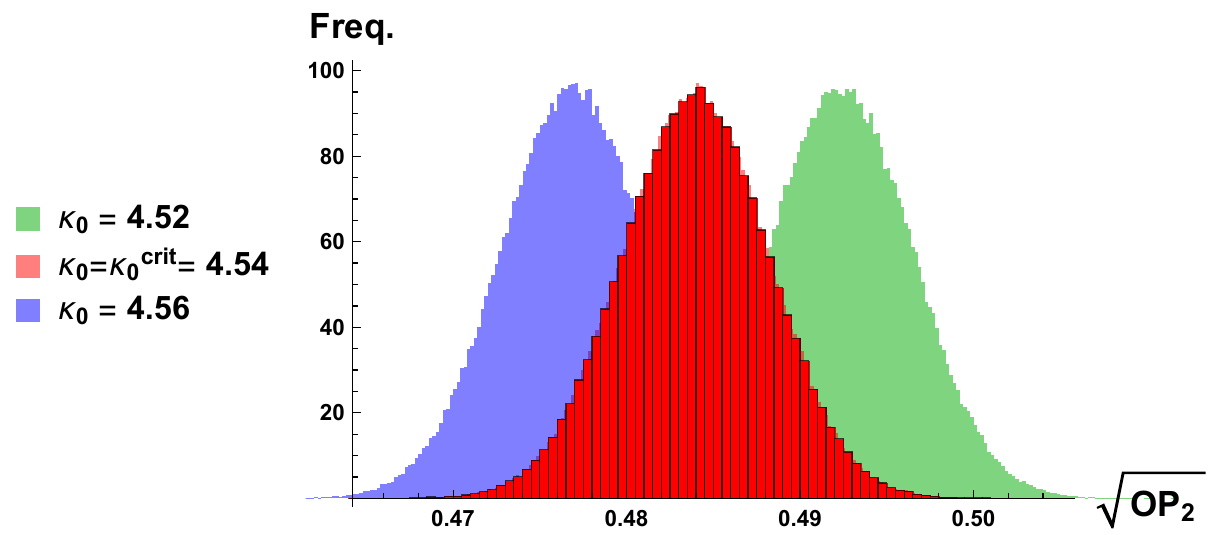}}
  \caption{\small Exemplary histograms of  $\sqrt{OP_2}\equiv \sqrt{{N_{3,2}/N_{4,1}}}$ at pseudo-critical $\kappa_{0}^{crit}=4.54$ value in toroidal CDT with $T=40$ time slices and with fixed $N_{4,1}=160k$ and also local volume fixing $n_1=1k$ and $n_{21}=7k$. 
The  histogram plotted in red is exactly at the transition point  measured using the susceptibility peak method (see Fig. \ref{FigT31}), while the green / blue data are for histograms measured for a slightly lower / higher value of $\kappa_0$ than the critical value. The red, green and blue histograms overlap showing that the transition is  smooth, allowing no space for the existence of any more than one state at the transition point.}
\label{FigT33}
\end{figure}

\end{subsubsection}

\end{subsection}

\begin{subsection}{Summary of the toroidal topology}
{Summing up this part, using the $OP_2\equiv N_{3,2}/N_{4,1}$ order parameter we have analysed  in detail the $A$-$C$ transition for a system with toroidal spatial topology, $T=4$ time slices and $N_{4,1}$ volume fixing. We have  confirmed two of  three signatures of 
the first order transition (see Table \ref{Table1}), namely the shift exponent $\gamma $ consistent with one, and scaling of the Binder cumulant $B_{OP_2}^{min}$,  which diverges from zero when  the lattice volume is increased. Unfortunately, we were not able to observe $OP_2$ histograms with two distinct peaks.  We have then checked the impact of the number of time slices used in numerical Monte Carlo simulations and we found universal scaling relations of $\langle OP_2 \rangle$ and $\chi_{OP_2}$ with the lattice volume $N_{4,1}$ and $T$, suggesting that the $OP_2$ can be modelled as an average of  $T$ statistically independent identical random variables $\widetilde{OP}_2(n_t)$, dependent only on $ \langle n_t \rangle = N_{4,1} /T $. This kind of behaviour is indeed expected inside phase $A$, where  one can show that triangulations  consist of $T$ independent (or at least uncorrelated) time slices, but, surprisingly, it also seems to hold (at least not too deep)  inside phase $C$, where various time slices are correlated \cite{Ambjorn:2016fbd}. The study led to the scaling relation of Eq. \rf{ChiN32Scaling} which strongly supports the first order nature of the $A-C$ transition.
Finally, we have investigated the impact of  the volume fixing method,   either by changing the global volume fixing to that of $N_4$ fixed, or by adding a local volume fixing term yielding a non-trivial spatial volume dependence. Unfortunately, in none of the cases were we able to observe the metastable state switching at the $A-C$ transition points during our Monte Carlo runs. We suspect that this is caused by much larger finite size effects in CDT with the toroidal spatial topology, compared to the spherical topology case.}

\end{subsection}

\end{section}

\begin{section}{Discussion and conclusions}\label{Discussion}

{We have investigated the $A-C$ phase transition in CDT with spherical and toroidal spatial topology.}
{For the spherical topology, }
fixing the number of $N_{4,1}$ simplices and keeping the number of time slices at $T=80$ we have determined the pseudo-critical $\kappa_{0}$ value for 8 different lattice volumes, finding a shift exponent of {$\gamma=1.16 \pm 0.07$} that strongly supports the first order nature of the $A$-$C$ transition. This finding is further supported by calculations of the Binder cumulant, which is found to move away from 
zero with increasing lattice volume, which also suggests a first order transition.
In the case of spherical topology, a double 
{peaked} histogram appears at pseudo-critical transition points regardless of the particular volume fixing method, namely it does not seem to matter whether one fixes the total number of simplices $N_{4}$ or just the number of $N_{4,1}$ simplices. Varying the number of time slices between $T=80$ and $T=4$ at the pseudo-critical $\kappa_{0}$ value 
{has a noticeable impact}. Namely, 
{for $T=4$ we can not verify the presence of a double peak structure in the histogram data up to a resolution of three 
decimal places in $\kappa_0$. However, an analysis of the Monte Carlo time evolution either side of the putative 
transition suggests a double peak structure is likely to emerge for a greater resolution of $\kappa_0$. 
This is the 
consequence of the very sharp separation of the states at both sides of the transition, further supporting the first order 
nature of the $A-C$ transition, consistent with the earlier finding of ref.\ \cite{Ambjorn:2012ij}.}

{For toroidal topology, fixing the number of $N_{4,1}$ simplices and keeping the number of time slices at $T=4$ we have determined the pseudo-critical $\kappa_{0}$ value for 11 different lattice volumes, finding a shift exponent of $\gamma=1.22 \pm 0.08$, also consistent with the first order transition. We have also investigated  the Binder cumulant, which is found to move away from 
zero with increasing lattice volume, also supporting the first order nature of the transition. We were not able to observe double peaks in histograms of the order parameter measured at the transition points. Varying the number of time slices or changing the volume fixing method also does not lead to the metastable state switching during Monte Carlo simulations at the transition points in toroidal CDT. A detailed analysis of the scaling of susceptibility at the transition points leads to the discovery of the universal scaling relation \rf{ChiN32Scaling}:
$$
\chi^{crit}_{N_{3,2}} = {\chi^*}\cdot {N_{4,1}} \ , \ \chi^*=const.\ ,
$$
independent of the number of time slices $T$ and valid for any lattice volume $N_{4,1}$. Such a scaling, observed for finite lattices, typically translates into delta-like singularities of susceptibility in the infinite volume limit, thus strongly supporting the first order nature of the $A-C$ transition in toroidal CDT.}

{Although the $A-C$ transition seems to be much smoother for toroidal CDT, and therefore one couldn't observe the metastable state separation at the transition points, this might be attributed to much larger finite size effects for toroidal topology compared to  spherical topology. Therefore, all our results strongly suggest that the volume fixing method, the number of time slices used in the Monte Carlo simulations and the chosen topology of  spatial hypersurfaces of equal global proper time coordinates do not have an impact on the nature of the $A-C$ transition. Thus the behaviour is very universal.}

\end{section}


\section*{Acknowledgements}

JGS acknowledges support from the grant UMO-2016/23/ST2/00289 from the National Science centre, Poland.
JA and DC acknowledge support from the Danish Research Council grant {\it Quantum Geometry}. AG acknowledges 
support by the National Science Centre, Poland, under grant no. 2015/17/D/ST2/03479.



\end{document}